\documentclass[12pt]{article} 
\usepackage{amsfonts}
\usepackage{amsmath}
\usepackage{amssymb}
\usepackage{amsthm}
\usepackage{array}
\usepackage[nosort]{cite}
\usepackage{caption}
\usepackage{color}
\usepackage{dsfont}
\usepackage{float}      
\usepackage{framed}
\usepackage{graphicx}
\usepackage{indentfirst}
\usepackage{mathrsfs}
\usepackage{multirow}
\usepackage{setspace}
\usepackage{subdepth}
\usepackage{titlesec}
\usepackage[dotinlabels]{titletoc}
\usepackage{wrapfig}
\usepackage{layout}
\usepackage{url}
\usepackage{slashed}
\usepackage{authblk}
\usepackage{tikz}
\usepackage{diagbox}
\usepackage[hyphens]{xurl}
\usepackage{lipsum}
\usepackage{etoolbox}
\usepackage{xcolor}
\usepackage{import}
\usepackage{mathtools}
\usepackage{example}
\usepackage{parskip}
\usepackage{subcaption}
\usepackage{chngcntr}
\usepackage[section]{placeins}
\usepackage{multicol}

\usepackage{hyperref}
\hypersetup{colorlinks=true}
\hypersetup{linkcolor=black}
\hypersetup{citecolor=red}
\hypersetup{urlcolor=blue}

\numberwithin{equation}{section}

\usepackage[left=2.5cm,right=2.5cm,top=2.5cm,bottom=3cm]{geometry}
\def\tilde{\widetilde}
\def\t{\tilde}

\def\bar{\overline}
\def\b{\bar}

\def\half{{1 \over 2}}
\def\d{\partial}

\def\ep{\varepsilon}
\def\1{{\mathds 1}}

\newcommand{\Z}{{\mathbb Z}}

\newcommand{\R}{{\mathbb R}}

\def\SL{{\mathscr L}}

\def\CN{{\mathcal N}}
\def\CO{{\mathcal O}}

\def\CV{{\mathcal V}}

\DeclareFontShape{OT1}{cmr}{mx}{n}%
    {<->cmr10}{}
\newcommand{\mytitlefont}{\fontseries{mx}\selectfont}
\DeclareMathAlphabet{\titlemath}{OT1}{cmr}{mx}{n}

\begin{document}



\begin{titlepage}

\begin{center}

~\\[20pt]

{\fontsize{30pt}{0pt} \mytitlefont Phases of Giant \\[10pt] Magnetic Vortex  Strings} 

\vskip25pt

Thomas T.~Dumitrescu and Amey P.~Gaikwad

\vskip10pt

{\it Mani L.\,Bhaumik Institute for Theoretical Physics,\\[-2pt] Department of Physics and Astronomy,}\\[-2pt]
       {\it University of California, Los Angeles, CA 90095, USA}\\[4pt]

\bigskip

\end{center}

\bigskip
\bigskip

\noindent We consider Abrikosov-Nielsen–Olesen magnetic vortex strings in 3+1 dimensional Abelian Higgs models. We systematically analyze the giant vortex regime using a combination of analytic and numerical methods. In this regime the strings are infinitely long, axially symmetric, and support a large magnetic flux~$n$ along the symmetry axis in their core that causes them to spread out in the transverse directions. Extending previous observations, we show that the non-linear equations governing giant vortices can essentially be solved exactly. The solutions fall into different universality classes, reflecting the properties of the Higgs potential, that become sharply distinct phases in the large-$n$ limit. We use this understanding to shed light on the binding energies and stability of vortex strings in each universality class. 

\vfill

\begin{flushleft}
November 2025   
\end{flushleft}

\end{titlepage}


\tableofcontents

\bigskip

\section{Introduction}\label{sec:intro}

In this paper we consider magnetic vortex strings of Abrikosov-Nielsen–Olesen (ANO) type~\cite{Abrikosov:1956sx, Nielsen:1973cs} in 3+1 dimensional Abelian Higgs models (AHMs) with a single~$U(1)$ gauge field~$a_\mu$ and a single unit-charge Higgs field~$\phi$,\footnote{~Generalizations with additional charged and neutral fields (which naturally arise in supersymmetric versions of~\eqref{lagint}) will be analyzed in~\cite{DG}.}
\begin{equation}\label{lagint}
    \SL = -\frac{1}{4e^2} f_{\mu\nu} f^{\mu\nu} - |(\d_\mu - i a_\mu) \phi|^2 - V(\phi)~.
\end{equation}
Here~$V(\phi) = V(|\phi|) \geq 0$ is a gauge-invariant potential with a Higgs vacuum, where~$\phi  \neq 0$ and~$V = 0$.\footnote{~Our discussion in this paper will be entirely classical; our conventions are spelled out in section~\ref{sec:AHMs}.} The magnetic vortices that arise in this model -- which are of broad interest across many areas of physics -- are reviewed in many textbooks, see e.g.~\cite{Manton:2004tk,Shifman:2012zz, Weinberg:2012pjx}. The vortex equations that govern them (i.e.~the Euler-Lagrange equations of~\eqref{lagint}, supplemented with suitable boundary conditions) are highly non-linear, second-order PDEs, which are typically studied numerically. 

In this paper, we will analyze a regime in which the vortex equations simplify considerably and can essentially be solved exactly:
\begin{itemize}
    \item We consider vortices that are infinitely extended along the~$z$-direction.
    \item We assume axial symmetry, so that the vortices are suitably rotationally symmetric in the transverse~$xy$-plane; $r = \sqrt{x^2 + y^2}$ is the radial coordinate in that plane. 
    \item We take the magnetic flux in the~$xy$-plane to be very large,\footnote{~This conserved flux is the charge associated with the magnetic one-form symmetry of~\eqref{lagint}.}
    \begin{equation}\label{vortdef}
        n = \frac{1}{2\pi} \int (da)_{xy} \; \gg 1~.
    \end{equation}
    We will denote the tension of the lightest axially-symmetric string with magnetic flux~$n$ by~$T_n$. Below we will see that the vortices grow without bound in the transverse  radial direction~$r$ as we increase~$n$. In the large-$n$ limit we thus refer to them as giant strings or vortices (see~\cite{PhysRevLett.125.251601, Penin_2021, Cuomo:2022kio,Cuomo:2023vvd} for recent discussions of giant vortices in various contexts).
\end{itemize}
The idea that complicated non-linear equations should simplify in the limit of large charge (e.g.~large particle number) is realized in many systems. A class of solitons that share some features with the giant vortices in this paper is furnished by $Q$-balls~\cite{Coleman:1985ki}, which are comprised of interacting scalar fields and carry charge $Q$ under an ordinary zero-form global symmetry. Notably, they also become large and more tractable in the large-$Q$ limit.\footnote{~The relationship between vortices and $Q$-balls is particularly striking in 2+1 dimensions. The simplest $Q$-balls arise in the theory of a single complex scalar field with a suitable potential (i.e.~the 2+1d version of the setup analyzed in~\cite{Coleman:1985ki}). These are mapped by particle-vortex duality~\cite{Peskin:1977kp, Dasgupta:1981zz} to vortices in a 2+1d Abelian Higgs model~\eqref{lagint} with a suitable choice of potential~$V(\phi)$. Note that large $Q$-ball/vortex bound states only exist in the Type I regime of this duality, where like-charge particles/vortices attract. In 2+1 dimensions, the vorticity~\eqref{vortdef} is a 0-form charge, carried by the point-particle vortices, and the duality exchanges~$Q \leftrightarrow n$.} 

Let us sketch a simple picture of large-$n$ vortex strings that can be inferred using variational arguments~\cite{Bolognesi_2005, Bolognesi_2006} (see section~\ref{seclargen} for more detail). It is based on the idea that the string is characterized by a large transverse radius~$r_n$, subdividing it into various regions:
\begin{itemize}
    \item The core region~$0 \leq r \lesssim r_n$, where the Higgs field approximately vanishes, $\phi(r) \simeq 0$. The fact that~$\phi(r)$ must vanish as~$r \to 0$ is a standard feature of vortices (see section~\ref{sec:AHMs}), but in the large-$n$ limit it approximately vanishes throughout the core region. The reason for this is that the string must accommodate the large magnetic flux~\eqref{vortdef}. In the large-$n$ limit the magnetic field in the core is approximately constant, scaling as\footnote{~Here we limit the discussion to the $n$-scaling of various quantities of interest. See section~\ref{seclargen} for more detail.}  
    \begin{equation}\label{Bincore}
        B \sim (da)_{xy} \sim {n \over r_n^2}~.
    \end{equation}
    \item The boundary region~$r \sim r_n$ is of~$\CO(1)$ and remains fixed in the large-$n$ limit. In this region the fields smoothly transition from their values in the core to the Higgs vacuum in the exterior. 
    \item The exterior region~$r \gtrsim r_n$, where the fields are essentially in the Higgs vacuum. 
\end{itemize}
We can now estimate the core and boundary contributions to the string tension,\footnote{~Since the fields in the exterior region are approximately in the Higgs vacuum, they essentially do not contribute to the string tension.}
\begin{equation}\label{vartenint}
    T_n \sim r_n^2 \left(B^2 + V(\phi = 0)\right) + \sigma r_n~.
\end{equation}
Here the first two terms arise from the magnetic field and scalar potential in the core region, whose area is~$\pi r_n^2$, while the third term is proportional to the length~$2 \pi r_n$ of the boundary. Here~$\sigma$ is an~$\CO(1)$ surface tension arising from the variation of the fields in the boundary region. If we minimize~\eqref{vartenint}, subject to the constraint~\eqref{Bincore}, we find two qualitatively distinct scenarios:
\begin{itemize}
    \item[(B)] If~$V(\phi = 0) > 0$, then the point~$\phi = 0$ is not a vacuum. In this case we find that the string tension is dominated by the extensive core region and scales as follows,
    \begin{equation}\label{Tnbulk}
        T_n \sim r_n^2 \sim n~, \qquad n \gg 1~,
    \end{equation}
    while the surface term~$\sim \sigma r_n \sim \sqrt{n}$ is subleading. We refer to strings in this universality class as core-dominated or bulk strings. A representative example that we study in detail below is the conventional Abelian Higgs model, with quartic scalar potential
    \begin{equation}\label{v4int}
        V(\phi) = V_4(\phi) = {\lambda \over 2} (|\phi|^2 - v^2)^2~, \qquad \lambda~, v > 0~.
    \end{equation}
    An important special case arises when~$\lambda = e^2$, in which case the vortex strings in this model are BPS-saturated and satisfy a system of first-order equations. Their tension is then exactly linear in~$n$, consistent with~\eqref{Tnbulk}. 
    
    \item[(DW)] If~$V(\phi = 0) = 0$, then~$\phi = 0$ is a Coulomb vacuum that is exactly degenerate with the Higgs vacuum, where~$\phi \neq 0$. A representative example in this universality class that we analyze  below -- referred to as the degenerate Abelian Higgs model -- is obtained by taking the potential in~\eqref{lagint} to be
    \begin{equation}\label{v6int}
        V(\phi) = V_6(\phi) = {\lambda \over 2} |\phi|^2 \left(|\phi|^2 - v^2\right)^2~, \qquad \lambda~, v > 0~.
    \end{equation}
    In this case minimizing~\eqref{vartenint} subject to~\eqref{Bincore} leads to a sub-extensive scaling of the string tension, 
    \begin{equation}
        T_n \sim r_n \sim n^{2/3}~,
    \end{equation}
    because core and boundary regions contribute comparable amounts. We refer to the strings in this universality class as domain-wall strings, because their large-$n$ scaling is determined by the boundary term~$\sigma r_n$ in~\eqref{vartenint}. Here~$\sigma > 0$ is the tension of the domain wall interpolating between Coulomb and Higgs vacua. 
\end{itemize}

\noindent In the remainder of this paper, we elaborate on the simple picture of magnetic vortex strings sketched above, making it completely rigorous in the large-$n$ limit. 

In section~\ref{sec:AHMs} we begin by reviewing the Abelian Higgs model~\eqref{lagint} with  scalar potentials~\eqref{v4int}, \eqref{v6int} and spell out the non-linear equations governing axially symmetric vortices. 

In section~\ref{seclargen}, we elaborate on the variational arguments above and obtain precise formulas for~$T_n$ and~$r_n$ in the large-$n$ limit. We also emphasize that bulk and domain-wall strings should be viewed as sharply distinct large-$n$ phases. 

In section~\ref{bulklargensec}, we solve the axially-symmetric vortex equations in the conventional Abelian Higgs model, with scalar potential~\eqref{v4int}, when~$n \gg 1$. We use the method of matched asymptotic expansions (see~\cite{PhysRevLett.125.251601, Penin_2021} for previous work), which requires a refinement of the various string regions introduced above. The solution in the core region is obtained analytically using the WKB approximation. The boundary region must be studied numerically, but the equations nevertheless simplify in the large-$n$ limit. We explain in detail how to match the numerical solution in the boundary region to the core and exterior regions. In the special case~$\lambda = e^2$ discussed below~\eqref{v4int}, we show that our large-$n$ solutions correctly reduce to solutions of the first-order BPS equations. 

In section~\ref{walllargensec}, we find the large-$n$ solutions for strings in the degenerate Abelian Higgs model, with scalar potential~\eqref{v6int}. In this case we are able to provide a fully analytic solution in all regions of the string. A prominent role is played by the domain wall interpolating between the Coulomb and Higgs vacua of~\eqref{v6int}, which surrounds the string core, and its translational zero mode, which is fixed by small residual interactions with the magnetic field~$B \sim n^{-1/3}$ in~\eqref{Bincore}.

In section~\ref{binding}, we use our understanding of large-$n$ vortex strings to shed light on their binding energies, stability, and interactions (see also appendix~\ref{sec:ahminteraction}). 

Throughout the paper, numerical solutions of the vortex equations play an important role. The numerical approach used to generate these solutions is described in appendix~\ref{appnum}. Conversely, the large-$n$ solutions presented in this paper can serve as useful seeds for accurate numerical solutions at small values of $n$.


\section{Magnetic Vortex Strings in Abelian Higgs Models}\label{sec:AHMs}

In this section, we review some basic aspects of magnetically charged vortex strings in $3+1$ dimensional Abelian Higgs models (AHMs). Much of this material is standard, see e.g.~\cite{Manton:2004tk,Weinberg:2012pjx,Shifman:2012zz} for a textbook treatment.

\subsection{Lagrangian and Choice of Scalar Potential}\label{sec:ahmlag}

Throughout, we study the simplest class of Abelian Higgs models in 3+1 dimensions: $U(1)$ gauge theory, with gauge field $a^{(1)}= a_\mu dx^\mu$ and field strength $f^{(2)} = da^{(1)}$, coupled to a single complex scalar field~$\phi$ of unit charge. The Lagrangian is given by\footnote{~Our convention for the Minkowski metric is $\eta_{\mu\nu} = \text{diag}(- +++)$.}
\begin{equation}\label{eq:ahmlagrangian}
    \SL = -\frac{1}{4e^2} f_{\mu\nu} f^{\mu\nu} - \left|(\partial_\mu -ia_\mu)\phi\right|^2  -  V(\phi)~.
\end{equation}
Several comments are in order:
\begin{itemize}
    \item The electromagnetic gauge coupling $e > 0$ is dimensionless. 
    \item The potential $V(\phi) = V(|\phi|)$ is a gauge-invariant polynomial in $|\phi|^2$. Below we will consider different choices of~$V(\phi)$.
    \item We will analyze the Lagrangian \eqref{eq:ahmlagrangian} classically, even though quantum corrections can sometimes change the conclusions significantly. 
\end{itemize}

The theory~\eqref{eq:ahmlagrangian} enjoys a $U(1)_m^{(1)}$ magnetic flux 1-form symmetry~\cite{Gaiotto:2014kfa} (see also the reviews~\cite{McGreevy:2022oyu, Cordova:2022ruw}), because the field-strength two-form $f^{(2)}$ is closed and has quantized periods on any compact 2-cycle $\Sigma_2$,
\begin{equation}
        d f^{(2)} = 0~, \qquad {1 \over 2 \pi} \int_{\Sigma_2} f^{(2)} \in \Z~.
    \end{equation}
The associated magnetically charged defects are 't Hooft lines. The realization of~$U(1)_m^{(1)}$ in the IR depends on the potential $V(\phi)$. We distinguish two cases:
\begin{itemize}
    \item[(C)] If the potential has a minimum at $\phi = 0$, the $U(1)_m^{(1)}$ symmetry is spontaneously broken and the massless photon is the associated Nambu-Goldstone Boson. We refer to this as a Coulomb vacuum of the model. 
    \item[(H)] If the potential has a minimum at $\phi \neq 0$, the photon is massive and the $U(1)_m^{(1)}$  symmetry is linearly realized. We refer to this as a Higgs vacuum. The dynamical excitations that carry $U(1)_m^{(1)}$ charge are precisely the magnetic vortex strings that we are interested in. For this reason we will assume that $V(\phi)$ admits at least one Higgs vacuum. 
\end{itemize}
We will now describe the choices of $V(\phi)$ that we will consider in detail.

\subsubsection{The Conventional Abelian Higgs Model with Quartic Potential}\label{sec:quarticV}

The conventional, renormalizable Abelian Higgs Model (AHM) has the following quartic scalar potential,
\begin{equation}\label{eq:ahmpot}
  V(\phi) =   V_4(\lambda, \phi) \equiv \frac{\lambda}{2} \left(|\phi|^2 - v^2\right)^2~.
\end{equation}
Here $\lambda > 0$ is the dimensionless quartic coupling, and $v > 0$ has mass-dimension one. This model has a unique vacuum, where
$\phi = v$ (up to gauge transformations), i.e.~$v$ is the vev of $\phi$. The $U(1)$ gauge group is completely Higgsed, and phase of $\phi$ is eaten by the photon to produce a spin-$1$ vector boson of mass $m_V$; the modulus of $\phi$ describes the spin-$0$ scalar Higgs field of mass $m_H$. It follows from~\eqref{eq:ahmlagrangian} and~\eqref{eq:ahmpot} that
\begin{equation}\label{eq:ahmmass}
    m_H^2 = 2\lambda v^2~, \qquad m_V^2 = 2e^2 v^2~.
\end{equation}

It is useful to define the following dimensionless ratio,
\begin{equation}\label{eq:betadef}
  \beta \equiv \frac{m_H^2}{m_V^2} = \frac{\lambda}{e^2}~.
\end{equation}
We will follow the standard practice of further subdividing the AHMs we study into type I and type II, depending on the mass hierarchy between the Higgs and the vector bosons,
\begin{equation}\label{eq:types}
    \begin{split}
       & \text{Type I} : \quad\; \beta <   1~, \quad m_H < m_V ~, \qquad  \\
      &  \text{Type II} : \quad \beta >   1~, \quad m_H > m_V ~.\qquad \\ 
    \end{split}
\end{equation}
As we will review below, some basic properties of the vortex strings -- such as whether well-separated strings attract or repel -- only depend on the mass hierarchy between the Higgs and the vector boson, i.e., whether the model is type I or type II.

The point $\beta = 1$ separating the type-I and type-II regimes is marginal and requires separate treatment. Classically, the conventional AHM with quartic potential~\eqref{eq:ahmpot} and~$\beta = 1$ can be embedded into an $\CN=1$ supersymmetric model, in which the vortex strings turn out to be~$\half$-BPS.\footnote{~The simplest supersymmetric completion is given by an $\CN=1$ Abelian vector superfield $\CV$, containing the photon $a_\mu$,  and an~$\CN=1$ chiral superfield $\Phi$ (with bottom component~$\phi$) that carries unit charge under~$\CV$. There is also a Fayet-Iliopoulos term for $\CV$, which is needed in order for the magnetic vortex strings to be BPS saturated (see for instance~\cite{Dumitrescu:2011iu} and references therein). The superspace Lagrangian is
\begin{equation}
    \SL = \frac{1}{e^2} \int d^2 \theta \, W^\alpha W_\alpha + (\text{h.c}) + \int d^4\theta \, \left( \b{\Phi} e^{-2\CV} \Phi + \xi \CV\right)~.
\end{equation} 
Here $W_\alpha$ is the field-strength superfield of $\CV$. In components, the bosonic part of this Lagrangian is precisely given by~\eqref{eq:ahmlagrangian} with quartic potential~\eqref{eq:ahmpot}, upon equating~$\lambda = e^2$ and~$\xi = v^2$. Note that this theory suffers from a $U(1)$ gauge anomaly, and hence it does not make sense quantum mechanically; this can be cured by adding more fields.} In particular, as we will review below, static strings in the AHM with $\beta = 1$ do not exert any forces on one another, as is typical of BPS solitons. We will therefore refer to the~$\beta = 1$ point of this model as its BPS point.  

Since our analysis in this paper will be classical, we will allow ourselves to contemplate all values of the mass ratio~$\beta \equiv {m_H^2 / m_V^2}$ defined in~\eqref{eq:betadef}. However, the Higgs vacuum that is always present classically is only stable to quantum corrections if~$\beta$ is bounded from below~\cite{Linde:1975sw,Weinberg:1976pe},  
\begin{equation}
    \lambda \gtrsim e^4 \quad \Longleftrightarrow \quad    \beta  \gtrsim {e^2 }~.
\end{equation}

\subsubsection{The Degenerate Model with Sextic Potential and its Domain Wall}\label{degAHMandDW}

The conventional AHM described above can be generalized by allowing higher powers of~$|\phi|^2$ in the potential. Even though the resulting theory is not renormalizable, the additional non-renormalizable parameters in the potential can have important consequences for the phase structure of the model, and for its strings. By tuning these parameters, we can reach additional phases and transitions while maintaining the existence of at least one Higgs vacuum that we require for the existence of strings.

The simplest model of this kind has Lagrangian~\eqref{eq:ahmlagrangian} with a sextic scalar potential (bounded from below) of the schematic form
\begin{equation}\label{gensextic}
    V(\phi) \sim c_2 |\phi|^2 + c_4 |\phi|^4 + \lambda |\phi|^6~, \qquad c_{2,4} \in \R~, \quad  \lambda > 0~.
\end{equation}
For generic values of~$c_2, c_4$, the model has a unique vacuum, which can be of either Coulomb type without strings, or of Higgs type with strings. As we will explain in later sections, the Higgs-vacuum strings are (in a sense that we will make precise) qualitatively similar to the strings of the conventional Abelian Higgs model. In fact, this conclusion extends to arbitrary potential~$V(\phi)$, as long as it has a unique Higgs vacuum. 

An interesting exception occurs when the parameters~$c_2, c_4$ in~\eqref{gensextic} are tuned to a first-order phase transition. This requires one tuning, i.e.~it occurs along a curve in the~$c_2$-$c_4$ plane. Along this curve, the sextic potential~\eqref{gensextic} takes the following special form, 
\begin{equation}\label{eq:sexticV}
   V(\phi) =  V_6(\lambda, \phi) \equiv {\lambda \over 2} |\phi|^2 \left(|\phi|^2 - v^2\right)^2~, \qquad \lambda > 0~.
\end{equation}
Note that $\lambda \sim \Lambda^{-2}$ (with~$\Lambda$ a UV scale) has mass dimension $-2$. This model has two isolated, degenerate vacua, which are separated by a potential barrier, and for this reason we refer to the Abelian Higgs model with potential~\eqref{eq:sexticV} as the degenerate model. 

We now describe in more detail the vacua of the degenerate Abelian Higgs model: 
\begin{itemize}
    \item[(H)] A Higgs vacuum at $|\phi| = v$, with Higgs and vector boson masses given by
    \begin{equation}\label{eq:degmasses}
        m_H^2 = 2 \lambda v^4~, \qquad m_V^2 = 2 e^2 v^2~.
    \end{equation}
    The dimensionless ratio $\beta$, which we continue to define as in~\eqref{eq:betadef}, is thus given by
    \begin{equation}\label{betadeg}
        \beta \equiv {m_H^2 \over m_V^2} = {\lambda v^2 \over e^2}~.
    \end{equation}
    We also continue to use the classification into type-I and type-II models defined in~\eqref{eq:types}. However, the point $\beta = 1$ is  no longer associated with an embedding into a supersymmetric theory in 3+1 dimensions;\footnote{~This conclusion can be evaded in~$2+1$ dimensions, but it requires adding a Chern-Simons term~\cite{Lee:1990it,Jackiw:1990pr}.} consequently we do not refer to it as the BPS point. This fact is also reflected in the properties of the strings at~$\beta = 1$.  
    \item[(C)] A Coulomb vacuum at $\phi = 0$, with a massless photon. In this vacuum the scalar field~$\phi$ describes a charged particle of mass
    \begin{equation}\label{eq:mphi}
        m_\phi^2 = \half \lambda v^4 = \left({m_H \over 2}\right)^2~.
    \end{equation}
\end{itemize}

\noindent The presence of two degenerate vacua implies the existence of a domain wall interpolating between them. In unitary gauge, where $\phi(x) \geq 0$, the domain wall profile $\phi(z)$ does not involve the gauge field and only depends on the coordinate $z = x_3$ transverse to the wall. Finding $\phi(z)$ generically requires solving a second-order ODE. 

Our discussion is simplified by the fact that the scalar potential~\eqref{eq:sexticV} of the degenerate model admits a superpotential $W(\phi)$, such that $V_6(\phi) = (W'(\phi))^2$. The domain wall tension $\sigma_\text{DW}$ and the profile~$\phi(z)$ can thus be obtained by minimizing the following functional,
\begin{equation}\label{DWtension}
    \sigma_\text{DW} = \text{min} \left( \int_{-\infty}^{\infty} dz \, \left( (\phi'(z))^2 + \left(W'(\phi)\right)^2 \right) \right)~, \qquad W'(\phi) = \sqrt{\lambda \over 2} \phi(v^2 - \phi^2)~.
\end{equation}
Here we minimize over non-negative functions $\phi(z) \geq 0$, subject to the following boundary conditions,
\begin{equation}\label{wallbcs}
    \phi(-\infty) = 0~, \qquad \phi(\infty) = v~,
\end{equation}
so that the wall interpolates from the Coulomb vacuum on its left to the Higgs vacuum on its right.

Applying a standard BPS completion argument to \eqref{DWtension} (see for instance~\cite{Shifman:2012zz}), we find that the domain-wall tension is 
\begin{equation}\label{walltension}
    \sigma_\text{DW} = 2 (W(v) - W(0)) = {\sqrt \lambda \over 2 \sqrt 2} v^4~. 
\end{equation}
The corresponding profile function satisfies the following first-order BPS equation,
\begin{equation}
    \phi'(z) = W'(\phi)~, 
\end{equation}
whose solution (with boundary conditions~\eqref{wallbcs}) is
\begin{equation}\label{wallsoln}
    \phi(z) = {v \over \sqrt{1 + e^{- \sqrt{ 2 \lambda}  v^2 (z - z_0) }}} = {v \over \sqrt{1 + e^{- m_H (z - z_0) }}}~.
\end{equation}
Let us make some comments:
\begin{itemize}
    \item The approach to the Higgs vacuum $\phi = v$ at $z \to \infty$ is $\CO(e^{-m_H z})$ and governed by the Higgs mass $m_H$ in~\eqref{eq:degmasses}. 

    \item The approach to the Coulomb vacuum $\phi = 0$ at $z \to -\infty$ is $\CO(e^{m_\phi z})$ and governed by the $\phi$-mass $m_\phi = \half m_H$ (see~\eqref{eq:mphi}) in that vacuum. 
    
    \item The integration constant $z_0$ is the center of mass position of the domain wall. It is an exact zero mode of the solution and will play an important role below. 
\end{itemize}

\subsection{Magnetic Vortex Strings}

We are interested in the magnetic vortex -- or Abrikosov-Nielsen-Olesen~\cite{Abrikosov:1956sx, Nielsen:1973cs} -- strings of the Abelian Higgs models~\eqref{eq:ahmlagrangian}, with quartic potential~$V = V_4$ in~\eqref{eq:ahmpot} (the conventional model) and sextic potential~$V = V_6$ in~\eqref{eq:sexticV} (the degenerate model). In order for these strings to exist as finite-tension excitations, we need to study the Higgs vacua~$|\phi| = v$ of these models, in which the~$U(1)_m^{(1)}$ magnetic-flux symmetry is unbroken and the magnetic flux of the strings is a conserved quantum number. 

\subsubsection{Magnetic Flux and Vorticity}

\begin{figure}[t!]
\centering
\includegraphics[width=0.5\textwidth]{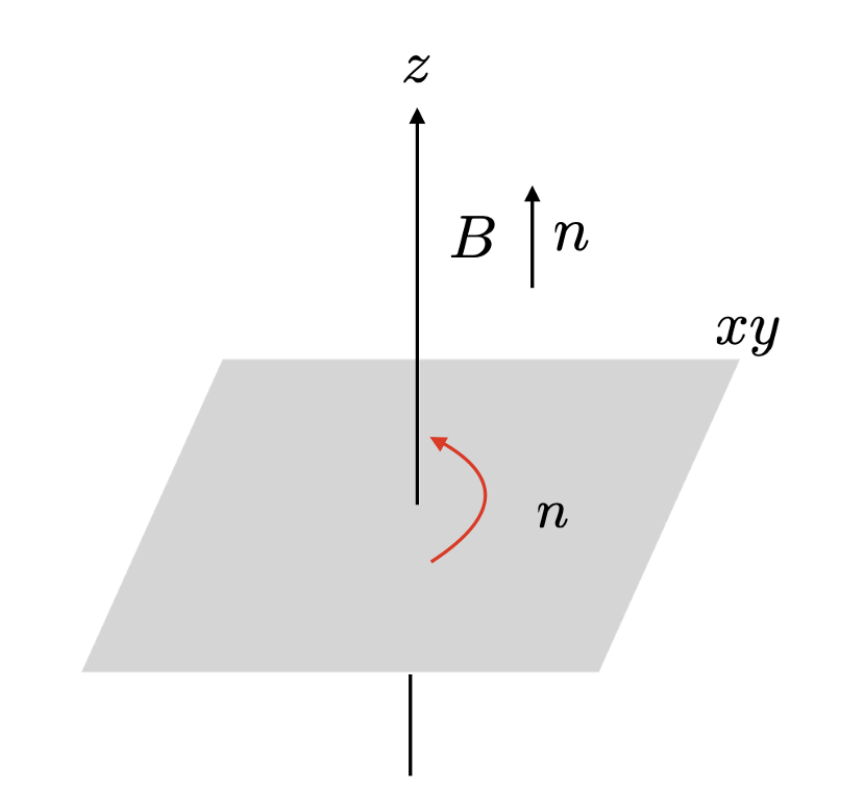}
\caption{\label{fig:vortexcartoon} A schematic depiction of an axially symmetric ANO vortex carrying~$n-$units of conserved magnetic flux along the~$z-$axis. The red arrow illustrates the winding of the Higgs field around the vortex. Requiring finite tension fixes the winding number to be~$n$, matching the total magnetic flux carried by the string. }
\end{figure}

Through this paper we will be interested in strings that carry~$n$ units of quantized magnetic flux in the~$xy$-plane~$\Sigma_{xy}$,
\begin{equation}\label{eq:nflux}
    n \equiv {1 \over 2 \pi} \int_{\Sigma_{xy}} da = {1 \over 2\pi} \oint_{\d \Sigma_{xy}} a \in \Z~.
\end{equation}
Using the charge-conjugation symmetry of~\eqref{eq:ahmlagrangian}, it is sufficient to consider strings of positive flux~$n \geq 1$. We will sometimes refer to them as~$n$-strings, and the~$n = 1$ case as a minimal or fundamental string. We see from~\eqref{eq:nflux} that~$n$ fixes the holonomy of~$a_\mu$ at the spatial boundary of the~$xy$-plane. Since it is conserved, it follows that the string extends to~$z = \pm \infty$ in the~$z$-direction. See figure~\ref{fig:vortexcartoon}.

The non-vanishing holonomy~\eqref{eq:nflux} around the cycle~$\d \Sigma_{xy}$ at spatial infinity also forces the phase of the Higgs field~$\phi$ to have winding number~$n$ around the same cycle. To see this, consider the total energy of the Lagrangian~\eqref{eq:ahmlagrangian}, 
\begin{equation}\label{totalE}
   E =  \int d^3 x \, \bigg(\frac{1}{2e^2} f_{0i}^2  + |(\d_0 - i a_0)\phi|^2  + \frac{1}{4e^2} f_{ij}^2 + |(\d_i - ia_i)\phi|^2 + V(\phi) \bigg)~,
\end{equation}
where~$i = 1,2,3$ is a spatial coordinate index. The requirement that the string have finite tension~$T = E/L$ (with~$L \to \infty$ the length of the string) implies that every term in the integrand must separately vanish as we approach~$\d\Sigma_{xy}$ in the directions transverse to the string, where~$|\phi| \to v$ approaches the Higgs vacuum and both potentials~$V(\phi) = V_{4, 6}(\phi) \to 0$ in~\eqref{eq:ahmpot}, \eqref{eq:sexticV} vanish. In particular, the requirement that~$(\d_i - ia_i)\phi \to 0$ implies that 
\begin{equation}\label{winding}
    a_i  = - i \d_i \log \phi  \qquad \text{at spatial infinity } \; \d \Sigma_{xy}~.
\end{equation}
Together with~\eqref{eq:nflux} this implies that the phase of~$\phi$ has winding number~$n$ around~$\d \Sigma_{xy}$. This winding number is also referred to as the vorticity of the complex Higgs field~$\phi$; in the models we study here it is synonymous with the magnetic flux of the string.\footnote{~In general, the requirement of finite string tension fixes the vorticity of all charged scalars to be their electric charge times the magnetic flux~$n$. In particular, electrically neutral scalar fields of non-zero vorticity always lead to infinite string tension, as is the case for superfluid vortex strings.} 

A general string configuration of vorticity~$n$ is described by solutions of the Euler-Lagrange equations obtained by varying~\eqref{eq:ahmlagrangian}, subject to~\eqref{eq:nflux} and the boundary conditions at $\d \Sigma_{xy}$ that ensure finite string tension, i.e.~the fact that~$|\phi| \to v$ and that its phase has vorticity~$n$ at~$\d \Sigma_{xy}$. These equations are highly non-linear PDEs and have been the subject of extensive study, both analytic (e.g.~\cite{Jaffe:1980mj})  and numerical (e.g.~\cite{Jacobs:1978ch}). Throughout, we will limit our analysis of these equations to the more tractable case with axial symmetry.

\subsubsection{Strings with Axial Symmetry}

We will use polar coordinate~$r, \theta$ in the~$xy$-plane. A naive imposition of axial symmetry (with all fields~$\theta$-independent) is incompatible with the winding constraint~\eqref{winding}. Instead, we will work with the following ansatz,
\begin{equation}\label{ansatz}
    \phi(x) \equiv v \varphi(t, r, z) e^{in\theta}~, \qquad a_\mu(x) = a_\mu(t, r, z)~.
\end{equation}
As~$r \to \infty$, the complex, dimensionless coefficient function~$\varphi(t,r, z)$ has unit norm~$|\varphi| \to 1$ to ensure that~$\phi(x)$ approaches the Higgs vacuum with the correct vorticity~$n$; similarly, $a_\theta(t, r, z)$ must have holonomy~$n$. Note that~\eqref{ansatz} is rotationally invariant under a diagonal subgroup of conventional rotations around the~$z$-axis and global~$U(1)$ gauge transformations. 

Throughout, we will be interested in the~$n$-vortex solution of minimal energy. It is straightforward to check that the energy~\eqref{totalE} is minimized by setting to zero the~$t, z$ gradient terms to zero,\footnote{~These terms are activated by excitations traveling along the string.} which is consistent with the constraints. The ansatz~\eqref{ansatz} then reduces to
\begin{equation}\label{ansatzii}
    \phi(x) = v \varphi(r) e^{in\theta}~, \qquad a_0 = a_1 = 0~, \qquad a_{r, \theta}(x) = a_{r, \theta}(r)~.
\end{equation}
Two further simplifications are possible:
\begin{itemize}
\item We use the remaining freedom to perform~$r$-dependent gauge transformation to fix radial gauge~$a_r(r) = 0$. 
\item In this gauge, the phase~$\text{Arg} \, \varphi$ of the complex field~$\varphi(r)$, which is not subject to any constraints, only contributes to the energy density~\eqref{totalE} by a term proportional to~$|\varphi|^2 (\d_r \text{Arg} \, \varphi)^2$. Thus the energy is minimized by taking the phase of~$\varphi(r)$ to be a constant, which can be absorbed using a global gauge transformation.\footnote{~Strictly speaking, minimizing the energy implies that the phase of~$\varphi(r)$ is a piecewise constant function that can jump when~$\varphi = 0$. This generalization does not occur in our solutions, all of which have non-vanishing~$\varphi(r) \neq 0$ for~$r > 0$, i.e.~the only zero of~$\varphi(r)$ is at the origin.}  
\end{itemize}
In summary, the only fields that remain in our ansatz are given by\footnote{~Note that~$a_\theta$ is such that the one-form~$a^{(1)} = a_\theta d\theta$. In particular, it is dimensionless.}
\begin{equation}\label{eq:ahmansatz}
    \phi(x) = v \varphi(r)e^{in\theta}~, \qquad a_\theta(r) \equiv n (1 - a(r))~,  \qquad \varphi(r) , a(r) : [0, \infty) \to \R~.
\end{equation}
 Here~$\varphi(r)$ and~$a(r)$ are real functions of~$r$, which are subject to the following boundary conditions,
\begin{equation}\label{eq:ahmbc}
        \varphi(r = 0)=0~, \qquad \varphi(r = \infty) =1~, \qquad a(r = 0)=1~, \qquad a(r = \infty) = 0~. 
\end{equation}
The boundary conditions at~$r = 0$ are dictated by regularity (i.e.~finiteness of the string tension), because both~$\phi(x)$ and~$a_\theta(r)$ have vorticity/holonomy around the origin. The boundary conditions at~$r = \infty$ arise from the requirement that~$|\phi(x)| \to v$ should approach the Higgs vacuum, subject to the vorticity/holonomy constraints~\eqref{eq:nflux} and~\eqref{winding}. 

Since we are now studying strings that are uniformly extended along the~$z$-direction, with length~$L \to \infty$, we will switch from the Energy~$E$ to the tension~$T = E/L$. We will be interested in finding the axially symmetric~$n$-strings of lowest tension~$T_n$, which can be found by minimizing the following functional (obtained by substituting the ansatz~\eqref{eq:ahmansatz} into the energy~\eqref{totalE} and dividing by~$L$),
\begin{equation}\label{Tndef}
    T_n \equiv \text{min} \left\{2 \pi \int_0^\infty r dr \, \left({n^2 \over 2 e^2 r^2} \left({da \over dr}\right)^2 + v^2 \left[\left({d \varphi \over dr}\right)^2 + {n^2 \over r^2} a^2 \varphi^2 \right] + V(v \varphi) \right)\right\}~.
\end{equation}
Here the minimization is over real, smooth functions~$\varphi(r), a(r)$ subject to the boundary conditions~\eqref{eq:ahmbc}. We will occasionally refer to the functional in curly braces as the (off-shell) string tension, with the understanding that only the minima of this functional correspond to actual (on-shell) string solutions. 

\begin{figure}[t!]
\centering
\includegraphics[width=0.49\textwidth]{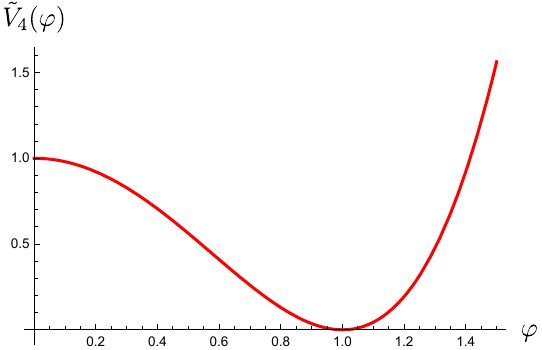}
\includegraphics[width=0.49\textwidth]{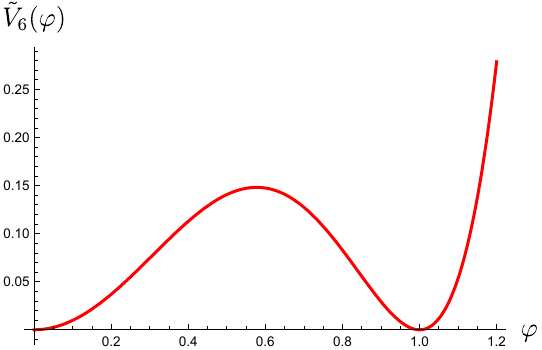}
\caption{\label{fig:vtplots} The dimensionless, rescaled potentials~$\widetilde{V}_4(\varphi)$ and~$\widetilde{V}_6(\varphi)$, given in~\eqref{rescaledV}, corresponding to the conventional and degenerate models, respectively. The curves are shown for a representative choice of the parameter~$\beta=1$.}

\end{figure}

We will further simplify the notation as follows:
\begin{itemize}
\item We introduce a dimensionless radial coordinate
\begin{equation}\label{udef}
    u \equiv e v r~,
\end{equation}
in terms of which the string tension~\eqref{Tndef} reads
\begin{equation}\label{eq:ahmtension}
    {T_n \over 2 \pi v^2} =  \int_0^\infty u du \, \left({n^2 \over 2 u^2} \left(a'(u)\right)^2 + \left(\varphi'(u)\right)^2 + {n^2 \over u^2} a^2(u) \varphi^2(u) + {\t V}(\varphi(u)) \right)~.
\end{equation}
Here primes indicate a~$u$-derivative and~${\t V}$ is a dimensionless rescaled potential,
\begin{equation}\label{rescaledV}
    {\t V}(\varphi) \equiv {1 \over v^4 e^2} V(v \varphi) = \begin{cases}
       \;   \t V_4(\varphi) = {\beta \over 2} (\varphi^2 -1)^2 \qquad \;\; \; \, (\text{conventional}) ~,\\
       \;   \t V_6(\varphi) =   {\beta \over 2} \varphi^2 (\varphi^2 -1)^2 \qquad (\text{degenerate})~.
    \end{cases}
\end{equation}
See figure~\ref{fig:vtplots} for an illustration of the rescaled potentials in the two models.
In both the conventional model, with quartic potential~$V = V_4$ in~\eqref{eq:ahmpot}, and the degenerate model, with sextic potential~$V = V_6$ in~\eqref{eq:sexticV}, the rescaled potentials~${\t V}_{4, 6}$ only depend on the ratio~$\beta = {m_H^2 / m_V^2}$ in~\eqref{eq:betadef}, \eqref{betadeg} that determines whether the model is Type I or Type II according to~\eqref{eq:types}. 

\item Throughout the paper, we set dimensions via
\begin{equation}\label{eq:ahmsimpl}
v^2=\sqrt{2}~.
\end{equation}
\end{itemize}

\begin{figure}[t!]
\centering
\includegraphics[width=0.49\textwidth]{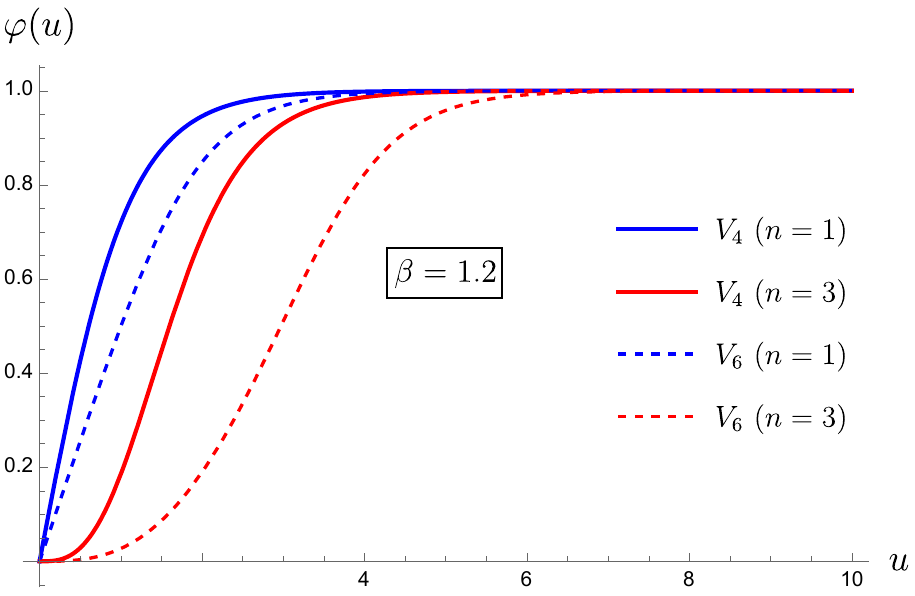}
\includegraphics[width=0.49\textwidth]{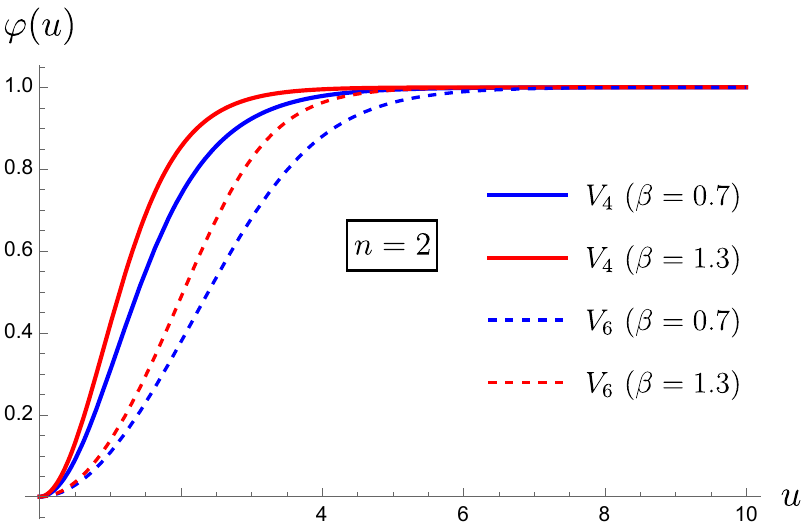}
\includegraphics[width=0.49\textwidth]{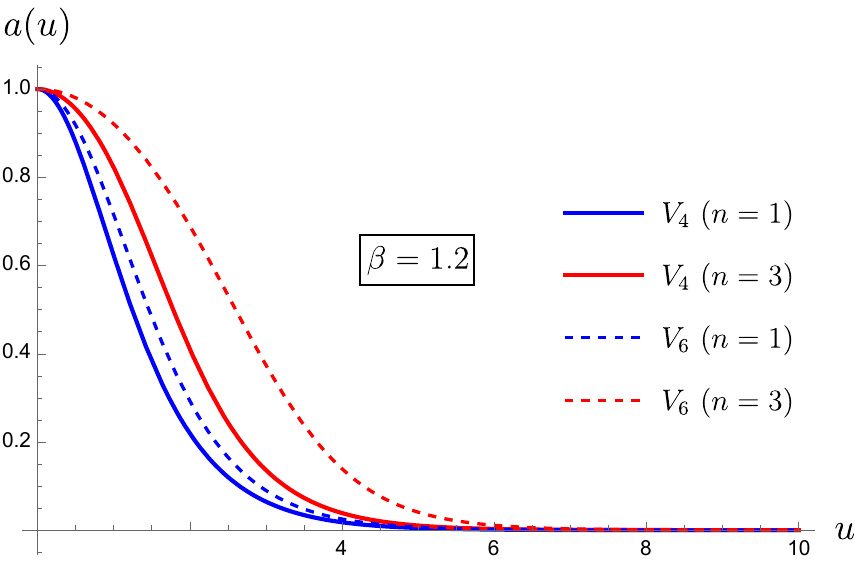}
\includegraphics[width=0.49\textwidth]{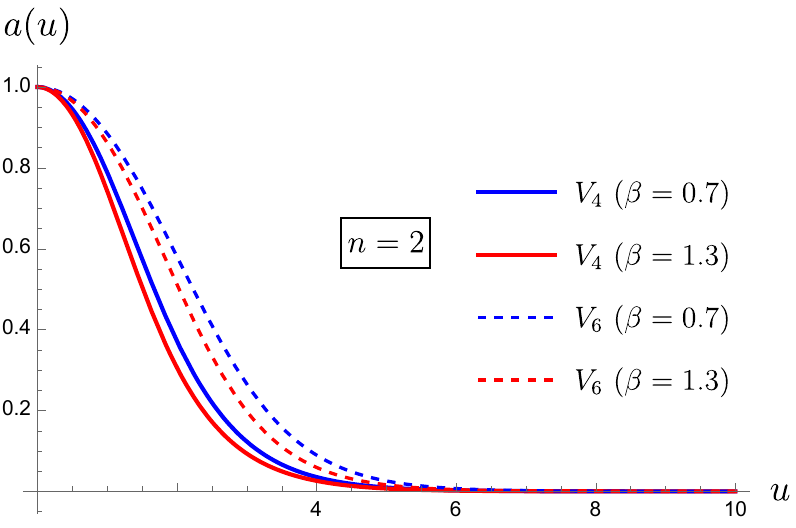}
\caption{\label{fig:compahmmodels} The figure depicts numerical solutions for the Higgs field~$\varphi(u)$ (top) and the gauge field~$a(u)$ (bottom) of the axially-symmetric vortex equations~\eqref{eqn:ahmeom} with conventional and degenerate Abelian-Higgs-model potentials~$V_{4,6}$ (shown in solid and dashed lines, respectively). On the left, we fix~$\beta=1.2$ and vary~$n = 1$ (blue) and~$n =3$ (red); on the right, we fix~$n = 2$ and consider~$\beta=0.7$ (type I, blue) and $\beta=1.3$ (type II, red). The asymptotics at~$u = 0, \infty$ are discussed in section~\ref{secasym}. Note that increasing~$n$ or decreasing~$\beta$ makes the solutions move to the right, i.e.~the vortices become wider. The same occurs at fixed~$n, \beta$ if we go from the conventional to the degenerate model. These qualitative observations will be made precise in section~\ref{seclargen}.}
\end{figure}

The minimum-energy axially symmetric $n$-string profiles can be found by solving the Euler-Lagrange equations that follow by extremizing the tension functional~$T_n$ in~\eqref{eq:ahmtension},
\begin{equation}\label{eqn:ahmeom}
    \begin{split}
       \varphi''(u) + \frac{1}{u} \varphi'(u) = ~ & \frac{n^2}{u^2} a^2(u) \varphi(u) + \half {d \t V \over d \varphi}~,
       \\
        a''(u) - \frac{1}{u} a'(u) = & ~ 2 a(u) \varphi^2(u) ~.\\
    \end{split}
\end{equation}
Here, the term in the first equation that depends on the rescaled potential~\eqref{rescaledV} takes the form
\begin{equation}\label{vtterm}
    \half {d \t V \over d \varphi} = \begin{cases} \beta \varphi(u)(\varphi^2(u)-1) \qquad  \qquad \qquad \; (\t V = \t V_4)~,\\ 
       {\beta \over 2} \varphi (\varphi^2 -1)^2 + \beta \varphi^3 (\varphi^2-1) \qquad (\t V = \t V_6)~.
       \end{cases}
\end{equation}
The equations~\eqref{eqn:ahmeom}, together with the boundary conditions~\eqref{eq:ahmbc}, are the vortex equations that we will study throughout the paper. They have no analytic solutions for generic values of~$n$ and~$\beta$, and hence we must solve them numerically (see appendix~\ref{appnum}). Some numerical solutions are shown in figure~\ref{fig:compahmmodels}.  As we will see, an analytic approach is possible for strings with very large magnetic flux~$n \gg 1$. 

For future use, we express the BPS domain wall solution~\eqref{wallsoln} of the degenerate model in our rescaled variables,
\begin{equation}\label{rescaledwall}
  \varphi(u) = {\phi(u) \over v} = \left(1 + e^{- \sqrt{2 \beta} \, (u-u_0)}\right)^{-\half}~.  
\end{equation}
Here,~$u_0$ is the translational modulus of the wall, which determines where it is centered.

\subsubsection{Asymptotics}\label{secasym}

Let us consider the asymptotic behavior of solutions~$a(u), \varphi(u)$ of the vortex equations~\eqref{eqn:ahmeom} (with potential term~\eqref{vtterm}) near~$u = 0, \infty$, where their values are fixed by~\eqref{eq:ahmbc}. 

The leading asymptotics near the origin~$u = 0$ take the following form,
\begin{equation}\label{eqn:ahmoriginasym}
    a(u) \simeq 1- \frac{u^2}{u_n^2}~, \qquad~ \varphi(u) \simeq  c_n u^n~, \qquad u \ll 1~.
    \end{equation}
Here~$u_n$ and~$c_n$ are two integration constants, which (unlike the exponents in~\eqref{eqn:ahmoriginasym}) not only depend on~$n$, but also on the choice of potential via~\eqref{vtterm}, and hence on~$\beta$. When needed, we indicate this by writing them as~$u_n(\beta), c_n(\beta)$. 

The approach to the gapped Higgs vacuum at~$u \to \infty$ is exponential,
\begin{equation}\label{eq:ahmasym}
    a(u) \simeq a_\infty \sqrt{u} e^{-\sqrt{2}u}~, \qquad \varphi(u) \simeq 1-  \frac{\varphi_\infty}{\sqrt{u}} e^{-\sqrt{2\beta} u}  - \frac{n^2 a_\infty^2  e^{-2\sqrt{2}u}}{2(\beta-4) u}~, \qquad u \gg 1~,
    \end{equation}
    where~$a_\infty, \varphi_\infty$ are integration constants that depend on~$n$ and the choice of potential. In~\eqref{eq:ahmasym} we indicate the asymptotics for generic values of~$\beta \neq 4$, see \cite{Plohr:1981cy,Perivolaropoulos_1993,Manton:2004tk} for details. When~$\beta < 4$, the leading~$\varphi$-asymptotics are dominated by the~$\CO(\varphi_\infty)$ term in~\eqref{eq:ahmasym}, while the~$\CO(a_\infty^2)$ term is subleading. This is the familiar case where the asymptotics are obtained by linearizing the vortex equations~\eqref{eqn:ahmeom} around the Higgs vacuum. (The result is the same for the two choices of potential in~\eqref{vtterm}.) In this case, the leading exponentials in~\eqref{eq:ahmasym} are~$a(r) \sim e^{- m_V r}$  and~$\varphi(r) \sim e^{- m_H r}$ (with masses given by~\eqref{eq:ahmmass} or~\eqref{eq:degmasses}) in our rescaled units~\eqref{udef}, \eqref{eq:ahmsimpl}.  

    By contrast, when~$\beta > 4$, the~$\CO(\varphi_\infty)$ term in the~$\varphi$-asymptotics is subleading, and the leading~$\CO(a_\infty^2)$ term is obtained by substituting the~$a$-asymptotics in~\eqref{eq:ahmasym} into the non-linear~$\CO(a^2 \varphi)$ term on the right side of the first equation in~\eqref{eqn:ahmeom}, which must be retained when solving for the large-$u$ behavior of~$\varphi(u)$.

\subsubsection{BPS Strings  in the Conventional Abelian Higgs Model}

As already mentioned below~\eqref{eq:types}, the conventional Abelian Higgs model has special properties at the BPS point~$\beta = 1$, at which~$m_H = m_V$, that can be understood by embedding the model into a supersymmetric theory. (This is not true for the degenerate model.) At~$\beta = 1$ the vortex equations also have special properties. For axially-symmetric vortices, this can be seen by BPS-completing the string tension~\eqref{eq:ahmtension} for the case of the conventional quartic potential~$\t V_4$ in~\eqref{rescaledV}, with~$\beta = 1$. This leads to
\begin{equation}\label{BPScompTn}
\begin{split}
    \frac{T_n}{2\pi v^2} = \int_0^\infty u \, du \, \bigg( & {1 \over 2} \Big[\frac{n}{u}a'(u) -(\varphi^2(u)-1)\Big]^2 \\
    & +  \Big[\varphi'(u) -\frac{n}{u} \varphi(u) a(u)  \Big]^2 -  \frac{n}{u} \,a'(u)\bigg) \geq n~,
\end{split}
\end{equation}
where we have used~$v^2 = \sqrt{2}$ to set units, as in~\eqref{eq:ahmsimpl}. This gives a lower bound on the string tension,
\begin{equation}\label{BPST}
    T_n \geq T_{\text{BPS}, n} \equiv 2   \pi v^2 \, n~, \qquad v^2 = \sqrt{2}~,
\end{equation}
which is saturated if and only if the bracketed expressions in~\eqref{BPScompTn} vanish, 
\begin{equation}\label{eqn:ahmBPSeqns}
     \frac{n}{u} a'(u) =  \varphi^2(u)-1~, \qquad   \varphi'(u) =  \frac{n}{u} \varphi(u) a(u)~. 
\end{equation}
These first-order BPS equations imply the full second-order vortex equations~\eqref{eqn:ahmeom} when $\t V = \t V_4$ and $\beta=1$ (the converse is not true).\footnote{At the BPS point,~\cite{Taubes:1979ps,Jaffe:1980mj} proved an existence theorem for the BPS equations~\eqref{eqn:ahmBPSeqns}. Further,  they proved that no solutions exist to the full second order equations~\eqref{eqn:ahmeom} which are not solutions to the first order BPS equations~\eqref{eqn:ahmBPSeqns}.} Somewhat surprisingly, even the axially-symmetric BPS equations~\eqref{eqn:ahmBPSeqns} do not have any known analytic solutions. However, these equations can be used to simplify the asymptotics, e.g.~substituting the leading~$u \to \infty$ asymptotics~\eqref{eq:ahmasym} into~\eqref{eqn:ahmBPSeqns}, we find that the coefficients~$a_\infty,\varphi_\infty$ are related via
\begin{equation}\label{eqn:bpsasym}
    a_\infty = {\sqrt{2} \over n} \, \varphi_\infty~.
\end{equation}


\section{Strings in the Large Magnetic Flux Limit}\label{seclargen}

In this section, we begin our discussion of vortex strings in the limit of large magnetic flux~$n \to \infty$.\footnote{~We take the large-$n$ limit while keeping the Higgs potential~$\t V$ appearing in the vortex equations, i.e.~the parameter~$\beta$ in~\eqref{vtterm}, fixed as~$n \to \infty$   .} We will sketch a very simple, intuitive picture of the strings in this limit, somewhat similar to the one advocated in~\cite{Bolognesi_2005,Bolognesi_2006}, and use it to estimate their size and tension. Crucially, we find very different qualitative behaviors in the conventional and degenerate Abelian Higgs models, and we explain why they should be thought of as different, sharply defined large-$n$ phases of vortex strings. We will elaborate on these results in sections~\ref{bulklargensec} and~\ref{walllargensec}, where we rigorously analyze the axially-symmetric vortex equations~\eqref{eqn:ahmeom}, with quartic and sextic potentials~\eqref{vtterm}, in the large-$n$ limit.

\subsection{Giant Vortex Strings at Large $n$}\label{sec:intuition}

The picture that we will develop is that strings are characterized by a transverse size~$u_n$ in the radial~$xy$-direction (here~$u$ is the dimensionless radial coordinate in~\eqref{udef}), which diverges in the large-flux limit, i.e. the strings become giant magnetic vortices in this limit, 
\begin{equation}\label{untoinfty}
    u_n \to \infty \qquad \text{as} \qquad n \to \infty~.
\end{equation}
The tendency of strings to broaden with increasing~$n$ is already visible at small values of~$n$, such as those shown in figure~\ref{fig:compahmmodels}. 

Thanks to the boundary conditions~\eqref{eq:ahmbc}, the Higgs field~$\varphi(0) = 0$ vanishes at the origin. Roughly speaking, the radius~$u_n$ is defined so that 
\begin{equation}\label{verysmallphi}
\varphi(u) \simeq 0 \qquad \text{for} \qquad  u \lesssim u_n~.
\end{equation}
We refer to the region~$u \lesssim u_n$ as the core of the string. The fact that such a core region exists can already be gleaned by considering the small-$u$ asymptotics in~\eqref{eqn:ahmoriginasym},
\begin{equation}
    \varphi(u) \sim u^n = e^{n\log{u}}~,
\end{equation}
which (for sufficiently small, fixed~$u$) decays exponentially as~$n \to \infty$. 

In the core region~\eqref{verysmallphi}, where~$\varphi(u) \simeq 0$, the vortex equations~\eqref{eqn:ahmeom} can be integrated to find the gauge field profile~$a(u)$, 
\begin{equation}\label{aincore}
a(u) \simeq 1- {u^2 \over u_n^2} \qquad \text{for} \qquad  u \lesssim u_n~,
\end{equation}
which describes a constant magnetic field,\footnote{~Here~$B(u)$ is related to the conventional magnetic field~$f_{xy} = (da)_{xy}$ appearing in~\eqref{eq:nflux} via
\begin{equation}
    B(u) = {1 \over (ev)^2} \, f_{xy}(r)~, \qquad u = evr~.
\end{equation}
Note that~$B(u)$ is dimensionless, and that~$f_{xy} \, r dr = B(u) \, u du$.}
\begin{equation}\label{Bconstcore}
    B(u) \equiv -\frac{n}{u}a'(u) \simeq \frac{2n}{u_n^2} = \text{constant}~\qquad \text{for} \qquad u \lesssim u_n~.
\end{equation}
Note that this nearly constant magnetic field in the core region of the string essentially saturates its magnetic flux, 
\begin{equation}
    \int_0^{u_n}u \, du \,  B(u)  \simeq {B \over 2} u_n^2 = n~. 
\end{equation}
Thus the magnetic field must vanish quite rapidly outside the core region; the fields in the complementary exterior region have essentially already reached the Higgs vacuum, 
\begin{equation}\label{outsidecore}
    B(u) \simeq 0~,  \quad \varphi(u) \simeq 1 \qquad \text{for} \qquad u \gtrsim u_n~.
\end{equation}
The transition from the nearly constant values~$\varphi \simeq 0$ (see~\eqref{verysmallphi}) and~$B \simeq {2 n /  u_n^2}$ (see~\eqref{Bconstcore}) in the core region~$u \lesssim u_n$ to their Higgs-vacuum values in the exterior region~\eqref{outsidecore} occurs around~$u \simeq u_n$. As we will see below, the width of this boundary region is~$\CO(1)$ as~$n \to \infty$, while the core region grows as in~\eqref{untoinfty}.

To understand why the core region wants to spread out in the large-$n$ limit, let us consider the contribution of the constant magnetic field~\eqref{Bconstcore} to the string tension~\eqref{eq:ahmtension}, 
\begin{equation}\label{magstatE}
    {T_n \over 2 \pi v^2} \supset  \half \int_0^{u_n} u \, du \, B^2 =   \frac{n^2}{u_n^2}~.
\end{equation}
This magneto-static energy can be made arbitrarily small by increasing~$u_n$ without bound -- a runaway that is eventually stabilized by the energetics of the Higgs field~$\varphi$. The latter are sensitive to the shape of the scalar potential, and they turn out to be qualitatively very different in the conventional and the degenerate Abelian Higgs models. This leads to different scalings of the core radius~$u_n$, and hence also the string tension~$T_n$, as~$n \to \infty$, which we will now determine.

\subsection{Strings in a Large-$n$ Bulk Phase}

We now specialize to axially symmetric strings in the conventional Abelian Higgs model, with tension~$T_n$ given by~\eqref{eq:ahmtension} and quartic potential~$\t V(\varphi) = \t V_4(\varphi)$ in~\eqref{rescaledV}. 

In the large-$n$ limit, the discussion of section~\ref{sec:intuition} above amounts to the statement that the magnetic field~$B$ and the Higgs field~$\varphi$ in the core and exterior regions of the string approximately take the following constant values,
\begin{equation}\label{eq:ahmvarfieldprofile}
    B(u) \simeq \begin{cases}
        {2 n \over u_n^2} \qquad u \lesssim u_n \quad (\text{core}) \\ 0 \qquad \;\;\, u \gtrsim u_n \quad (\text{exterior}) \\
    \end{cases}~, \qquad  \varphi(u) \simeq \begin{cases}
        0 \qquad u\lesssim u_n \quad (\text{core}) \\ 1 \qquad  u \gtrsim u_n \quad (\text{exterior})\\
    \end{cases}
\end{equation}
When writing these formulas, we are omitting an~$\CO(1)$ region around~$u \simeq u_n$ in which the fields transition from their small-$u$ core values to their large-$u$ Higgs-vacuum values. We will estimate the contribution of this region below, and show that it is subleading at large $n$. 

Since the Higgs vacuum in the exterior region~$u \gtrsim u_n$ in~\eqref{eq:ahmvarfieldprofile} essentially does not contribute to the string tension~$T_n$ in~\eqref{eq:ahmtension}, it follows that the tension is entirely given by integrating over the core region~$u \lesssim u_n$,
\begin{equation}\label{eq:ahmtensionvarlargen}
        \frac{T_{n , \text{ core}}}{2\pi v^2} \simeq \frac{n^2}{u_n^2} +  \frac{\beta u_n^2}{4}~, \qquad v^2 = \sqrt{2}~.
\end{equation}
Here, the first term is the magneto-static energy already discussed in~\eqref{magstatE}, which favors large~$u_n$. By contrast, the second term, which is extensive (it scales like the area~$\sim u_n^2$ of the core region) arises from the Higgs potential in~\eqref{rescaledV},  which contributes~$\t V_4(\varphi = 0) = \beta / 2$. Here, it is crucial that the point~$\varphi = 0$ is not a minimum of the potential; instead, it contributes a positive energy density in the core, which penalizes large~$u_n$. 

Minimizing~\eqref{eq:ahmtensionvarlargen} with respect to~$u_n$ amounts to balancing the runaway magnetic-static term and the extensive potential term. This fixes~$u_n$ and the string tension -- notably their scaling with~$n$ -- in the large-$n$ limit, 
\begin{equation}\label{eq:ahmvarradiustension}
    u_n^2(\beta) \simeq \frac{2n}{\sqrt{\beta}}~, \qquad  \frac{T_n(\beta)}{2\pi} \simeq \sqrt{2\beta} \, n~, \qquad n \to \infty~. 
\end{equation}
Several comments are in order:
\begin{itemize}
\item The radius~$u_n \sim \sqrt{n} \to \infty$ grows without bound in the large-$n$ limit, consistent with~\eqref{untoinfty}. Thus, the string broadens indefinitely, and it effectively opens up a 3+1 dimensional bulk region in its core. However, this bulk region is not in the vacuum. Rather, it harbors a finite, non-zero magnetic field (see~\eqref{eq:ahmvarfieldprofile}),
\begin{equation}\label{bulkB}
    B\left(u \lesssim u_n\right) = \sqrt{\beta} > 0 \qquad \text{as} \qquad n \to \infty~,
\end{equation}
which pins the Higgs field to~$\varphi = 0$. For this reason we will refer to it as a bulk Coulomb phase, even though the point~$\varphi = 0$ is not a minimum of the quartic potential~$\t V_4(\varphi)$. 
\item The string tension~\eqref{eq:ahmvarradiustension} is extensive in the large-$n$ limit, with a finite, positive energy density given by
\begin{equation}\label{extensive}
    {T_n \over  \pi u_n^2} \simeq \sqrt{2} \beta~, \qquad u_n \sim \sqrt{n} \to \infty~,
\end{equation}
which characterizes the bulk Coulomb phase (with constant magnetic field~\eqref{bulkB}) in the interior of the string. For this reason we say that the large-$n$ strings in the conventional Abelian Higgs model are in a bulk phase. This can be generalized to other models for which the point~$\varphi = 0$ is not a minimum of the potential.

\item When~$\beta = 1$ is at the BPS point, the leading large-$n$ tension~\eqref{eq:ahmvarradiustension} precisely matches the exact BPS tension in~\eqref{BPST}. Thus there are no~$1/n$ corrections at this point. 

\item In the above discussion we have neglected the boundary region around~$u \simeq u_n$ that separates the core region of the string from the Higgs vacuum in the exterior region. As we will show in section~\ref{bulklargensec} below, the width of this region remains~$\CO(1)$ in the large-$n$ limit, while the radius~$u_n \to \infty$ of the bulk region that it surrounds grows without bound. For this reason we can neglect the curvature of the boundary, and estimate its contribution to the string tension as
\begin{equation}\label{bdysigma}
    T_{n, \text{ boundary}} \simeq (2 \pi u_n) \,  \sigma(\beta)~,
\end{equation}
where~$\sigma(\beta)$ is the rescaled,\footnote{~\label{fn:dwtension}If~$\sigma_\text{DW}$ is the conventional physical surface tension, which has units of~$(\text{energy})^3$, then the contribution of a cylindrical interface of physical radius~$r$ to the physical string tension~$T$ (with units of~$(\text{energy)}^2$) is~$\Delta T = 2 \pi r \sigma_\text{DW}$. Since~$u = e v r$ and we are setting units via~$v^2 = \sqrt{2}$, we conclude that in these units
\begin{equation}
    \sigma = {\sigma_\text{DW} \big|_{v^2 = \sqrt{2}}  \over 2^{1/4} e }~.
\end{equation}
For instance, the domain wall of the degenerate model, with tension~$\sigma_\text{DW}$ in~\eqref{walltension}, has~$\sigma = \sqrt{\beta}/2$.} dimensionless~$\CO(1)$ tension of the straight codimension-one surface that separates the bulk Coulomb phase stabilized by the constant magnetic field in the core of the string from the Higgs vacuum in the exterior region. Since~\eqref{bdysigma} scales as the perimeter of the core region rather than its area, it is sub-extensive and hence subleading compared to the extensive large-$n$ core contribution~\eqref{extensive}. 
\end{itemize}

For future use we define an improved estimate of the core radius~$u_n$ that arises from minimizing both the leading bulk and the subleading boundary contributions to the string tension,
\begin{equation}\label{unimp}
    u_n^\text{imp} = {\sqrt{2n} \over \beta^{1/4}} - {\sigma \over 2 \sqrt{2} \beta}~.
\end{equation}
Note that the subleading correction is~$\CO(1)$, and that its sign is fixed by the sign of~$\sigma$. It is a well-known fact (which we will verify below) that $\sigma > 0$ in the type-I regime, where~$\beta < 1$, while~$\sigma < 0$ in the type-II regime, where~$\beta > 1$. This implies that we expect the large-$n$ solution to be to the right of the true numerical solution in the former case, and to its left in the latter, consistent with figure~\ref{flux30comparision}.

\subsection{Strings in a Large-$n$ Domain-Wall Phase}
\label{sec:largendwphase}

Let us now discuss how the preceding discussion of giant vortex strings -- with a core of radius~$\sim u_n$ surrounded by an exterior Higgs-vacuum region -- is modified in the degenerate Abelian Higgs model, with sextic potential~$\t V(\varphi) = \t V_6(\varphi)$ in~\eqref{rescaledV}. (Again, the discussion is somewhat heuristic; see section~\ref{walllargensec} for a rigorous treatment.) The defining feature of this model is that the Higgs vacuum at~$\varphi = 1$ is exactly degenerate with a Coulomb vacuum at~$\varphi = 0$, with the domain wall~\eqref{wallsoln} of tension~$\sigma_\text{DW}$ in~\eqref{walltension} interpolating between them. This has two important consequences for the energy budget of strings in the large-$n$ limit:
\begin{itemize}
\item Since the (approximately constant) values of~$B(u)$ and~$\varphi(u)$ in the core and exterior of the string are still given by~\eqref{eq:ahmvarfieldprofile}, it follows that the scalar potential~$\t V_6(\varphi)$ attains its minimum in both regions; hence, it does not contribute extensively to the string tension. Thus the entire contribution from the core of the string is due to the runaway magnetic energy~\eqref{magstatE},
\begin{equation}\label{coredwrun}
    T_{n, \, \text{core}} \simeq (2 \pi \sqrt{2}) \, {n^2 \over u_n^2}~.
\end{equation}

\item In the degenerate model, the runaway~\eqref{coredwrun} is stabilized by a non-extensive boundary term of the form~\eqref{bdysigma},
\begin{equation}\label{dwbdybis}
    T_{n, \, \text{boundary}} \simeq (2 \pi u_n) \sigma(\beta)~, \qquad \sigma(\beta) = {\sigma_\text{DW} \over 2^{1/4} e} = {\sqrt{\beta} \over 2}~.
\end{equation}
Here~$\sigma(\beta)$ is the tension $\sigma_\text{DW}$ (see~\eqref{walltension}) of the straight domain wall in the degenerate model, translated to our rescaled units (see footnote~\ref{fn:dwtension}).
\end{itemize}
The total string tension is given by adding the core~\eqref{coredwrun} and boundary~\eqref{dwbdybis} contributions,
\begin{equation}
    T_n \simeq 2 \pi \left( {\sqrt 2 n^2 \over u_n^2} + \sigma(\beta) u_n\right)~.
\end{equation}
Minimizing with respect to~$u_n$, we find that the core radius and the string tension are
\begin{equation}\label{uTdw}
    u_n = {\sqrt{2}  n^{2/3} \over \sigma^{1/3} }~, \qquad T_n = (2 \pi u_n) \, {3 \over 2} \sigma~, \qquad \sigma = {\sqrt{\beta} \over 2}~.
\end{equation}
Here~$\sigma$ is the domain-wall tension in~\eqref{dwbdybis}.

Some comments are in order:
\begin{itemize}
    \item In the large-$n$ limit~$T_n \sim u_n \sim n^{2/3} \to \infty$, so that the strings are still giant in this limit. The string tension scales sub-extensively, i.e. the bulk energy density vanishes,
    \begin{equation}
        {T_n \over \pi u_n^2} = 0 \qquad \text{as} \qquad n \to \infty~.
    \end{equation}
    This is consistent with the fact that the bulk phase in the core of the string realizes the Coulomb vacuum of the degenerate Abelian Higgs model, without a magnetic field (see~\eqref{eq:ahmvarfieldprofile}),
    \begin{equation}\label{wallB}
        B\left(u \lesssim u_n\right) =  \left({\beta \over 4 n}\right)^{1/3} \to 0 \qquad \text{as} \qquad n \to \infty~.
    \end{equation}
    This constitutes a sharply distinct large-$n$ phase from the bulk phase of the large-$n$ strings in the conventional AHM with quartic potential, which have positive-energy-density~\eqref{extensive} and a non-zero magnetic field~\eqref{bulkB}.
    
    \item The string tension scales as a surface term~$T_n = (2\pi u_n)\sigma_\text{eff}$, with effective surface tension
    \begin{equation}
        \sigma_\text{eff} = {3 \over 2} \sigma~.
    \end{equation}
    Here~$\sigma >0$ is the tension of the straight domain wall interpolating between the Coulomb and Higgs vacua of the degenerate Abelian Higgs model. For this reason we say that the large-$n$ strings in that model realize a domain-wall phase. The fact that~$\sigma_\text{eff} > \sigma$ is due to the fact that~$\sigma_\text{eff}$ also includes the magnetic energy of the string. 
\end{itemize}

\subsection{Comparing the Bulk and Domain-Wall Phases}

\begin{figure}[t!]
\centering
\includegraphics[width=0.49\textwidth]{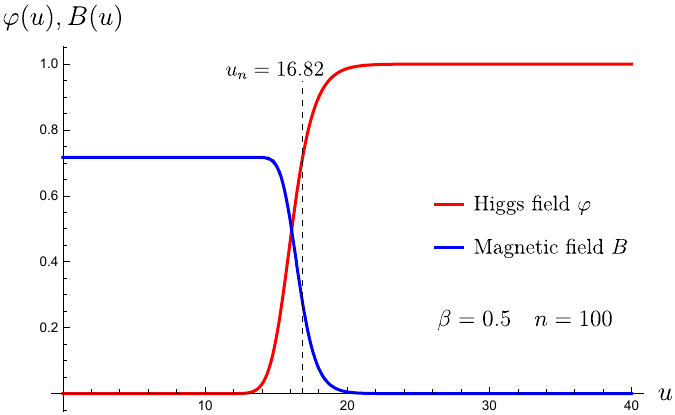}
\includegraphics[width=0.49\textwidth]{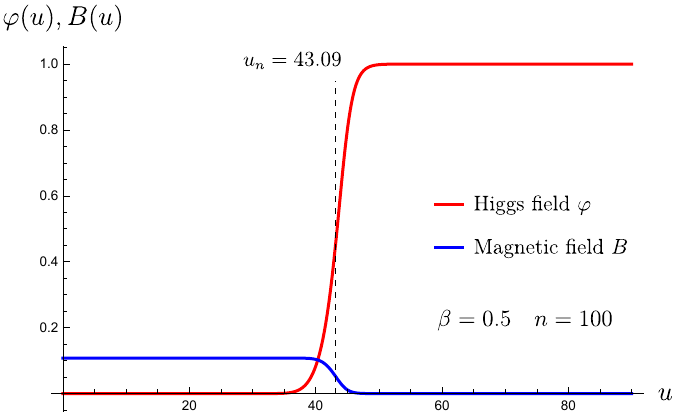}
\caption{\label{fig:bulkwallcomp} Higgs and magnetic field profiles for a large string of magnetic flux~$n = 100$ in the conventional (left panel) and degenerate (right panel) Abelian Higgs models.  The dimensionless mass ratio~$\beta = 0.5$ is the same for both models.}
\end{figure}

Let us compare the bulk and domain wall phases,  see figure~\ref{fig:bulkwallcomp} for a representative case. In both cases the core radius~$u_n \to \infty$ at large~$n$, but the rate is different (see~\eqref{eq:ahmvarradiustension}, \eqref{uTdw}),
\begin{equation}
    u_n \big|_\text{bulk} \sim n^{1/2}~, \qquad u_n \big|_\text{domain-wall} \sim n^{2/3} \qquad \text{as} \qquad n \to \infty~.
\end{equation}
The basic reason why the string is much more spread out in the domain-wall phase is that its core does not pay an extensive energy penalty for remaining in the Coulomb vacuum~$\varphi = 0$ of the degenerate sextic potential~$\t V_6$, while it does pay such a penalty for the conventional quartic potential~$\t V_4$, for which~$\varphi  = 0$ is not a minimum.

The fact that the core of the domain-wall-phase strings is more dilute is also visible in other observables, e.g. the magnetic field in the core (see~\eqref{bulkB}, \eqref{wallB}),
\begin{equation}
    B(u \lesssim u_n)\big|_\text{bulk} = \sqrt{\beta}~, \qquad B(u \lesssim u_n)\big|_\text{domain wall} \sim {1 \over n^{1/3}} \qquad \text{as} \qquad n \to \infty~,
\end{equation}
which is finite in the bulk phase, but vanishes in the domain-wall phase. Relatedly, the energy density, which scales as~$B^2$, is finite in the bulk phase (see~\eqref{extensive}), but vanishes in the domain-wall phase,
\begin{equation}
    {T_n\big|_\text{bulk} \over \pi u_n^2} = \sqrt 2 \beta~, \qquad {T_n\big|_\text{domain wall} \over \pi u_n^2} \sim {1 \over n^{2/3}} \qquad \text{as} \qquad n \to \infty~.
\end{equation}
Thus both the magnetic field and the energy density serve as good order parameters that distinguish these large-$n$ phases. By contrast, the Higgs field~$\varphi$ always vanishes in the string core, and thus it is not a good order parameter. 

It is easy to engineer a large-$n$ transition between these two phases by taking the potential~$\t V_6$ of the degenerate model (depicted in figure~\ref{fig:vtplots}) and perturbing it slightly by lifting the vacuum at the origin~$\varphi = 0$, so that~$\Delta \t V_6(\varphi = 0) > 0$, while maintaining the Higgs vacuum (i.e.~$\Delta \t V_6(\varphi = 1) = 0$). Then the large-$n$ strings will discontinuously jump from being in the domain-wall phase, with vanishing bulk magnetic field and energy density, to being in the bulk phase, with non-zero bulk magnetic field and energy density. Of course we are also losing the bulk Coulomb vacuum at~$\varphi = 0$, since these two phenomena are intertwined. However, the large-$n$ transition on the strings described above can be measured by an observer residing entirely in the Higgs vacuum of the model, which evolves smoothly (i.e.~it does not undergo a bulk phase transition) under the perturbation~$\Delta \t V_6$.


\section{Large-$n$ Solutions for Bulk Strings}\label{bulklargensec}

In this section we improve on the variational discussion of large-$n$ strings in section~\ref{seclargen}. Here we focus on the strings in the conventional Abelian Higgs model with quartic potential~$\t V_4$.  We will solve the axially-symmetric vortex equations~\eqref{eqn:ahmeom}, repeated here,
\begin{equation}\label{eqn:vorteqbis}
    \begin{split}
       \varphi''(u) + \frac{1}{u} \varphi'(u) = ~ & \frac{n^2}{u^2} a^2(u) \varphi(u) + \beta \varphi(u) (\varphi^2(u)-1)~,
       \\
        a''(u) - \frac{1}{u} a'(u) = & ~ 2 a(u) \varphi^2(u)~, \\
    \end{split}
\end{equation}
in the large-$n$ limit -- with fixed~$\beta$ -- using the method of matched asymptotic expansions. Aspects of this problem were previously considered in~\cite{PhysRevLett.125.251601,Penin_2021}.

\subsection{Regions of the Large-$n$ String}\label{regions}

\begin{figure}[t!]
\centering
\includegraphics[width=\textwidth]{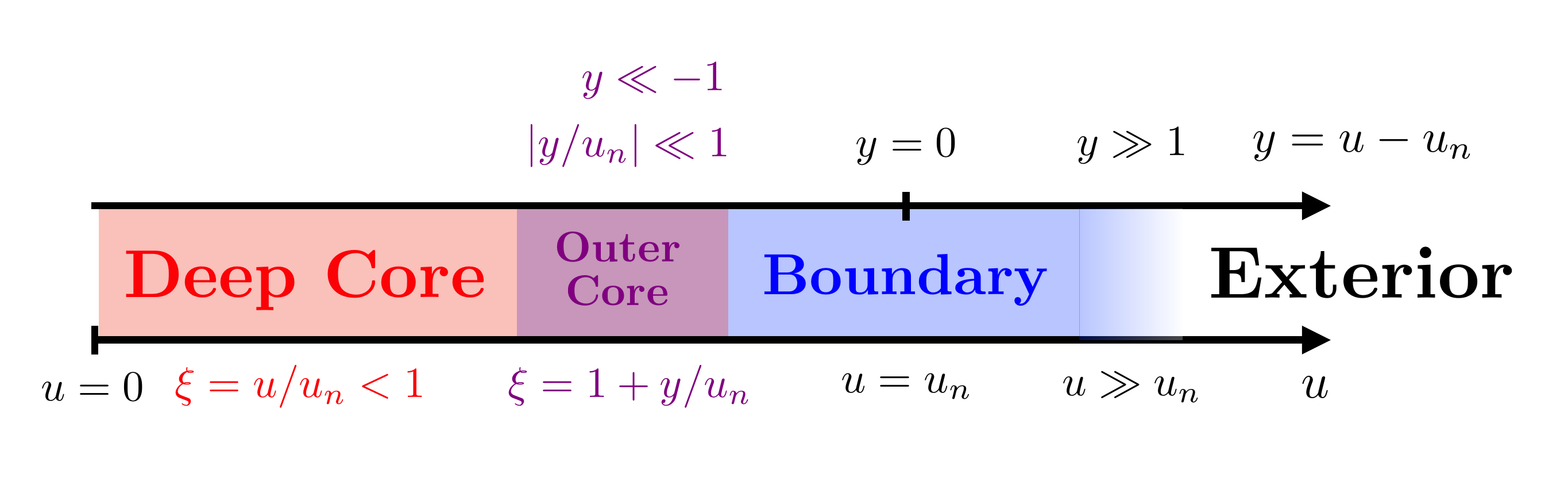}
\caption{\label{fig:regions} The deep and outer core, boundary, and exterior regions of large-$n$ strings, as well as their shaded overlaps. The coordinate~$\xi = u/u_n$ is particularly useful in the deep core. By contrast, the~$y$-coordinate interpolates between the outer core, the boundary, and the exterior.}
\end{figure}

In section~\ref{seclargen} we referred to the region~$u \lesssim u_n$, as the core of the string, $u \sim u_n$ as the boundary, and~$u \gtrsim u_n$ as the exterior. In the leading large-$n$ limit, the radius~$u_n$ of the string is given by~\eqref{eq:ahmvarradiustension}, which we repeat here,
\begin{equation}\label{eq:bulkradbis}
    {u_n}(\beta) = {\sqrt{2 n} \over \beta^{1/4}} \sim \sqrt{n}~, \qquad n \to \infty~.
\end{equation}
We will now make these regions and their scalings with~$n$ more precise. To this end, it is useful to introduce other coordinate systems, depicted in figure~\ref{fig:regions}, for the string: a shifted coordinate centered on~$u = u_n$, 
\begin{equation}\label{ydef}
    y \equiv u - u_n~,
\end{equation}
and a rescaling of~$u$ by~$u_n$,
\begin{equation}\label{xidef}
    \xi = {u \over u_n}~.
\end{equation}
In terms of these coordinates we can define the regions of the string as follows (see figure~\ref{fig:regions}):
\begin{itemize}
    \item[(C)] The defining feature of the string core region is that~$\varphi(u) \ll 1$ is small. As we will see below, this requires\footnote{~By this we mean that~$y$ is a sufficiently negative~$\CO(1)$ number that need not grow in the large-$n$ limit. The precise criterion~\eqref{ycond} depends on~$\beta$, which remains fixed in this limit.}
    \begin{equation}\label{eq:coredef}
 \text{Core:}\quad y \ll -1~.
\end{equation}
We will further distinguish between the deep core of the string, where~$\xi$ approaches some fixed number~$\xi < 1$ in the large-$n$ limit,\footnote{~Around~\eqref{phiorigin} we refer to the region where~$\xi = u/u_n \to 0$ in the large-$n$ limit as the deepest core.} 
\begin{equation}\label{dc}
    \text{Deep Core:} \quad  
      \xi  \in [0, 1)~, \quad   y = (\xi-1) u_n~, \quad n \to \infty~,
\end{equation}
and the outer core, where 
\begin{equation}\label{oc}
  \text{Outer Core:} \quad \xi = 1 + {y \over u_n} \to 1~, \quad -u_n \ll y \ll -1~, \quad n \to \infty~.
\end{equation}
In the outer core~$y/u_n \to 0$ in the large-$n$ limit, i.e.~$y$ grows more slowly than~$u_n \sim \sqrt{n}$.

\item[(B)] The boundary region is defined by
\begin{equation}\label{bdy}
    \text{Boundary}: \quad - u_n \ll y \ll u_n~, \qquad n \to \infty~.
\end{equation}
This overlaps with the outer core (shown in purple in figure~\ref{fig:regions}), where we will match the core and boundary solutions. 

\item[(E)] The exterior of the string is the region
\begin{equation}\label{ext}
   \text{Exterior:} \quad y \gg 1~.
\end{equation}
The overlap of the exterior with the boundary will also be used to match solutions in these regions. 

\end{itemize}

\subsection{Core Solution via the WKB Approximation}

As we will verify self-consistently below, the Higgs field~$\varphi(u) \ll 1$ is small throughout the core region~\eqref{eq:coredef}, and hence the vortex equations~\eqref{eqn:vorteqbis} linearize,
\begin{equation}\label{convlinvort}
    \varphi'' + {1 \over u} \varphi' = \left({n^2 a^2(u) \over u^2} - \beta\right) \varphi~, \qquad a(u) = 1-{u^2 \over u_n^2(\beta)}~.
\end{equation}
Let us remove the $\varphi'$-term by defining a new function
\begin{equation}\label{psidef}
    \varphi(u) = {\psi(u) \over \sqrt{u}}~,
\end{equation}
which satisfies
\begin{equation}\label{eq:schroed}
    \psi''(u) = p^2(u) \psi(u)~, \qquad p^2(u) \equiv {n^2 a^2(u) - {1 \over 4}  \over u^2} - \beta~. 
\end{equation}
As we will see~$p^2(u) > 0$ is positive throughout the core. We can thus think of~\eqref{eq:schroed} as a Schr\"odinger equation in the classically forbidden region, where the wavefunction~$\psi(u)$ varies exponentially. Below we will freely use quantum-mechanical terminology when discussing this equation. 

In this analogy, the large-$n$ limit is the standard semiclassical limit~$\hbar \sim {1 \over n} \to 0$. We will therefore attempt to solve the Schr\"odinger equation~\eqref{eq:schroed} using the WKB approximation (see~\cite{Berry:1972na} for an excellent review with references). The two linearly independent WKB solutions are
\begin{equation}\label{eq:wkbsol}
    \psi_{\text{WKB}\pm}(u) \simeq {1 \over \sqrt{p(u)}} \exp\left( \pm \int^u du' \, p(u')\right)~.
\end{equation}
The WKB approximation is valid as long as the following error term is uniformly small,
\begin{equation}\label{eq:err}
    \ep(u) \equiv {1 \over p^{3/2}(u)} {d^2 \over du^2} \left(p^{-1/2}(u) \right)~, \qquad \left|\ep(u) \right| \ll 1~.
\end{equation}
As usual, this condition is violated near classical turning points, where~$p(u)$ vanishes, but we will show that~\eqref{eq:err} holds throughout the core region as long as~$n \gg 1$. 

Let us examine the function~$p^2(u)$ in more detail,\footnote{~Here we have used~$n \gg 1$ to simplify the centrifugal barrier term in~$p^2(u)$ as follows,
\begin{equation}\label{cent}
   {n^2 -{1 \over 4} \over u^2} \quad \to \quad {n^2 \over u^2}~.
\end{equation}
At finite, non-zero~$n$ the standard WKB approximation breaks down near~$u = 0$ due to the rapidly varying centrifugal term~\eqref{cent}, as can be seen by examining the error~\eqref{eq:err}. If the other terms in~$p^2(u)$ are suitably slowly varying, the prescription~\eqref{cent} (known as the Langer modification, see~\cite{Berry:1972na}) makes it possible to apply this modified version of the WKB approximation even when~$n$ is not large. 
}
\begin{equation}\label{psqfac}
    p^2(u) \simeq {n^2 a^2(u) \over u^2} - \beta = \left({n a(u) \over u} - \sqrt{\beta}\right)\left({n a(u) \over u} + \sqrt{\beta}\right)~. 
\end{equation}
Thus~$p^2(u) > 0$ is positive for all~$0 \leq u < u_*$, with~$u_*$ the classical turning point where the first factor in~\eqref{psqfac} vanishes,
\begin{equation}\label{turnpt}
 p(u_*) = 0~, \qquad u_* \simeq u_n - 1 + \CO\left({1 \over \sqrt{n}}\right) \qquad \Longleftrightarrow \qquad y_* \simeq -1 + \CO\left({1 \over \sqrt{n}}\right)~. 
\end{equation}
To analyze~$p(u)$ in the large-$n$ limit, we use the coordinate~$\xi = u/u_n$ defined in~\eqref{xidef},
\begin{equation}
    a(\xi) = 1 - \xi^2~, \qquad p^2(\xi) = {n \sqrt{\beta} \over 2} \left({a (\xi) \over \xi}\right)^2 - \beta~. 
\end{equation}
The deep core~\eqref{dc} and the outer core~\eqref{oc} require separate treatment:
\begin{itemize}
\item[(DC)] In the deep core, $\xi \in [0,1)$ is an~$\CO(1)$ number that is to be held fixed as~$n \to \infty$.  Then~$a(\xi) \neq 0$ and we are free to Taylor expand
\begin{equation}\label{pofxiexp}
    p(\xi) = \sqrt{n \sqrt{\beta} \over 2}  \left({1- \xi^2 \over \xi }\right)  \sqrt{ 1 - { 2 \sqrt{\beta} \,  \xi^2 \over n (1-\xi^2)^2 }}\simeq \sqrt{n \sqrt{\beta} \over 2}  \left({1- \xi^2 \over \xi }\right)  \left(1 - {  \sqrt{\beta} \,  \xi^2 \over n (1-\xi^2)^2 } + \cdots \right)~.
\end{equation}
The expansion parameter is~$1/n$ and the ellipsis in~\eqref{pofxiexp} is~$\CO(1/n^2)$, both of which are small at large $n$. Similarly, substituting~\eqref{pofxiexp} into~\eqref{eq:err} we find that the error in the deep core is given by
    \begin{equation}\label{errofxi}
        \ep(\xi) \simeq {3 \xi^4 + 10 \xi^2 -1\over 4 n^2 ( \xi^2-1)^4 }~.
    \end{equation}
Thus the error is~$\CO(1/n^2)$ throughout the deep core,\footnote{~Note that~$\ep(\xi)$ vanishes at~$\xi \simeq 0.3117$.} and hence the WKB approximation is uniformly valid there. 

\item[(OC)] In the outer core we have
    \begin{equation}
        \xi = 1 + {y \over u_n} \to 1~, \qquad n \to \infty~,
    \end{equation}
    so that~$y$ grows more slowly than~$u_n \sim \sqrt{n}$ in the large-$n$ limit. In particular, $y$ may be an~$\CO(1)$ constant that does not grow with~$n$ at all. In this case the expansion~\eqref{pofxiexp} is still valid, but now the expansion parameter is
    \begin{equation}
        {2 {\sqrt \beta} \xi^2 \over n (1-\xi)^2 (1 + \xi)^2} \simeq {1 \over y^2}  \ll 1~,
    \end{equation}
    and the ellipsis in~\eqref{pofxiexp} is~$\CO(1/y^4)$. Even though these need not be large-$n$ suppressed, they are nevertheless small as long as~$y \ll -1$. Similarly, the error term~\eqref{errofxi} in the outer core region is given by
    \begin{equation}
        \ep(y) \simeq  {3 \over 4 \beta y^4}~,
    \end{equation}
    so that the WKB approximation is valid as long as 
    \begin{equation}\label{ycond}
        y \ll - {1 \over \beta^{1/4}}~.
    \end{equation}
\end{itemize}
We conclude that the WKB approximation is uniformly valid throughout the core as long as~$n \gg 1$ and~$y$ is sufficiently far from the turning point, as dictated by~\eqref{ycond}.

Given these conditions on~$n$ and~$y$, we can simplify the WKB wavefunctions~\eqref{eq:wkbsol}, starting with the integral in the exponent,
\begin{equation}\label{wkbexp}
     u_n \int^\xi d\xi' \, p(\xi') = \frac{n}{2}  \left(2 \log \xi -\xi ^2\right) + \frac{\sqrt{\beta }}{2}  \log \left(1-\xi ^2\right) + \cdots~.
\end{equation}
In the deep core, the error term represented by the ellipsis is~$\CO(1/n)$, and we will omit it. However, we have kept the~$\CO(1)$ term since it is of the same order as the leading contribution from the pre-factor~$1/\sqrt{p(\xi)}$.\footnote{~The most familiar version of the WKB approximation involves functions~$p(u)$ whose~$\hbar$-dependence is given by a single overall factor of~$1 / \hbar$. The exponent of the WKB wavefunctions~\eqref{eq:wkbsol} is then~$\CO(1/\hbar)$ large and classical, while the~$1/\sqrt{p(u)}$ prefactor gives subleading~$\CO(\hbar^0)$ quantum corrections. In our problem, the function~$p(u)$ has a more complicated~$\hbar \sim 1/n$ dependence (see~\eqref{pofxiexp}), and hence the exponent~\eqref{wkbexp} of the WKB wavefunctions contains both classical~$\CO(n)$ terms and subleading quantum corrections; the latter contribute at the same order as the prefactor.} For consistency, we will therefore drop all subleading terms in the prefactor. Altogether, we find that the WKB wavefunction takes the form
\begin{equation}
    \psi_{\text{WKB}+}(\xi) \simeq \left({2 \over n \sqrt{\beta} }\right)^{1 \over 4} \xi^{n + \half} (1 - \xi^2)^{(\sqrt{\beta} -1)/2}  \exp\left(- {n \over 2} \, \xi^2\right)~.
\end{equation}
Note that~$\psi_{\text{WKB}+}$ has the required behavior~$\varphi(u) \sim u^{-\half} \psi_{\text{WKB}+}(u) \sim u^n$ as~$u \to 0$, while~$\psi_{\text{WKB}-}$ does not and must therefore be discarded. The full WKB solution for the Higgs field is thus 
\begin{equation}\label{wkbhiggs}
    \varphi(u) = {C_\text{WKB}(n, \beta) \over \sqrt{n}}  \left({u \over u_n}\right)^n \left(1 - {u^2 \over u_n^2}\right)^{\sqrt{\beta} -1 \over 2}    \exp\left(- {\sqrt{\beta} \over 4} u^2 \right)~.
\end{equation}
Here~$C_\text{WKB}$ is a constant that can depend on~$n$ and~$\beta$.

Let us consider~\eqref{wkbhiggs} near the origin~$u = 0$, in the deepest part of the deep core region, where~$u/u_n \to 0$ in the large-$n$ limit. In this regime, the WKB prefactor in~\eqref{wkbhiggs} is subleading, so that the Higgs field simplifies,
\begin{equation}\label{phiorigin}
    \varphi(u) \simeq {C_\text{WKB}(n, \beta) \over \sqrt{n}}  \left({u \over u_n}\right)^n  \exp\left(- {\sqrt{\beta} \over 4} u^2 \right)~, \qquad u/u_n \to 0 \text{ as } n \to \infty~.
\end{equation}
This allows us to read off the coefficient~$c_n$ of the leading~$\CO(u^n)$ term in the Taylor series~\eqref{eqn:ahmoriginasym} of~$\varphi(u)$ around~$u = 0$,
\begin{equation}\label{cofn}
    c_n(\beta) = {C_\text{WKB}(n, \beta) \over \sqrt{n} u_n^n}~.
\end{equation}
The Higgs field~\eqref{phiorigin} was proposed in~\cite{PhysRevLett.125.251601,Penin_2021} throughout the entire core region, but as we see from~\eqref{wkbhiggs} this ceases to be correct once~$u/u_n$ -- and hence the WKB prefactor -- are~$\CO(1)$. (Importantly, this includes the outer core matching region with the boundary, see below.) An exception occurs at the BPS point~$\beta = 1$, as we will discuss in section~\ref{sec:bpslargen} below.

We can also expand~\eqref{wkbhiggs} in the outer core region~\eqref{oc}, using the~$y$-coordinate~\eqref{ydef}, 
\begin{equation}\label{phiofy}
    \varphi(y) =  C_\text{core/bdy}(\beta) \, |y|^{\sqrt{\beta} -1 \over 2}  \exp\left(-{\sqrt{\beta} \over 2} y^2\right)~, \quad -u_n \ll y \ll -1~,
\end{equation}
so that~$|y|/u_n \to 0$ as~$n \to \infty$. Here the prefactor is defined as follows,
\begin{equation}\label{Ccbdef}
   C_\text{core/bdy}(\beta) = {C_\text{WKB}(n, \beta) \over \sqrt{n}}  \left({2   \over u_n}\right)^{\sqrt{\beta} -1 \over 2} \exp\left(-{ n \over 2}\right)~.
\end{equation}
In section~\ref{bulkbdymatch} we will show that this constant only depends on~$\beta$ -- but not on~$n$ -- in the large-$n$ limit, by matching~\eqref{phiofy} with the Higgs field in the boundary region of the string. Thus~$\varphi(y)$ in~$\eqref{phiofy}$ attains the regime~$\varphi(y) \ll 1$ precisely when the condition~\eqref{ycond} for the validity of the WKB approximation is met. This self-consistently justifies our analysis of the linearized vortex equations~\eqref{convlinvort} in the core region.

Finally, eliminating~$C_\text{WKB}$ from~\eqref{cofn} and~\eqref{Ccbdef} leads to the following relation,
\begin{equation}
    c_n(\beta) = {e^{n/2}  \over u_n^{n}}  \left({2 \over u_n} \right)^{(1-\sqrt{\beta})/2} C_\text{core/bdy}(\beta)~.
\end{equation}
Since~$C_\text{core/bdy}$ does not depend on~$n$, this completely fixes the~$n$-dependence of~$c_n$.

\subsection{Boundary Solution and Matching}\label{bulkbdymatch}

We will now solve the vortex equations~\eqref{eqn:vorteqbis} in the boundary region~\eqref{bdy}, where $|y|/u_n \to 0$ as~$n \to \infty$. We will then match these solutions to the core and exterior regions, completing our large-$n$ vortex solution.

\subsubsection{Simplifying the Vortex Equations in the Boundary Region}

Since~$u \gg u_n \sim \sqrt{n}$ is uniformly large in the boundary region, we drop the curvature terms~$\sim 1/u$ on the left-hand side of~\eqref{eqn:vorteqbis} and work to leading order in small~$y/ u_n$, so that
\begin{equation}\label{eqn:bdyeom}
    \begin{split}
        \varphi''(y)  = & ~ \gamma^2 (y)\varphi(y) + \beta \varphi(y)(\varphi^2(y)-1)~, \\ 
        \gamma''(y) = & ~ 2 \gamma(y) \varphi^2(y)~. \\
    \end{split}
\end{equation}
Here, following \cite{Penin_2021}, we have defined a rescaled version of the gauge field~$a(y) \equiv a(u = u_n +y)$ that is~$\CO(1)$ throughout the boundary region,
\begin{equation}\label{gammadef}
    \gamma(y) = \frac{n}{u_n+y} a(y) \simeq {n \over u_n} a(y)~.
\end{equation}
The boundary equations~\eqref{eqn:bdyeom} have two features that are not shared by the full vortex equations~\eqref{eqn:vorteqbis}:
\begin{itemize}
    \item They are $n$-independent, and as we will see below, no $n$-dependence is introduced by boundary conditions (more precisely, by matching). Thus, the solutions~$\varphi(y), \gamma(y)$ are also~$n$-independent. In turn~$a(y) \sim \gamma(y) / \sqrt{n}$ is uniformly small at large $n$. 
\item They  have translational symmetry in~$y$. This arises because the curvature of the interface separating the Coulomb phase in the core and the Higgs phase in the exterior is negligible  in the large-$n$ limit. Translational symmetry is broken, and the associated would-be modulus of the boundary solution is fixed, by matching to the core (see below).
\end{itemize}

\subsubsection{Matching to the Core and Exterior Solutions}\label{bulkmatch}

Even though the boundary equations~\eqref{eqn:bdyeom} are valid as long as~$|y| \ll u_n$, it is convenient (both for analytical and numerical reasons) to perform the matching to the core and exterior regions for large~$|y| \gg 1$ that are fixed in the large-$n$ limit. 

In the outer core matching region, where~$y \ll -1$, the Higgs field is given by~\eqref{phiofy},
\begin{equation}\label{hrep}
    \varphi(y) = C_\text{core/bdy}(\beta) \, |y|^{\sqrt \beta -1 \over 2} \exp\left(- {\sqrt \beta \over 2} y^2\right)~, \qquad y \ll -1~,
\end{equation}
while the rescaled gauge field~\eqref{gammadef} takes the form
\begin{equation}\label{gmatch}
    \gamma(y) = - \sqrt{\beta} y~, \qquad y \ll -1~.
\end{equation}
In both equations, we have taken the large-$n$ limit at fixed~$y \ll -1$. (Recall that~$C_\text{core/bdy}$ is $n$-independent.) Since the Higgs field~\eqref{hrep} is small, $\varphi(y) \ll 1$, it can be obtained by solving the boundary equations~\eqref{eqn:bdyeom} in a regime where they linearize, as was the case in the core. Then~$\gamma(y)$ is an arbitrary quadratic function of~$y$, but matching with~\eqref{gmatch} fixes~$\gamma(y) = -\sqrt{\beta} y$. (As anticipated above, this breaks translation symmetry in~$y$.) The resulting linearized equation for~$\varphi(y)$ is a parabolic cylinder (or Weber) equation, but since we are interested in~$y \ll -1$, it suffices to use the WKB approximation (already discussed around~\eqref{ycond} for the outer core region) to recover~\eqref{hrep}.

In the exterior matching region~$y \gg 1$ (or~$u \gg u_n$), we must match the boundary solutions to the asymptotic formulas~\eqref{eq:ahmasym},  Taylor expanded in small~$y / u_n \ll 1$,
\begin{equation}\label{eq:ahmasymbis}
    \gamma(y) = \gamma_\text{bdy/ext}e^{-\sqrt{2}y}~, \qquad \varphi(y) \simeq 1- \varphi_\text{bdy/ext} e^{-\sqrt{2\beta}y}  - \frac{\gamma^2_\text{bdy/ext}  e^{-2\sqrt{2}y}}{2(\beta-4)}~, \qquad y \gg 1~,
    \end{equation}
where
\begin{equation}\label{eq:bdytailmatch}
     \gamma_\text{bdy/ext} = {n a_\infty \over \sqrt{u_n}} e^{-\sqrt{2} u_n}~, \qquad \varphi_\text{bdy/ext} = \frac{\varphi_\infty}{\sqrt{u_n}} e^{-\sqrt{2\beta} u_n}~, \qquad u_n = {\sqrt{2n} \over \beta^{1/4}}~.
\end{equation}
Since the boundary solution, and hence the constants~$\gamma_\text{bdy/ext}, \varphi_\text{bdy/ext}$ in the exterior matching region, are~$n$-independent, it follows that the~$n$-dependence of~$a_\infty, \varphi_\infty$ is entirely fixed by~\eqref{eq:bdytailmatch}. (The leading exponentials in these relations were already discussed in~\cite{Penin_2021}.) It can indeed be checked that~\eqref{eq:ahmasymbis} are asymptotic solutions of~\eqref{eqn:bdyeom} close to the Higgs vacuum~$\gamma = 0, \varphi = 1$ (see the closely related discussion around~\eqref{eq:ahmasym}).

\subsubsection{Numerical Solution of the Boundary Equations}\label{sec:numericsbdy}
In this section, we present a systematic numerical analysis of the vortex solutions and the associated constants that characterize their different spatial regions. Our goal is to verify the analytic relations derived in the large-$n$ limit and to quantify the accuracy and consistency of these analytics for finite values of the flux for various values of $\beta$.

\begin{itemize}
    \item We can solve the vortex equations in the boundary region, given in~\eqref{eqn:bdyeom}, numerically. For illustration in figure~\ref{fig:bdynumericalsolution}, we plot the field profiles for the Higgs field~$\varphi(y)$ and the gauge field~$\gamma(y)$ for~$\beta=0.5$ and~$\beta = 3$, obtained by solving the vortex equations~\eqref{eqn:bdyeom} numerically.
    \begin{figure}[t!]
        \centering
        \includegraphics[width=0.49\textwidth]{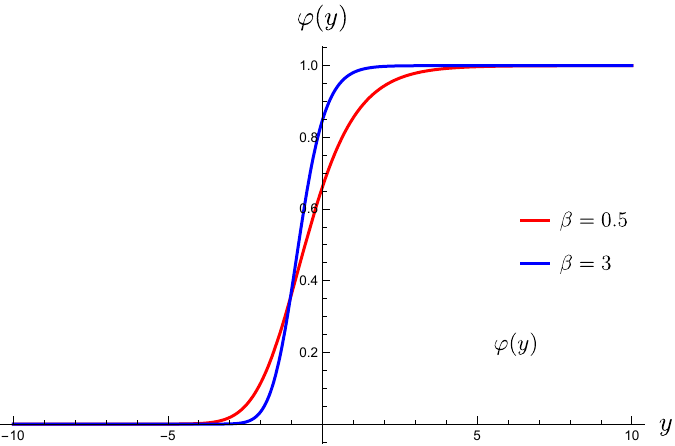}
        \includegraphics[width=0.49\textwidth]{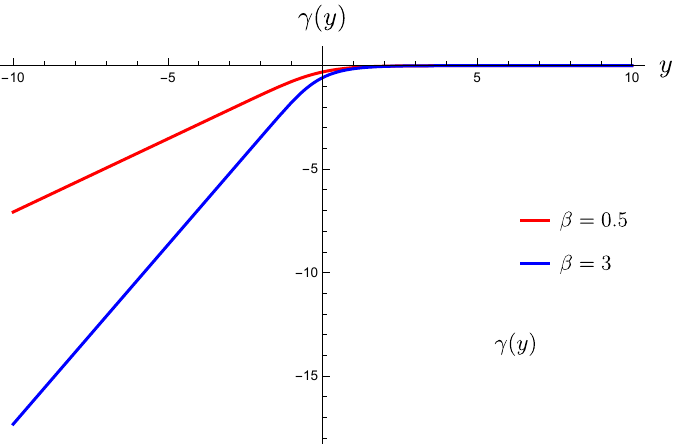}
        
        \caption{\label{fig:bdynumericalsolution} For two values of $\beta$, we depict the solutions $\varphi(y), ~\gamma(y)$ of the vortex equations in the boundary region, given in \eqref{eqn:bdyeom}. }
    \end{figure}

\item In the outer core matching region, where $y\ll-1$, we can match the core solution given in \eqref{hrep} with the numerically obtained boundary solution for the Higgs field. This determines the constant~$C_{\text{core/bdy}}(\beta)$. For various~$\beta$, the value of $C_{\text{core/bdy}}(\beta)$ is listed in table~\ref{fr}. For~$\beta = 1$ we find agreement with~\cite{Penin_2021}, who evaluated~$C_{\text{core/bdy}}(\beta=1) = \frac{1}{\sqrt{e}} = 0.61$.

To illustrate this explicitly, for two values of~$\beta = 0.5,~0.9$, we verify the matching between the boundary and the core solution \eqref{hrep} in figure~\ref{fig:logexhibit}.

\begin{table}[]
\centering
\setlength\extrarowheight{5pt}
\begin{tabular}{ |p{2.5cm}|p{2.5cm}|   } 
    \hline
    \multicolumn{2}{|c|}{ $C_{\text{core/bdy}}(\beta)$} \\
    \hline
        $\beta$ & $C_{\text{core/bdy}}(\beta)$\\
    \hline
        $ \beta = 0.1  $  & $0.41$   \\
        $ \beta = 0.2$  & $0.45$     \\
        $\beta = 0.3 $  & $0.48$    \\
        $ \beta = 0.4 $  & $0.50$    \\ 
        $ \beta = 0.5 $  & $0.53$    \\ 
        $ \beta = 0.6 $  & $0.54$    \\ 
        $ \beta = 0.7 $  & $0.56$    \\ 
    \hline
\end{tabular}
\quad \quad 
\begin{tabular}{ |p{2.5cm}|p{2.5cm}|   } 
    \hline
    \multicolumn{2}{|c|}{ $C_{\text{core/bdy}}(\beta)$} \\
    \hline
        $\beta$ & $C_{\text{core/bdy}}(\beta)$\\
     \hline
        $ \beta = 0.8$  & $0.58$     \\
        $\beta = 0.9$  & $0.59$    \\
        $ \beta = 1  $  & $ \frac{1}{\sqrt{e}} = 0.61$   \\
        $ \beta = 2 $  & $0.74$    \\ 
        $ \beta = 3 $  & $0.86$    \\ 
        $ \beta = 4 $  & $0.97$    \\ 
        $ \beta = 5 $  & $1.08$    \\ 
    \hline
\end{tabular}
\caption{Variation of~$C_{\text{core/bdy}}(\beta)$ with parameter~$\beta$. It monotonically increases with~$\beta$. At the BPS point~$\beta=1$,\cite{Penin_2021} calculate~$C_{\text{core/bdy}}(\beta=1)$ explicitly to be~$1/\sqrt{e}$. This is corroborated numerically in the table. }
\label{fr}
\end{table}

 \begin{figure}[t!]
    \centering
    \includegraphics[width=0.49\textwidth]{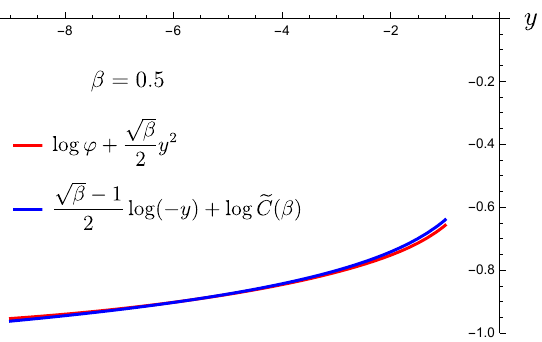}
    \includegraphics[width=0.49\textwidth]{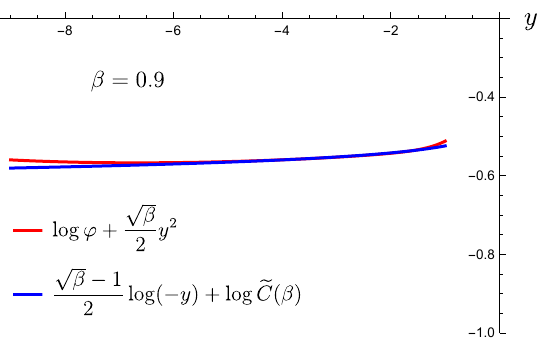}
    \includegraphics[width=0.49\textwidth]{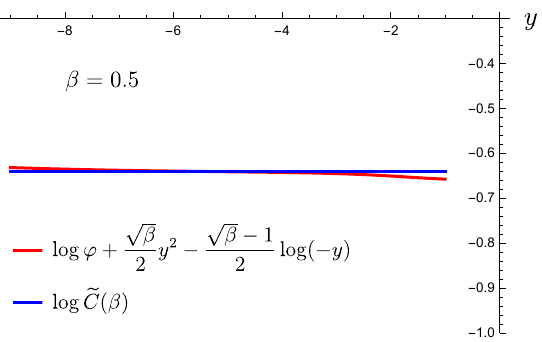}
    \includegraphics[width=0.49\textwidth]{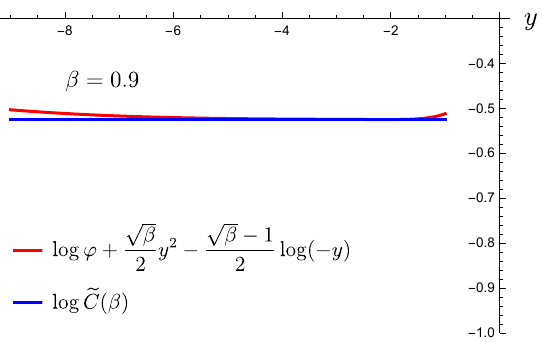}
        
    \caption{\label{fig:logexhibit} The boundary function~$\varphi(y)$ is computed numerically and matched with~\eqref{hrep} in the outer core matching region for~$y\ll-1$. We show two values of~$\beta=0.5,~0.9$.  This matching procedure yields the value of~$C_{\text{core/bdy}}(\beta)$.}
\end{figure}

\item With~$C_{\text{core/bdy}}(\beta)$ determined, we can directly compare the large-$n$ analytic solution to the full numerical solutions for sufficiently large flux. In figure~\ref{fig:type1V4comparesolution}, we show these comparisons for a representative value of~$\beta = 0.5$ and flux~$n=200$. We see that the agreement between the analytic profiles for the Higgs field $\varphi(u)$ and the gauge field $a(u)$ and the full numerical solutions to~\eqref{eqn:ahmeom} is excellent across the entire radial domain.

To assess the effectiveness of the large-$n$ expansion at moderately small flux, in figure~\ref{flux30comparision} we compare the numerical Higgs field~$\varphi(u)$ profiles with the analytic large-$n$ solution for two representative values of~$\beta$, one in the type-I regime~($\beta=0.1$) and one in the type-II regime~($\beta=3$). In addition to the core radius~$u_n$, given in ~\eqref{eq:bulkradbis}, obtained from minimizing the bulk energy, we also consider the improved radius~$u_n^{\text{imp}}$, given in~\eqref{unimp}, which includes the~$\CO(1)$ correction arising from the combined bulk and boundary contributions. The improved radius shifts the analytic profile towards small radii in the type-I regime (where~$\sigma>0$) and towards larger radii in the type-II regime (where~$\sigma<0$). In both regimes, this correction leads to a visibly improved agreement with the numerical solution even for flux as small as~$n=30$.

Finally, in figure~\ref{fig:fundamentalV4comparesolution}, we contrast the large-$n$ solution with the numerical profiles of the fundamental~$n=1$ string for~$\beta=0.5$. 

\begin{figure}[t!]
    \centering
    \includegraphics[width=0.49\textwidth]{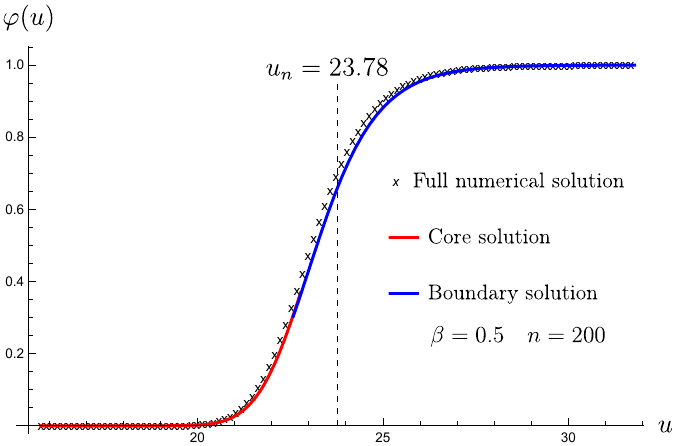}
    \includegraphics[width=0.49\textwidth]{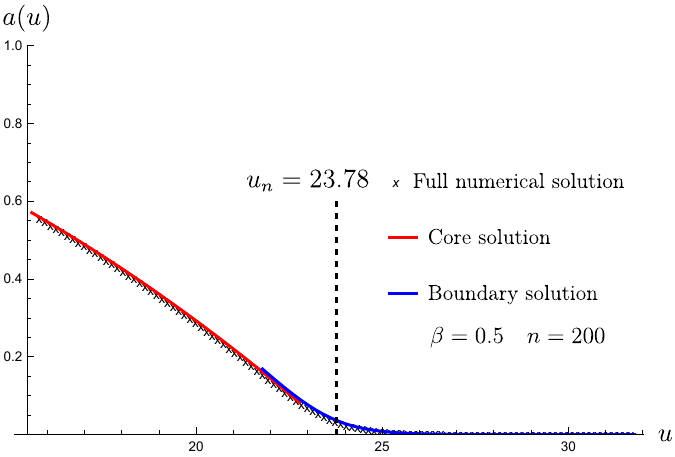}
    \caption{\label{fig:type1V4comparesolution}
    Comparison between the full numerical solution to the vortex equations~$\eqref{eqn:ahmeom}$ (shown as crosses) and the large-$n$ analytic solution for representative parameters~$\beta =0.5$ and~$n=200$. The core solution is shown in red and the boundary solution in blue.}
\end{figure}

\begin{figure}[t!]
    \centering
    \includegraphics[width=0.49\textwidth]{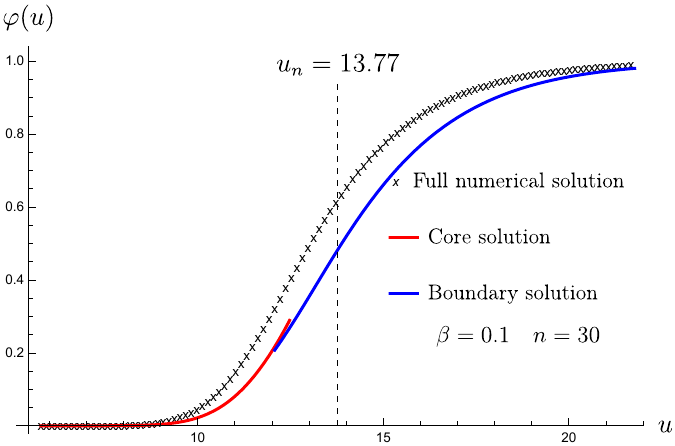}
    \includegraphics[width=0.49\textwidth]{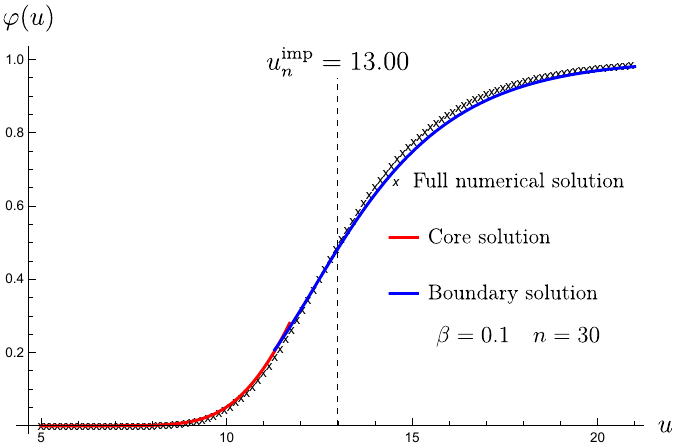}
    \includegraphics[width=0.49\textwidth]
    {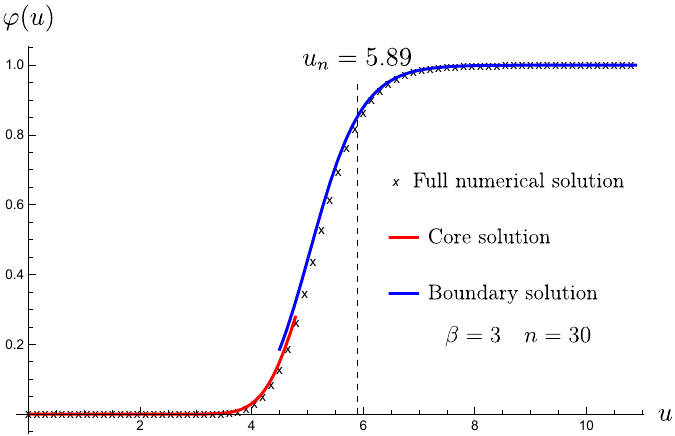}
    \includegraphics[width=0.49\textwidth]{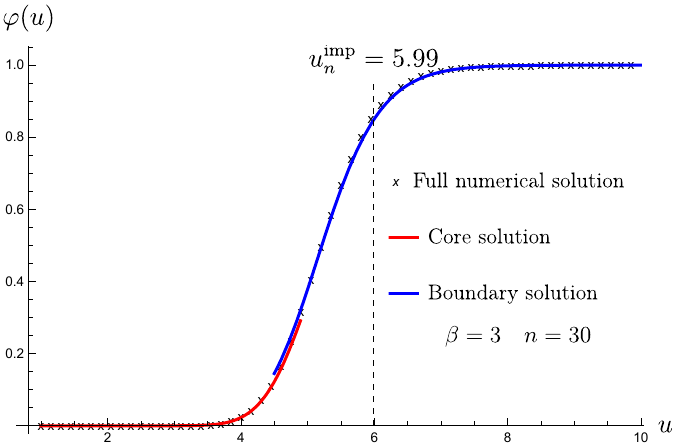}
    
    \caption{\label{flux30comparision}
    Comparison of the Higgs field profile obtained from the full numerical solution (depicted as crosses) of the vortex equations~\eqref{eqn:ahmeom} with the corresponding large-$n$ analytic answers (depicted as solid curves), shown for~$\beta=0.1$ in the type-I regime and~$\beta=3$ in the type-II regime, with~$n = 30$. We show results using the core radius~$u_n$ given in~\eqref{eq:bulkradbis} (left panels) and using the improved value~$u_n^{\text{imp}}$, given in~\eqref{unimp} (right panels). The improved radius~$u_n^{\text{imp}}$ incorporates the~$\CO(1)$ correction arising from minimizing the combined bulk and boundary energies. In the type-I regime ($\sigma>0$), this shift decreases the radius, moving the analytic profile to the left and improving the agreement with the numerics. In the type-II regime~($\sigma<0$), the opposite occurs. 
    }
\end{figure}

\begin{figure}[t!]
    \centering
    \includegraphics[width=0.49\textwidth]{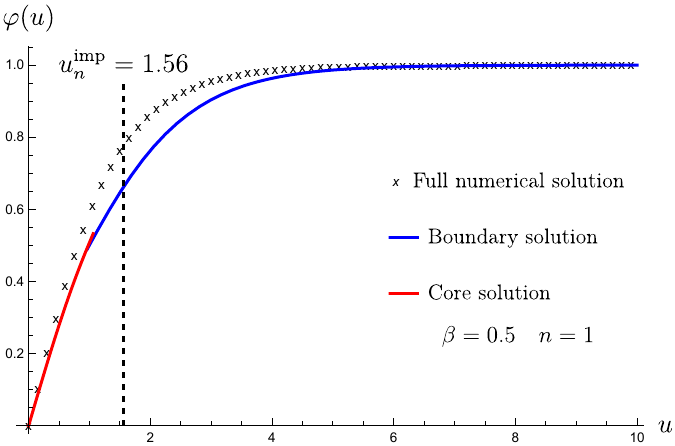}
    \includegraphics[width=0.49\textwidth]{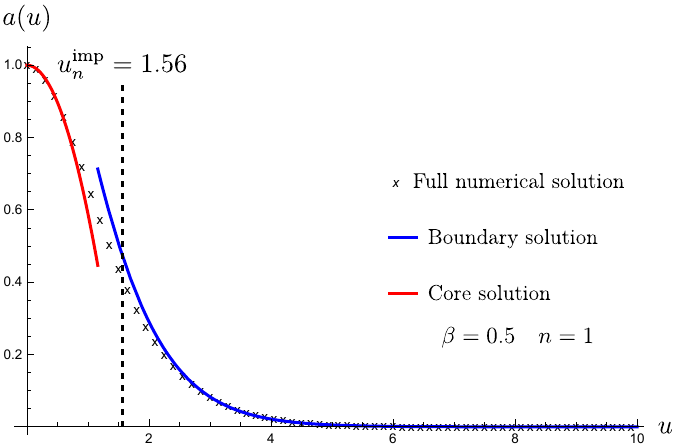}
    \caption{\label{fig:fundamentalV4comparesolution} Comparison between the full numerical  solution to the vortex equations~\eqref{eqn:ahmeom} (shown as crosses) and the large-$n$ analytic answer (depicted as solid curves) for the fundamental string~$n=1$ at a representative value of~$\beta=0.5$. The core and boundary analytic solutions are shown in red and blue, respectively. Even for such small flux, the analytic expressions reasonably capture the qualitative features of the full profile. }
\end{figure}
    
\item Following~\eqref{Ccbdef} and~\eqref{cofn}, we can explicitly verify the relationship between~$c_n$ and~$C_{\text{core/bdy}}$ for small flux~$n$. While this correspondence is expected to become exact in the infinite flux limit, it is instructive to perform a consistency check at small values of~$n$ to assess how far the large-$n$ approximation remains reliable. Numerically, the constant~$c_n$ can be extracted for small $n$ using the near origin behavior~$\varphi\sim u^n$ (see~\eqref{eqn:ahmoriginasym}). We carry out this comparison for~$n=3$ and~$\beta = 0.5,~0.6$, obtaining,
\begin{equation}
    \begin{split}
    \beta=0.5 \qquad & c_n^{\text{numeric}} = 0.11 \qquad C_{\text{core/bdy}}(0.5) = 0.53\implies c_n^{\text{large-n analytics}} = 0.09~, \\
    \beta =0.6 \qquad & c_n^{\text{numeric}} = 0.13  \qquad C_{\text{core/bdy}}(0.6) = 0.54\implies c_n^{\text{large-n analytics}} = 0.11~. \\ 
    \end{split}
\end{equation}
The agreement between the numerical results and the large-$n$ analytic prediction is remarkably good, even for such small fluxes.

\item Furthermore, we evaluate the constants~$\varphi_{\text{bdy/ext}}$ and~$\gamma_{\text{bdy/ext}}$ for a range of values of~$\beta$. The procedure follows the same steps as those used to determine~$C_{\text{core/bdy}}$ in the previous bullet points. The resulting values are summarized in table~\ref{bdyext}. At the BPS point,~$\beta=1$, these constants can be directly related to the parameter~$w_\infty$, introduced in~\cite{Penin_2021}. The correspondence takes the form,
\begin{equation}\label{bpsbdyextrelation}
    w_\infty = -0.33~, \qquad \gamma_{\text{bdy/ext}}(\beta =1 ) = - \sqrt{2}w_\infty = 0.47~, \qquad \varphi_{\text{bdy/ext}}(\beta=1) = - w_\infty~.
\end{equation}
As seen from table~\ref{bdyext}, our numerical results are in excellent agreement with these BPS results.

In addition, using~\eqref{eq:bdytailmatch}, we compare~$\varphi_{\text{bdy/ext}}$ and~$\gamma_{\text{bdy/ext}}$ with the asymptotic quantities~$a_\infty$ and~$\varphi_\infty$, which are obtained numerically from the full vortex solutions~\eqref{eqn:ahmeom}. For illustration, we carry out this comparison at a representative coupling~$\beta = 0.5$ and flux numbers~$n=10,~25,~100$.
\begin{equation}
    \begin{split}
        \log{a_\infty(n=10)} =  4.94  & \implies \log \gamma_\text{bdy/ext}^{\text{large-}n} = -1.11 \\ 
        \log{a_\infty(n=25)} = 8.70 & \implies \log \gamma_\text{bdy/ext}^{\text{large-}n} = -1.04 \\ 
        \log{a_\infty(n=100)} = 19.60 & \implies \log \gamma_\text{bdy/ext}^{\text{large-}n} = -0.99\\ 
    \end{split}
    \qquad 
    \log \gamma_{\text{bdy/ext}} = -0.8
\end{equation}
As expected, the agreement between the constants improves systematically with increasing~$n$.

\begin{table}[]
\centering
\setlength\extrarowheight{5pt}
\begin{tabular}{ |p{2.5cm}|p{2.5cm}|   } 
    \hline
    \multicolumn{2}{|c|}{ $\gamma_{\text{bdy/ext}}(\beta)$} \\
    \hline
        $\beta$ & $\gamma_{\text{bdy/ext}}(\beta)$\\
    \hline
        $ \beta = 0.1  $  & $0.94$   \\
        $\beta = 0.5 $  & $0.45$    \\ 
        $ \beta = 1 $  & $0.47$    \\ 
        $ \beta =  2 $  & $0.54$    \\ 
        $ \beta = 3 $  & $0.62$    \\ 
    \hline
\end{tabular}
\quad \quad
\begin{tabular}{ |p{2.5cm}|p{2.5cm}|   } 
    \hline
    \multicolumn{2}{|c|}{ $\varphi_{\text{bdy/ext}}(\beta)$} \\
    \hline
        $\beta$ & $\varphi_{\text{bdy/ext}}(\beta)$\\
    \hline
        $ \beta = 0.1  $  & $0.70$   \\
        $\beta = 0.5 $  & $0.43$    \\ 
        $ \beta = 1 $  & $0.33$    \\ 
        $ \beta = 2 $  & $0.28$    \\
        $ \beta =  3 $  & $0.35$   \\ 
    \hline
\end{tabular}
\caption{Values of constants~$\gamma_{\text{bdy/ext}}(\beta)$  and~$\varphi_{\text{bdy/ext}}(\beta)$ (see~\eqref{eq:ahmasymbis}) for various values of~$\beta$. Their variation is not monotonic with respect to~$\beta$. At the BPS point~$\beta=1$, the constants are related to the parameter~$w_\infty$ introduced in~\cite{Penin_2021}, which is related to the constants~$\gamma_{\text{bdy/ext}}(\beta)$  and~$\varphi_{\text{bdy/ext}}(\beta)$ via~\eqref{bpsbdyextrelation}. We see that numerics are in agreement with this result at the BPS point.}
\label{bdyext}
\end{table}

\item We briefly comment on the qualitative accuracy  associated with the determination of the various constants considered in this section, namely~$C_{\text{core/bdy}}(\beta),~\gamma_{\text{bdy/ext}}(\beta)$, and~$\varphi_{\text{bdy/ext}}(\beta)$. While we do not attempt a systematic quantitative error analysis, we can identify the sources of uncertainty at a qualitative level. The first arises from the finite spatial domain used in the numerical computation which introduces effects due to the limited box size. The second stems from numerical precision and floating point limitations inherent to the discretization scheme.

See, for instance, figure~\ref{fig:logexhibit} and $\beta=0.5$. The numerical solution is obtained on a finite spatial domain,~$y\in [-10,10]$. As shown, the constant behavior begins to deviate slightly near the right boundary. This effect originates from the imposed hard Dirichlet boundary conditions,~$\varphi(y=-10)=0$, which cause the logarithmic term to diverge at the edge of the domain. In other words, the artificial boundary truncation forces the field to vanish too rapidly. The deviations become noticeable once the logarithm approaches machine precision. This prevents the solution from maintaining its ideal constant profile. Thus, the variations observed near the boundary are not physical but rather reflect these combined effects of finite box size and numerical precision.

\item  For various values of~$\beta$, we define the surface tension~$\sigma(\beta)$ of the planar interface between the Coulomb region in the interior and the Higgs vacuum in the exterior by~\eqref{eq:tensionbdy}, which will be derived in section~\ref{sec:tension}, and quantifies the boundary contribution to the string tension. See also~\eqref{bdysigma} and the discussion around it.  The corresponding values of~$\sigma(\beta)$ are tabulated in table~\ref{tab:sigma}. Numerically, we find that~$\sigma(\beta)>0$ for type-I strings and~$\sigma(\beta)<0$ for type-II strings. At the BPS point,~$\sigma(\beta=1)=0$. 

\begin{table}[]
\centering
\setlength\extrarowheight{5pt}
\begin{tabular}{ |p{2.5cm}|p{2.5cm}|   } 
    \hline
    \multicolumn{2}{|c|}{ Surface tension $\sigma(\beta)$} \\
    \hline
        $\beta$ & $\sigma(\beta)$\\
    \hline
        $ \beta = 0.1  $  & $0.22$   \\
        $\beta = 0.5 $  & $0.17$    \\ 
        $ \beta = 1 $  & $0.00$    \\ 
        $ \beta =  2 $  & $ -0.43$    \\ 
        $ \beta = 3 $  & $-0.91$    \\ 
    \hline
\end{tabular}
\caption{Surface tension~$\sigma$ of the planar interface between the Coulomb region and the Higgs vacuum. Note that~$\sigma(\beta)>0$ in the type-I regime and~$\sigma(\beta)<0$ in the type-II regime; it vanishes at the BPS point~$\beta=1$.}
\label{tab:sigma}
\end{table}

\end{itemize}

\subsection{String Tension}\label{sec:tension}

We now use our large-$n$ solutions above to compute the string tension~\eqref{eq:ahmtension}, 
\begin{equation}\label{eq:ahmtensionbis}
    {T_n \over 2 \pi \sqrt{2}} =  \int_0^\infty u du \, \left({n^2 \over 2 u^2} \left(a'(u)\right)^2 + \left(\varphi'(u)\right)^2 + {n^2 \over u^2} a^2(u) \varphi^2(u) + {\beta \over 2} (\varphi^2(u)-1)^2 \right)~.
\end{equation}
Throughout we neglect exponentially small contributions to the string tension, e.g.~we neglect the exterior region, where the fields are exponentially close to the Higgs vacuum. 

We will split the integral in~\eqref{eq:ahmtensionbis} in the outer core region (see figure~\ref{fig:regions}) that is shared between core and the boundary, at a cutoff point~$y_\text{c} \ll -1$ that is fixed as~$n \to \infty$. This allows us to separately consider the contributions from the core and from the boundary:
\begin{itemize}
\item[(C)] We know from~\eqref{phiofy} that~$\varphi(y)$ is exponentially small for all~$y \leq y_c$, so that we can approximate the core contribution to the string tension by setting~$\varphi(u) \simeq 0$ in~\eqref{eq:ahmtensionbis} and integrating the remaining approximately constant energy density up to the cutoff,
\begin{equation}\label{eq:convcoretension}
   {T_n^\text{core} \over 2 \pi \sqrt{2}} =  \int_0^{u_n + y_c} u du \, \left({n^2 \over 2 u^2} \left(a'(u)\right)^2 +  {\beta \over 2} \right) = {\beta \over 2} (u_n + y_c)^2~.
\end{equation}
This is a more rigorous, cutoff-dependent version of the variational core energy previously obtained in~\eqref{eq:ahmvarradiustension}.

\item[(B)] Throughout the boundary region we use~$u = u_n + y$, where~$y$ is fixed in the large-$n$ limit. Working to leading order in this limit, and using~\eqref{gammadef}, we obtain the following boundary contribution from~\eqref{eq:ahmtensionbis}, 
\begin{equation}\label{eq:bdytension}
    {T_n^\text{bdy} \over 2 \pi \sqrt{2}} =  u_n \int_{y_c}^\infty dy \, \left(\half \left(\gamma'(y)\right)^2 + \left(\varphi'(y)\right)^2 +  \gamma^2(y) \varphi^2(y) + {\beta \over 2} (\varphi^2(y)-1)^2 \right)~.
\end{equation}
Up to exponentially small terms in~$y_c$, this evaluates to\footnote{~The~$y_c$-dependence of~$T_n^\text{bdy}$ is deduced by taking a~$y_c$-derivative on both sides of~\eqref{eq:bdytension}, together with the fact that~$\varphi(y_c) \ll 1$ is exponentially small and~$\gamma(y_c) \simeq - \sqrt \beta y_c$ is given by~\eqref{gmatch}. This determines~$T_n^\text{bdy}$ up to an integration constant that we call~$\sigma$.}
\begin{equation}\label{eq:tensionbdy}
        {T_n^\text{bdy} \over 2 \pi \sqrt{2}} 
        =  - \beta u_n y_c + {\sigma(\beta) u_n \over \sqrt 2}~. 
\end{equation}
Here~$\sigma$ is a constant that depends on neither~$n$ nor~$y_c$. It represents the surface tension of a planar interface between the Coulomb region in the interior of the string -- which is not a vacuum -- and the Higgs vacuum in its exterior. At large~$y_c \ll -1$, this surface tension is subleading to the bulk vacuum energy~$-\beta u_n y_c$ contributed by the Coulomb non-vacuum region.
\end{itemize}

Adding the core~\eqref{eq:convcoretension} and boundary~\eqref{eq:tensionbdy} contributions, we obtain the full string tension in the large-$n$ limit, 
\begin{equation}\label{finaltension}
    {T_n \over 2 \pi} = {T_n^\text{core} \over 2\pi} + {T_n^\text{bdy} \over 2\pi} = {\beta \over \sqrt{2}} u_n^2 + \sigma(\beta) u_n + \CO(y_c^2)~, \qquad u_n = {\sqrt{2n} \over \beta^{1/4}}~.
\end{equation}
Several comments are in order:
\begin{itemize}
\item[(i)] The extensive core contribution~$\sim u_n^2 \sim n$ to the string tension precisely matches the variational prediction~\eqref{extensive}.

\item[(ii)] The subleading contribution~$\sim u_n \sim \sqrt n$ involves the surface tension~$\sigma(\beta)$ of a planar interface between the Coulomb region in the core and the Higgs vacuum in the exterior, as anticipated in~\eqref{bdysigma}. It was obtained numerically in section ~\ref{sec:numericsbdy}. As was discussed there,~$\sigma(\beta) > 0$ for type I strings (with~$\beta < 1$) and~$\sigma(\beta) < 0$ for type II strings (with~$\beta > 1$), with vanishing~$\sigma(\beta = 1) = 0$. As we will discuss in section~\ref{sec:stability}, the sign of~$\sigma(\beta)$ determines whether large-$n$ strings in the bulk phase form bound states.

\item[(iii)] The cutoff-dependent~$\CO(u_n y_c)$ term has canceled between~\eqref{eq:convcoretension} and~\eqref{eq:tensionbdy}, while the~$\CO(y_c^2)$ term in~\eqref{eq:convcoretension} remains. Since~$y_c$ is fixed in the large-$n$ limit, this term is smaller than the leading core and boundary contributions discussed in~(i) and (ii) above, both of which grow with~$n$. The~$\CO(y_c^2)$ term also cancels if we expand the boundary problem to next order in~$y/u_n$; we will not do it here.  
\end{itemize}

\subsection{Comparison with  the BPS equation at~$\beta = 1$}

\label{sec:bpslargen} 

At the BPS point~$\beta = 1$, every solution of the second-order vortex equations~\eqref{eqn:vorteqbis} is in fact also a solution of the first-order BPS equations~\eqref{eqn:ahmBPSeqns},

\begin{equation}\label{bpseqbis}
    {n \over u} a'(u) = \varphi^2(u)-1~, \qquad \varphi'(u) = {n \over u} \varphi(u) a(u)~.
\end{equation}
Here we briefly discuss, following~\cite{PhysRevLett.125.251601,Penin_2021}, how these equations are solved in the large-$n$ limit, and compare with the~$\beta = 1$ limit of our general large-$n$ solutions discussed above:
\begin{itemize}
    \item In the core we take~$\varphi(u) \ll 1$ and linearize the BPS equations~\eqref{bpseqbis} to obtain
    \begin{equation}\label{bpscore}
        a(u) = 1 - {u^2 \over 2n}~, \qquad \varphi(u) =( \text{const.}) \; u^n \exp\left(-{u^2 \over 4}\right)~.
    \end{equation}
We find exact agreement with our large-$n$ solutions in~\eqref{convlinvort} and~\eqref{wkbhiggs} throughout the core region upon substituting~$\beta = 1$ into those expressions. In particular, the WKB prefactor in the Higgs field~\eqref{wkbhiggs} fortuitously cancels at~$\beta = 1$.  As discussed around~\eqref{cofn}, the expression for the Higgs field in~\eqref{bpscore} is valid in the deepest core (where~$u/u_n \to 0$ at large $n$) for any value of~$\beta$, but unless~$\beta = 1$ it does not correctly describe the remainder of the core. 
\item The solution of the BPS equations in the boundary region, and their matching to the core and the exterior are very similar to the discussion above (see~\cite{PhysRevLett.125.251601,Penin_2021} for details). 
\item The BPS tension~\eqref{BPST}, which is exactly linear in~$n$, is completely saturated by the core contribution in~\eqref{finaltension} evaluated at~$\beta = 1$. Consequently the interface tension must exactly vanish at the BPS point,
\begin{equation}\label{sigbps}
    \sigma(\beta = 1) = 0~,
\end{equation}
since it would otherwise contribute at~$\CO(\sqrt{n})$ to the BPS tension. 
\item The matching constant for the Higgs field in~\eqref{hrep} was exactly determined to be~\cite{PhysRevLett.125.251601,Penin_2021}
\begin{equation}\label{Cmatchbps}
    C_\text{core/bdy}(\beta = 1) = {1 \over \sqrt{e}}~.
\end{equation}
\end{itemize}
Note that both~\eqref{sigbps} and~\eqref{Cmatchbps} are consistent with the numerical results summarized in section \ref{sec:numericsbdy}.


\section{Large-$n$ Solutions for Domain-Wall Strings}\label{walllargensec}

In this section we generalize the large-$n$ solutions of bulk strings in the conventional Abelian Higgs model described in section~\ref{bulklargensec} above to strings in the degenerate Abelian Higgs model (see section~\ref{degAHMandDW}) with sextic potential~$\t V_6 = {\beta \over 2} \varphi^2 (\varphi^2-1)^2$ in~\eqref{rescaledV}. As we will show rigorously, these strings exhibit the domain wall phase we previously explored using variational arguments in section~\ref{seclargen}. In the degenerate model, the axially-symmetric vortex equations~\eqref{eqn:ahmeom} read
\begin{equation}\label{eqn:vorteqdw}
    \begin{split}
       \varphi''(u) + \frac{1}{u} \varphi'(u) = ~ & \frac{n^2}{u^2} a^2(u) \varphi(u) + {\beta \over 2} \varphi(u) 
       \left(\varphi^2(u) -1\right)^2 + \beta \varphi^3(u) \left(\varphi^2(u)-1\right)~,
       \\
        a''(u) - \frac{1}{u} a'(u) = & ~ 2 a(u) \varphi^2(u)~. \\
    \end{split}
\end{equation}
We will again solve these equations in the large-$n$ limit (as before, $\beta$ remains fixed) using matched asymptotic expansions. An important difference is that the potential~$\t V_6$ admits a Coulomb vacuum at~$\varphi = 0$. The domain wall solution interpolating between the Coulomb and the Higgs vacua was given analytically in~\eqref{rescaledwall}; it will replace the numerical boundary solutions that we obtained in the conventional Abelian Higgs model, leading to fully analytic large-$n$ solutions in the degenerate model. 

We will continue to use the taxonomy for different regions of large-$n$ strings introduced in section~\ref{regions} (see in particular figure~\ref{fig:regions}), except that the core radius~$u_n$ of the string is now given by~\eqref{uTdw},
\begin{equation}\label{dwunbis}
    u_n = {\sqrt 2 n^{2/3} \over \sigma^{1/3}}~, \qquad  \sigma = {\sqrt \beta \over 2}~.
\end{equation}
Here~$\sigma$ is the tension of the Coulomb-to-Higgs-vacuum domain wall in our rescaled units, as discussed around~\eqref{dwbdybis}.

\subsection{Core Region}

In the core region~\eqref{eq:coredef}, the Higgs field~$\varphi(u) \ll 1$ is small, so that the vortex equations~\eqref{eqn:vorteqdw} linearize, 
\begin{equation}\label{linvort}
    \varphi'' + {1 \over u} \varphi' = \left({n^2 a^2(u) \over u^2} +{ \beta \over 2}\right) \varphi~, \qquad a(u) = 1-{u^2 \over u_n^2(\beta)}~,
\end{equation}
where the core radius~$u_n$ is now given by~\eqref{dwunbis}. As before, this leads to a Schr\"odinger equation for the wavefunction~$\psi(u) = \sqrt{u} \varphi(u)$, which takes the following form in the large-$n$ limit,
\begin{equation}\label{schodeqDW}
    \psi''(u) = p^2(u) \psi(u)~, \qquad p^2(u) \simeq {n^2 a^2(u) \over u^2} + {\beta \over 2}~.
\end{equation}
Note that this equation can be obtained from~\eqref{psqfac} by replacing~$- \beta \to + \beta / 2$. This sign change directly reflects the Coulomb vacuum in the degenerate model. As before, we will solve~\eqref{schodeqDW} in the large-$n$ limit using the WKB approximation,

\subsubsection{WKB Error Analysis}

Note that~$p^2(u)$ in~\eqref{schodeqDW} is strictly positive, i.e.~there is no classical turning point. As we will see, this implies that the WKB approximation works even better for domain-wall strings than for strings in the bulk phase. To this end, let us examine the WKB error~\eqref{eq:err} in the variable~$\xi = u/u_n$ defined in~\eqref{xidef}; we distinguish two regimes:
\begin{itemize}
    \item[(DC)] In the deep core~$\xi$ has a fixed large-$n$ limit strictly within the interval~$[0, 1)$, leading to an error scaling as~$\ep(\xi < 1) \sim 1/n^2$ in this region. This is similar to our previous discussion around~\eqref{errofxi}.
\item[(OC)] In the outer core we take~$\xi \to 1$ in the large-$n$ limit, except in factors of~$\xi -1 = y/u_n$, which we retain. This leads to the following error function in the outer core region, where~$|y| \ll u_n$ (with~$u_n \sim n^{2/3}$),
\begin{equation}\label{fullepofyDW}
    \ep(y) =  { \sigma ^{4/3} \over n^{2/3} }   \frac{(6  \sigma ^{4/3} ) n^{-2/3} y^2-2 \beta  }{\left((2 \sigma ^{4/3})n^{-2/3}  y^2 + \beta\right)^3}~, \qquad  |y|/n^{2/3} \to 0 \quad \text{as} \quad n \to \infty~.
\end{equation}
Let us examine this function in different regions:
\begin{itemize}
    \item[(i)]  When~$n^{1/3} \ll |y| \ll n^{2/3}$ in the large-$n$ limit,\footnote{~By this we mean that~$|y|/n^{1/3} \to \infty$ and~$|y|/n^{2/3} \to 0$ as~$n \to \infty$.} we can approximate~$\ep(y) \simeq n^{2/3}/y^4$. This function monotonically interpolates between the~$\ep(y) \sim 1/n^2$ scaling of the deep core (see above) for the largest allowed values of~$|y| \sim n^{2/3}$ in this range, to a different scaling~$\ep(y) \sim 1/{n^{2/3}}$ for the smallest~$|y| \sim n^{1/3}$ in this range. 
    \item[(ii)] When~$|y| = \CO(n^{1/3})$ (so that~$|y|/n^{1/3}$ has a finite, non-zero large-$n$ limit) then we must use~\eqref{fullepofyDW}, since all terms have the same large-$n$ scaling. In this regime the error function vanishes at~$y = \pm \sqrt\beta n^{1/3} / (\sqrt3 \sigma^{2/3})$, and it has local maxima at
    \begin{equation}\label{epmax}
        \ep\left(\pm \frac{\sqrt{3\beta } n^{1/3} }{2 \sigma ^{2/3}}\right) = \frac{4 \sigma ^{4/3}}{25 \beta ^2 n^{2/3}}~.
    \end{equation}
    \item[(iii)] When~$|y| \ll n^{1/3}$ (so that~$|y|/n^{1/3} \to 0$ in the large-$n$ limit) the~$y$-dependence of~\eqref{fullepofyDW} is subleading, so that the error is approximately constant
    \begin{equation}
        \ep(y) \simeq - {2 \sigma^{4/3} \over n^{2/3} \beta^2}~, \qquad |y|/n^{1/3} \to 0 \quad \text{as} \quad n \to \infty~.
    \end{equation}
    Note that this is larger in absolute value than~\eqref{epmax}, so that the largest WKB errors occur near~$y = 0$.
\end{itemize}
\end{itemize}
The upshot of the preceding discussion is that the WKB error is uniformly small in the large-$n$ limit throughout the entire core region,
\begin{equation}\label{wkbrangeDW}
    u = u_n + y~, \qquad -u_n \leq y \ll u_n~.
\end{equation}
Here the left inequality means that we can cover the entire core~$0 \leq u \leq u_n$ for negative~$y$, while the right inequality means that we can explore positive~$y$ as long as~$y / u_n \to 0$ in the large-$n$ limit. As anticipated, this is a significantly larger range than the $y$-values~\eqref{ycond} allowed for bulk strings.  

\subsubsection{WKB Solution}

As before, only the growing WKB solution in~\eqref{eq:wkbsol} satisfies the boundary condition~$\varphi(u) \sim u^n$ as~$u \to 0$, 
\begin{equation}\label{eq:wkbsolbis}
    \psi_{\text{WKB}+}(u) \simeq {1 \over \sqrt{p(u)}} \exp\left( + \int^u du' \, p(u')\right)~.
\end{equation}
In terms of the~$\xi$-coordinate, and recalling~\eqref{linvort}, \eqref{schodeqDW}, we find
\begin{equation}
    p(\xi) = \sqrt{{\beta \over 2}  +\frac{n^{2/3} \left(\xi ^2-1\right)^2 \sigma ^{2/3}}{2 \xi ^2}}~.
\end{equation}
We will consider this function in the same subregions we already encountered in our WKB error analysis above:  
\begin{itemize}
    \item[(DC)] In the deep core we can approximate
    \begin{equation}\label{DCpofxiDW}
        p(\xi) \simeq \frac{\left(1- \xi ^2\right) n^{1/3} \sigma^{1/3}}{\sqrt{2} \xi } + \frac{\beta  \xi }{2 \sqrt{2} \left(1-\xi ^2\right) n^{1/3} \sigma^{1/3} } + \CO(1/n)~.
    \end{equation}
    Upon integration, we find the exponent of the WKB wavefunction~\eqref{eq:wkbsolbis},
    \begin{equation}
        u_n \int^\xi d\xi' \, p(\xi') =  n \left( \log \xi-\half \xi ^2\right)-\frac{\beta n^{1/3}  \log \left(1-\xi ^2\right)}{4 \sigma ^{2/3}} + \CO(1/n^{1/3})~.
    \end{equation}
    Consistency demands that we only retain the~$\sqrt\xi$ from the WKB prefactor in the large-$n$ limit, so that the WKB wavefunction in the deep core reads
    \begin{equation}
       \psi_{\text{WKB}+}(\xi) =  {\xi ^{n+\frac{1}{2}} \over \left(1-\xi ^2\right)^{\beta n^{1/3} \over 4 \sigma ^{2/3} } } \exp\left(-\frac{n}{2}  \xi ^2 \right)~.
    \end{equation}
    Thus the full Higgs field in the deep core is given by
    \begin{equation}\label{phiDCDW}
        \varphi(u) = C_\text{DC}(n, \beta) \,  u^n \, \left(1-\frac{u^2}{u_n^2}\right)^{ -  \frac{\beta n^{1/3}}{4 \sigma ^{2/3}}} \, \exp\left({-\frac{\sigma ^{2/3} u^2}{4 n^{1/3}}}\right)~,
    \end{equation}
    where~$C_\text{DC}$ is a constant that can depend on~$n$ and~$\beta$.
    
    \item[(OC)] In the outer core region we take~$\xi \to 1$ and~$\xi -1 \to y/u_n$, so that
    \begin{equation}\label{pofyOC}
        p(y) = \sqrt{\frac{\beta }{2}+\frac{\sigma ^{4/3} y^2}{n^{2/3}}}~, \qquad |y|/n^{2/3} \to 0 \quad \text{as} \quad n \to \infty~.
    \end{equation}
Substituting this into the WKB formula~\eqref{eq:wkbsolbis}, we obtain an expression for the Higgs field which interpolates between the deep core result~\eqref{phiDCDW} for~$y \sim n^{2/3}$ and the small-$y$ regime that we now consider. This regime sets in for~$|y| \lesssim n^{1/3}$, where we can approximate
    \begin{equation}\label{smallyp}
        p(y) \simeq \sqrt{\beta \over 2}~, \qquad |y|/n^{1/3} \to 0 \quad \text{as} \quad n \to \infty~,
    \end{equation}
    so that the Higgs field takes the simple form
    \begin{equation}\label{smallyphi}
        \varphi(u) = C_\text{core/bdy}(n, \beta) \exp\left(y \sqrt{\beta / 2} \right)~.
    \end{equation}
    Here the constant~$C_\text{core/bdy}$ is proportional to~$C_\text{DC}$ in~\eqref{phiDCDW}. The exact relation can be obtained by matching the full WKB solutions in the deep and outer core for~$y \sim n^{2/3}$ (we will not show these formulas explicitly). We will fix~$C_\text{core/bdy}$ below by matching to the boundary region. This will also make it clear for what values of~$y$ the Higgs field~$\varphi(y) \ll 1$ is small, and show that our linearized WKB analysis above is self-consistent. 
\end{itemize}

\subsection{Boundary Region}

As in section~\ref{bulkbdymatch}, the boundary region is given by~$|y| \ll u_n$ (see~\eqref{bdy}), where  we  can simplify the vortex equations~\eqref{eqn:vorteqdw} of the degenerate model by dropping the curvature terms. This leads to the translationally invariant boundary equations
\begin{equation}\label{eqn:bdyeomDW}
    \begin{split}
        \varphi''(y)  = & ~ \gamma^2 (y)\varphi(y) +{\beta \over 2} \varphi(y) 
       \left(\varphi^2(y) -1\right)^2 + \beta \varphi^3(y) \left(\varphi^2(y)-1\right)~, \\ 
        \gamma''(y) = & ~ 2 \gamma(y) \varphi^2(y)~. \\
    \end{split}
\end{equation}
where the rescaled boundary gauge field is given by~\eqref{gammadef}, 
\begin{equation}\label{gammadefbis}
    \gamma(y) = \frac{n}{u_n+y} a(y) \simeq {n \over u_n} a(y)~.
\end{equation}
We will choose to match with the core region described above in a regime where~$y < 0$ is a large~$\CO(1)$ number as~$n \to \infty$. In this region we find from~\eqref{smallyphi} and~\eqref{linvort} that
\begin{equation}\label{matchphigam}
    \varphi(y) = C_\text{core/bdy}(n, \beta) \exp\left(y \sqrt{\beta/  2}\right)~, \qquad \gamma(y) = -{ y \sigma^{2/3} \over n^{1/3} }~.
\end{equation}
Since the boundary equations~\eqref{eqn:bdyeomDW} are~$n$-independent and~$\gamma(y)$ vanishes at both ends of the boundary in the large-$n$ limit, we will self-consistently solve the equations by setting~$\gamma(y) \simeq 0$ at leading order in the large-$n$ limit.

The resulting translationally-invariant second-order equation for the Higgs field is solved by the planar BPS domain wall~\eqref{rescaledwall} that interpolates between the Coulomb and Higgs vacua of the degenerate model,
\begin{equation}\label{dwbisbis}
    \varphi(y) = {1 \over \sqrt{1 + \exp\left(- \sqrt{2 \beta} (y - y_0) \right)}}~.
\end{equation}
Here~$y_0$ is the center of mass of the wall, which is an exact zero mode at leading order. In order for this solution to match~\eqref{matchphigam}, we must have 
\begin{equation}\label{y0ccoreby}
    y \ll y_0 = -\sqrt{\frac{2}{\beta}}\log{C_{\text{core/bdy}}}~.
\end{equation}
This is precisely the regime in which the Higgs field is small, and the linearized analysis in the core is valid. Below we will determine~$y_0(\beta)$ via a next-to-leading order calculation in the large-$n$ limit, which establishes that it is an~$n$-independent constant. Therefore the domain wall is centered in the outer core region~\eqref{smallyp}. The matching with the exterior of the string occurs in the region~$y \gg y_0$, where we find~\eqref{eq:ahmasymbis} with~$\varphi_\text{bdy/ext} = \half \exp\left(\sqrt{2 \beta} y_0\right)$. 

The fact that the domain-wall center~$y_0$ remains unfixed is due to our neglect of the boundary gauge field~$\gamma(y)$, which is subleading in~$n$ but explicitly breaks translation symmetry via its boundary conditions~\eqref{matchphigam}. We can determine the resulting small~$\gamma(y) \sim n^{-1/3}$ by substituting~\eqref{dwbisbis} into~\eqref{eqn:bdyeomDW},
\begin{equation}\label{gode}
    \gamma''(y) = {2 \gamma(y) \over 1 + \exp\left(- \sqrt{2 \beta} (y - y_0) \right)}~.
\end{equation}
This second-order, linear ODE has two linearly independent solutions:\footnote{~This analysis was first carried out using {\tt ChatGPT5}, and then independently verified using {\tt Mathematica}.}
\begin{itemize}
    \item A hypergeometric solution,
    \begin{equation}\label{g1}
        \gamma_1(y) = { _2F_1\left(1+\frac{1}{\sqrt{\beta }}~, \frac{1}{\sqrt{\beta }}~; 1~; \frac{1}{1 + \exp\left(- \sqrt{2 \beta} (y - y_0) \right)}\right) \over \left(1 + \exp\left( \sqrt{2 \beta} (y - y_0) \right)\right)^{\frac{1}{\sqrt{\beta }}}}~,
    \end{equation}
    whose asymptotic behavior at large positive~$y$ is given by
    \begin{equation}\label{g1asym}
        \gamma_1(y) \simeq \frac{\sqrt{\beta } \Gamma \left(2/\sqrt{\beta }\right) }{\Gamma \left({1 / \sqrt{\beta }}\right)^2} \, \exp \left(\sqrt{2} (y-y_0)\right)~, \qquad y \to \infty~.
    \end{equation}
    Since we require~$\gamma(y)$ to decay in this limit, we discard this solution, but we will nevertheless need the functional form of~$\gamma_1(y)$ below. To this end, we record its asymptotics at large negative~$y$,
    \begin{equation}\label{g1minusinfty}
        \gamma_1(y) \simeq 1 + {1 \over \beta} \exp\left(\sqrt{2\beta} (y -y_0)\right) + \CO\left(\exp\left(2 \sqrt{2\beta} (y -y_0)\right)\right)~, \qquad y \to -\infty~.
    \end{equation}

\item A second linearly independent solution~$\gamma_2(y)$ can be obtained by suitably integrating~$\gamma_1(y)$,
\begin{equation}\label{g2def}
    \gamma_2(y) = \gamma_1(y) \int_y^\infty {dt \over \gamma_1(t)^2}~.
\end{equation}
The fact that~$\gamma_2(y)$ satisfies~\eqref{gode} follows immediately from the fact that~$\gamma_1(y)$ is a solution.\footnote{~Alternatively, we can deduce~\eqref{g2def} by noting that the Wronskian of~$\gamma_{1}, \gamma_2$ must be a non-zero constant.} Moreover the upper limit of the~$t$-integral ensures that~$\gamma_2(y)$ vanishes as~$y \to \infty$, at a rate that can be determined by substituting~\eqref{g1asym} into~\eqref{g2def},
\begin{equation}\label{g2yplusinf}
    \gamma_2(y) \simeq \frac{\Gamma \left(1 /{\sqrt{\beta }}\right)^2 }{2 \sqrt{2\beta} \Gamma \left(2/{\sqrt{\beta }}\right)} \; e^{-\sqrt{2} (y-y_0)}~, \qquad y \to \infty~.
\end{equation}
This is exactly the expected large-$y$ asymptotics discussed around~\eqref{eq:ahmasymbis}.
\end{itemize}

\begin{figure}[t!]
    \centering
    \includegraphics[width=0.6\textwidth]{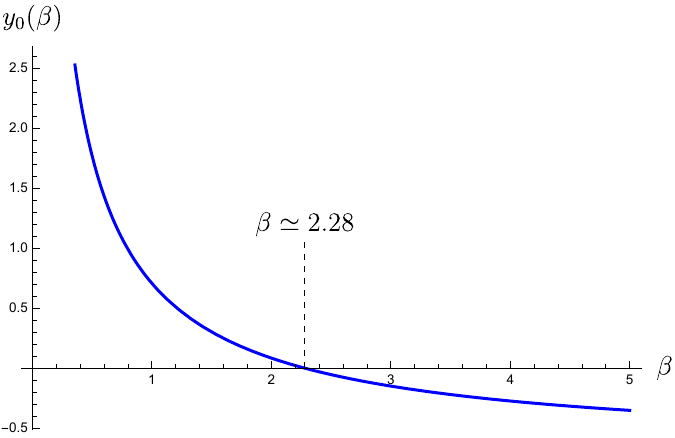} \caption{\label{fig:y0beta} The center of mass~$y_0(\beta)$ of the domain wall, defined in~\eqref{y0final}. Its various features are discussed in the main text below.}
\end{figure}

To evaluate~$\gamma_2(y)$ for large negative~$y$, it is useful to simplify the integral in~\eqref{g2def} by introducing a new variable~$z$, which is defined to be the last argument of the hypergeometric function in~\eqref{g1}, so that
\begin{equation}\label{zdef}  
z = \frac{1}{1 + \exp\left(- \sqrt{2 \beta} (y - y_0) \right)}~, \qquad \gamma_1(z) = (1-z)^{\frac{1}{\sqrt{\beta }}} \, _2F_1\left(1+\frac{1}{\sqrt{\beta }} \, ,\frac{1}{\sqrt{\beta }} \, ; 1 \, ; z\right)~.
\end{equation}
After this change of variables, the integral in~\eqref{g2def} reads 
\begin{equation}\label{zpint}
    \int_y^\infty {dt \over \gamma_1(t)^2} = {1 \over \sqrt{2 \beta}} \int_{z}^1 {dz' \over z'} \left({(1-z')^{-(2/\sqrt{\beta} +1) } \over \left(_2F_1(1 + 1/\sqrt{\beta} \, , 1/\sqrt\beta \, ; 1\, ; z')\right)^2} -1\right) -{1 \over \sqrt{2 \beta}} \log z~.
\end{equation}
Here we have added and subtracted~$1$ inside the parenthesis of the~$z'$-integrand to explicitly exhibit the~$\log z$ term. It is now straightforward to take~$y \to -\infty$, or~$z \to 0$, in which case the~$z'$ integral in~\eqref{zpint} converges, while the logarithmic term gives exactly~$-(y-y_0)$. Since~\eqref{zpint} reduces to~$\gamma(y)$ in this limit (where~$\gamma_1(y) \simeq 1$, up to exponentially small terms), the matching condition~$\gamma(y) \simeq -y$ in~\eqref{matchphigam} can only be satisfied if~$y_0$ is chosen to precisely cancel the~$z'$ integral in~\eqref{zpint},\footnote{~\label{fn:convint}Expanding the expression inside the parentheses in~\eqref{y0final} around~$z = 0$ we find
\begin{equation}\label{inttozero}
    \left(\cdots\right) = \frac{(2-\beta ) z}{\beta } + \CO(z^2)~,
\end{equation}
so that the integral in~\eqref{y0final} converges at~$z = 0$. (Recall that~$\beta > 0$.) It also converges at~$z = 1$, thanks to the asymptotics of the hypergeometric function,
\begin{equation}\label{zto1hyperg}
    _2F_1(1 + 1/\sqrt{\beta} \, , 1/\sqrt\beta \, ; 1\, ; z) \simeq \frac{\sqrt{\beta } \, \Gamma \left(2 / {\sqrt{\beta }}\right)}{\Gamma \left(1/{\sqrt{\beta }}\right)^2 \, (1-z)^{2/{\sqrt{\beta }}}}  + \CO(1)~, \qquad z \to 1~.
\end{equation}
}
\begin{equation}\label{y0final}
    y_0(\beta) = {1 \over \sqrt{2 \beta}} \int_{0}^1 {dz \over z} \left(1-{(1-z)^{-(2/\sqrt{\beta} +1) } \over \left(_2F_1(1 + 1/\sqrt{\beta} \, , 1/\sqrt\beta \, ; 1\, ; z)\right)^2}\right)~.
\end{equation}
We plot this function in figure~\ref{fig:y0beta}. 

\begin{figure}[t!]
    \centering
    \includegraphics[width=0.49\textwidth]{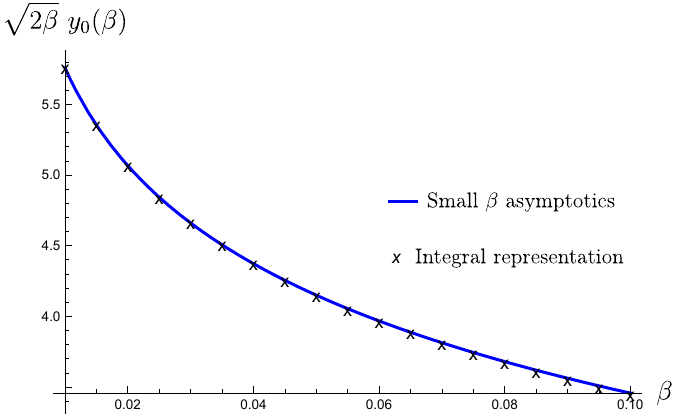} \includegraphics[width=0.49\textwidth]{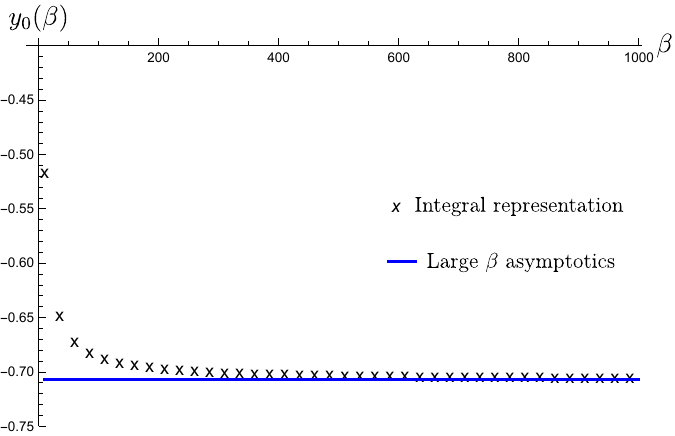} \caption{\label{fig:y0betaasym} Comparison of the small-$\beta$ (left panel) and large-$\beta$ (right panel) asymptotics of~$y_0(\beta)$, given by~\eqref{smally0betaasym} and~\eqref{largey0betaasym} respectively, with the full integral expression~\eqref{y0final}. In the left panel we plot~$\sqrt{2\beta} \, y_0(\beta)$ to facilitate the comparison.}
    \end{figure}

\noindent Let us comment on some features of the function~$y_0(\beta)$ in~\eqref{y0final}: 
\begin{itemize}
\item It is positive at small~$\beta$ and negative at large~$\beta$, with a zero at~$\beta \simeq 2.28$. 

\item  At $\beta=1$, the hypergeometric function simplifies as~$_2F_1(2 \, , 1 \, ; 1\, ; z) = 1/{(1-z)^2}$, so that~\eqref{y0final} evaluates to
    \begin{equation}
        y_0(\beta=1) = \frac{1}{\sqrt{2}}~.
    \end{equation}
    \item In the extreme type-I regime~$\beta \rightarrow 0$, the asymptotic form of $y_0(\beta)$ is given by,\footnote{~By comparing their Taylor expansions around~$z = 0$, one can show that the hypergeometric function in~\eqref{y0final} reduces to a modified Bessel function of the first kind, 
    \begin{equation}
        _2 F_1(1 + 1/\sqrt{\beta},  1/\sqrt{\beta}; 1; z) \to I_0(2 \sqrt{z/\beta})~, \qquad \beta \to 0~.
    \end{equation}
    Since~$I_0(2 t)$ grows exponentially for large positive~$t = \sqrt{z/ \beta}$, the hypergeometric term in the integrand~\eqref{y0final} is suppressed unless~$z \lesssim \beta$, so that we can approximate
    \begin{equation}\label{y0fnint}
       \sqrt{2\beta} \, y_0(\beta) \simeq  2 \int_0^{1/\sqrt{\beta}} {dt \over t} \, \left(1 - {1 \over I_0(2t)^2}\right) = \log (1/\beta) + 2 \gamma_E + \CO(\exp(-4/\sqrt\beta)) ~, \qquad \beta \to 0~.
    \end{equation}
    The logarithmic term is obvious; the subleading terms were obtained using {\tt AsymptoticIntegrate} in {\tt Mathematica}.
    The~$\CO(\beta)$ correction in~\eqref{smally0betaasym} arises from expanding the~$(1-z)^{-(2/\sqrt{\beta} + 1)}$ factor in~\eqref{y0final} beyond leading order in small~$z$.}
    \begin{equation}\label{smally0betaasym}
        y_0(\beta) \simeq \frac{1}{\sqrt{2\beta}} \left(\log{\left(1/\beta\right) + 2 \gamma_E} + \CO(\beta) \right) \quad , \quad \beta \rightarrow 0~,
    \end{equation}
    where $\gamma_E \simeq 0.5772$ is the Euler-Mascheroni constant. In figure~\ref{fig:y0betaasym} (left panel) we show the excellent agreement between the small-$\beta$ asymptotics~\eqref{smally0betaasym} and the full integral~\eqref{y0final}. The fact that~$y_0$ becomes large and positive as~$\beta \to 0$ is intuitive: in this limit the scalar potential, and hence the domain-wall tension~$\sigma \sim \sqrt{\beta}$, vanish so that the magnetic flux can spread out more.\footnote{~Recall that we take the limit~$\beta \to 0$ after we take~$n \to \infty$, so that the leading terms in~\eqref{uTdw} always dominate in the large-$n$ limit.}

    \item In the extreme type-II regime~$\beta\rightarrow \infty$, $y_0(\beta)$ approaches\footnote{~We have verified this statement numerically (see figure~\ref{fig:y0betaasym}). Analytically, it follows from the discussion in footnote~\ref{fn:convint} that the integral~\eqref{y0final} diverges at~$z = 1$ when~$\beta \to \infty$. Consequently, the leading contribution at large~$\beta \gg 1$ comes from a small region around~$z = 1$, which can be captured by changing variables to~$z = 1 - \exp(- \sqrt{\beta} s)$ with~$s \in [0, \infty)$. We will not describe the details here. 
    }
    \begin{equation}\label{largey0betaasym}
        y_0(\beta) \simeq -\frac{1}{\sqrt{2}}~, \qquad \beta \rightarrow \infty~.
    \end{equation}
    Figure~\ref{fig:y0betaasym} (right panel) shows the approach of~\eqref{y0final} to the large-$\beta$ limiting value~\eqref{largey0betaasym}. Note that~$y_0$ remains~$\CO(1)$ in this limit. 
\end{itemize}

Once~$y_0 = y_0(\beta)$ is fixed as in~\eqref{y0final}, it follows that~$\gamma_2(y) \simeq -y$ as~$y \to -\infty$. Comparing with~\eqref{matchphigam} we see that the correctly normalized solution for the boundary gauge field is
\begin{equation}
    \gamma(y) = {\sigma^{2/3} \over n^{1/3}} \, \gamma_2(y)~,
\end{equation}
where~$\gamma_2(y)$ was defined in~\eqref{g2def}. An efficient way to evaluate this function numerically is to use the~$z$-variable, as in~\eqref{zdef}, \eqref{zpint}. Using the~$y \to \infty$ asymptotics in~\eqref{g2yplusinf}, we conclude that the constant~$\gamma_\text{bdy/ext}$ that governs the~$y \to \infty$ asymptotics of~$\gamma(y)$ in~\eqref{eq:ahmasymbis} is given by
\begin{equation}
    \gamma_\text{bdy/ext} = {\sigma^{2/3} \over n^{1/3}} \cdot \frac{\Gamma \left(1 /{\sqrt{\beta }}\right)^2 }{2 \sqrt{2\beta} \Gamma \left(2/{\sqrt{\beta }}\right)} \; e^{\sqrt{2} y_0}~,
\end{equation}
while the constant~$\varphi_\text{bdy/ext}$ that governs the large-$y$ behavior of~$\varphi(y)$ in~\eqref{eq:ahmasymbis} is 
\begin{equation}
    \varphi_\text{bdy/ext} = \half e^{\sqrt{2 \beta} \, y_0}~.
\end{equation}
In turn, these constants are related to the large-$u$ asymptotic constants~$a_\infty, \varphi_\infty$ in the exterior of the string via~\eqref{eq:bdytailmatch}.

\subsection{String Tension}

We now use our large-$n$ solutions above to compute the string tension~\eqref{eq:ahmtension}, 
\begin{equation}\label{eq:ahmtensionbisDW}
    {T_n \over 2 \pi \sqrt{2}} =  \int_0^\infty u du \, \left({n^2 \over 2 u^2} \left(a'(u)\right)^2 + \left(\varphi'(u)\right)^2 + {n^2 \over u^2} a^2(u) \varphi^2(u) + {\beta \over 2} \varphi^2(u) (\varphi^2(u)-1)^2 \right)~.
\end{equation}
As in our discussion of bulk strings, we introduce a cutoff point~$y_c$ that is shared between the core and boundary regions and remains fixed as~$n \to \infty$. Recall from the discussion around~\eqref{matchphigam} that~$\varphi(y) = \exp\left((y-y_0) \sqrt{\beta/  2}\right)$ is exponentially small for such~$y$ as long as~$y \ll y_0$ is sufficiently to the left of the center of mass~$y_0$ of the domain wall. We will take~$y_c$ to lie in this region, so that~$\varphi(y)$ is exponentially small for all~$y \leq y_c$. 

As before, we separate core and boundary contributions to the string tension:
\begin{itemize}
    \item[(C)] Dropping exponentially small terms, we find 
    \begin{equation}\label{eq:coretension}
   {T_n^\text{core} \over 2 \pi \sqrt{2}} \simeq  \int_0^{u_n + y_c} u du \, \left({n^2 \over 2 u^2} \left(a'(u)\right)^2  \right) = {n^2  \over u_n^4} (u_n + y_c)^2~.
\end{equation}

\item[(B)] Following the discussion around~\eqref{eq:bdytension}, we take the boundary contribution to the tension of a domain-wall string to be
\begin{equation}\label{eq:bdytensionDW}
    {T_n^\text{bdy} \over 2 \pi \sqrt{2}} \simeq  u_n \int_{y_c}^\infty dy \, \left( \left(\varphi'(y)\right)^2  + {\beta \over 2} \varphi^2(y) (\varphi^2(y)-1)^2 \right) = {\sigma u_n\over \sqrt2}~, \qquad \sigma = {\sqrt\beta \over 2}~,
\end{equation}
up to exponentially small terms (thanks to~$y_c \ll y_0$, see above). Here we strictly work to leading order in the large-$n$ limit, which leads to several simplifications relative to the analogous bulk-string formula~\eqref{eq:bdytension}:
\begin{itemize}
    \item[$\bullet$] We have dropped the boundary gauge field~$\gamma \sim n^{-1/3}$, which would only contribute a subleading~$\CO(1)$ amount to the boundary string tension. A similar contribution would arise from the subleading backreaction~$\Delta \varphi \sim \CO(n^{-2/3})$ on~$\varphi$ that would result from substituting $\gamma$ into~\eqref{eqn:bdyeomDW}, which we did not determine. 
\item[$\bullet$] Since the domain wall is centered at~$y_0$, we should have multiplied the integral in~\eqref{eq:bdytensionDW} by~$u_n + y_0$, but since~$y_0$ is~$\CO(1)$ in the large-$n$ limit, the difference is also subleading in the string tension.
\end{itemize}
\end{itemize}
\noindent Combining~\eqref{eq:coretension} and~\eqref{eq:bdytensionDW}, we find that the total string tension is given by
\begin{equation}
    {T_n \over 2\pi} = \left(\half + 1\right) \sigma u_n + \CO(1)~, \qquad u_n = {\sqrt{2} n^{2/3} \over \sigma^{1/3}}~, \qquad \sigma = {\sqrt{\beta} \over 2}~.
\end{equation}
Here the first term in parenthesis is contributed by the core, and the second term by the boundary. We thus find perfect agreement with the variational expression~\eqref{uTdw} at leading order in the large-$n$ limit.

\subsection{Comparison with Exact Numerical Solutions}

In this section we compare our analytic large-$n$ solution above to numerical solutions of the full vortex equations~\eqref{eqn:vorteqdw} (see appendix~\ref{appnum}). 

We begin in figure~\ref{fig:y0betadwprofile30flux}, where we plot -- for a relatively modest value~$n = 30$ of the flux -- and two values of~$\beta$ the analytic boundary domain-wall solution~\eqref{dwbisbis} for the Higgs field, centered at~$y_0(\beta)$~determined in~\eqref{y0final}. We show the comparison with the full numerical solution, finding rather good agreement. 

\begin{figure}[h!]
    \centering
    \includegraphics[width=0.49\textwidth]{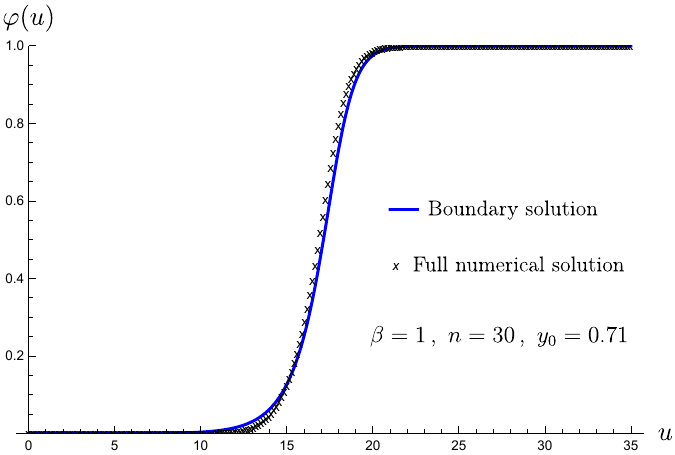}
    \includegraphics[width=0.49\textwidth]{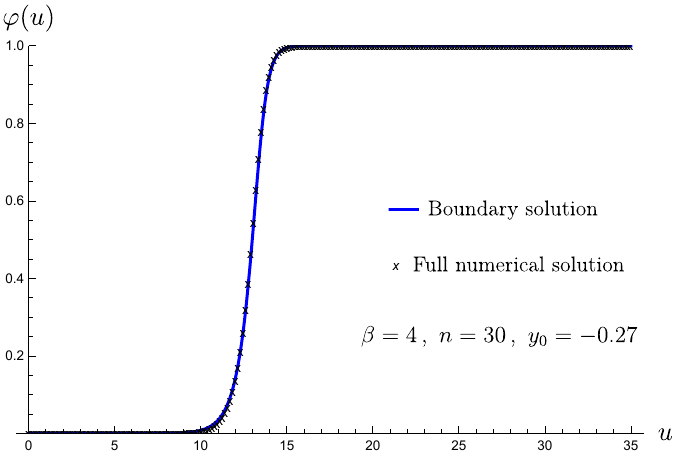}
    \caption{\label{fig:y0betadwprofile30flux} In solid blue we show the analytic boundary Higgs field~$\varphi(u)$ in~\eqref{dwbisbis}, centered at~$y_0(\beta)$ (see~\eqref{y0final}). The numerical Higgs profile obtained by solving~\eqref{eqn:ahmeom} is shown using black crosses. We fix~$n = 30$ and show~$\beta = 1$ (left panel) and~$\beta = 4$ (right panel).}
\end{figure}

\begin{figure}[h!]
    \centering
    \includegraphics[width=0.49\textwidth]{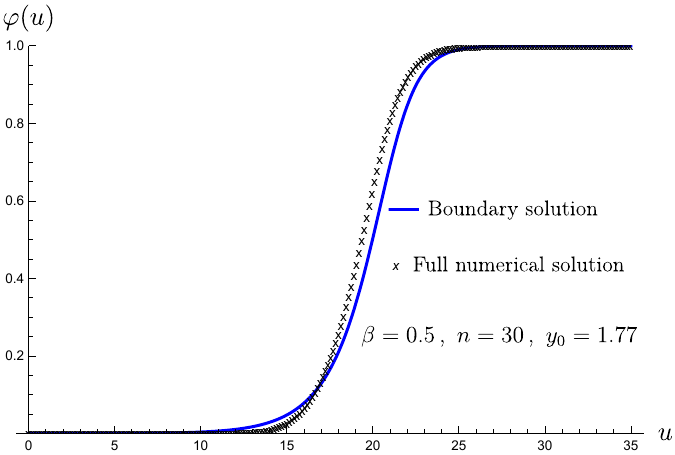}
    \includegraphics[width=0.49\textwidth]{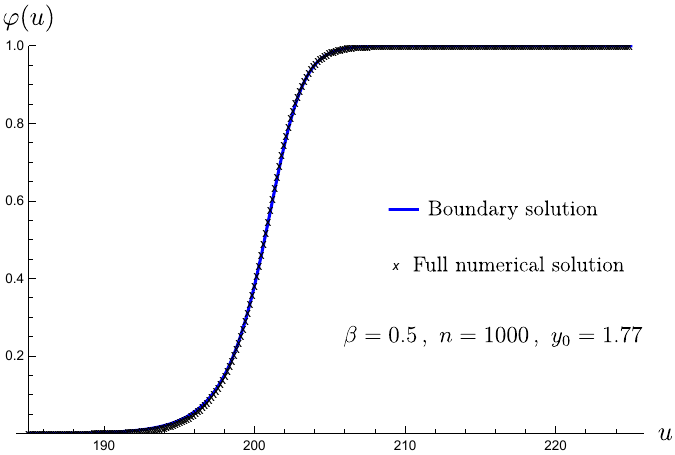}
\caption{\label{fig:y0betaflux301000} We show the analytic boundary Higgs field~$\varphi(u)$ (in blue), given in~\eqref{dwbisbis}, centered at~$y_0(\beta)$, defined in~\eqref{y0final}, and compare to the full numerical solution to the vortex equations~\eqref{eqn:ahmeom} (shown as crosses). We fix~$\beta = 0.5$ and compare~$n = 30$ (left panel) to~$n = 1000$ (right panel). The agreement improves significantly with increasing~$n$.}
\end{figure}

We then fix~$\beta = 0.5$ and compare~$n = 30$ and~$n = 1000$, finding significantly improved agreement with the full numerics in the latter case. The results are shown in figure~\ref{fig:y0betaflux301000}. 

\begin{figure}[t!]
    \centering
    \includegraphics[width=0.49\textwidth]{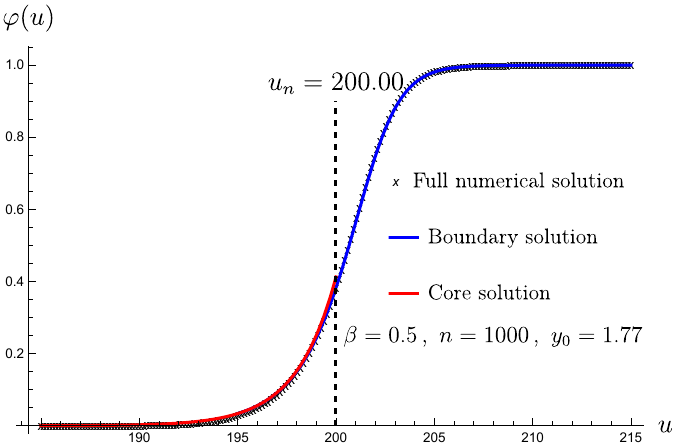}
    \includegraphics[width=0.49\textwidth]{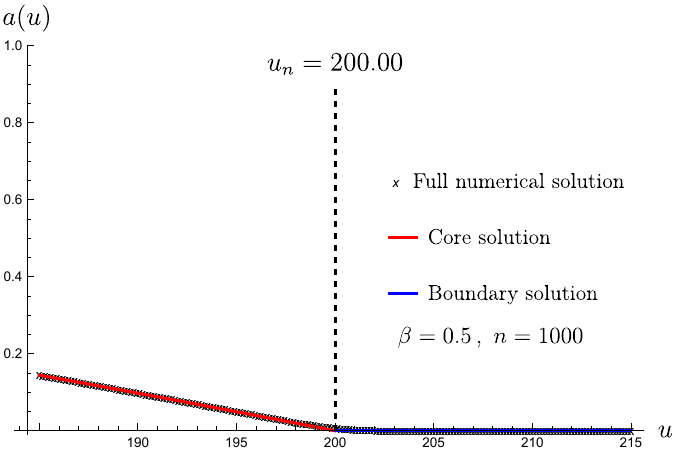}
    \caption{\label{fig:V6comparesolution}Comparison between the full numerical solution (shown in black crosses) and the analytic large-$n$ solution, for~$\beta=0.5$ and~$n=1000$. The core solution (given by~\eqref{phiDCDW} for the Higgs field in the left panel and~\eqref{linvort} for the gauge field in the right panel) is shown in red. The boundary solution (given by~\eqref{dwbisbis} for the Higgs field, and the~$\CO(n^{-1/3})$ solution~\eqref{g2def} for the gauge field) are shown in blue. We find excellent agreement in all regions of the string.}
\end{figure}

Finally, we compare the full numerical solutions to our analytic WKB solution for the Higgs field~\eqref{phiDCDW} in the core (with gauge field given by~\eqref{linvort}), as well as the boundary solution already mentioned above. The WKB solution in the core depends on an undetermined constant (see e.g.~\eqref{smallyphi}), which is fixed in terms of the boundary domain-wall center~$y_0(\beta)$ via~\eqref{y0ccoreby}. This leads to a fully analytic solution in all regions, which we compare to a full, high-precision numerical solution in figure~\ref{fig:V6comparesolution}, for~$\beta = 0.5$ and~$n = 1000$. We see that the agreement is excellent. 

\section{Stability and Interactions of Vortex Strings}
\label{binding}

In this section, we examine two closely related questions: first, whether the axially symmetric vortex solutions of magnetic flux~$n$ found above (i.e.~the $n$-strings of the AHM) are stable; second, what is the force between two parallel fundamental (i.e.~$n = 1$) strings as a function of their transverse separation. The main results of this section are summarized in table~\ref{tab:a2}, with further elaborations below.  

\renewcommand{\arraystretch}{1.4}  
\begin{table}[ht]
  \centering
    \begin{subtable}{\textwidth}
    \centering
    \caption{Conventional Abelian Higgs Model}
        \begin{tabular}{c|c|c}
        \hline
        \textbf{$\beta$} & Comparison between~$T_n$ and~$nT_1$ & Interaction force at large separations \\
        \hline
        $\beta < 1$ & $T_{n} < n \, T_{1}$  & Attractive\\
        $\beta = 1$ & $T_{n} = n \, T_{1}$  & Neutral\\
        $\beta > 1$ & $T_{n} > n \, T_{1}$ & Repulsive \\
        \hline
        \end{tabular}
    \end{subtable}
    \vspace{1em}
    \begin{subtable}{\textwidth}
    \centering
    \caption{Degenerate Abelian Higgs Model}
        \begin{tabular}{c|c|c}
        \hline
        \textbf{$\beta$} & Comparison between~$T_n$ and~$nT_1$& Interaction force at large separations \\
        \hline
         $\beta \leq 1 $ & $T_{n} < n \, T_{1}$ & Attractive \\
        $1 <\beta < \beta^{n;c}$ & $T_{n} < n \, T_{1}$ & Repulsive\\
        $\beta = \beta^{n;c}$ & $T_{n} = n \, T_{1}$ & Repulsive\\
        $\beta > \beta^{n;c}$ & $T_{n} > n \, T_{1}$ & Repulsive\\
        \hline
        \end{tabular}
    \end{subtable}
    \caption{As a function of~$\beta$ (left column), we compare the~$n$-string tension~$T_n$ with the energy of~$n$ fundamental strings (middle column), and indicate whether two parallel widely separated~$n = 1$ strings attract or repel (right column). The top panel (a) shows the conventional AHM. For the degenerate AHM in the bottom panel (b), we use $\beta = \beta^{n;c}$ to denote the critical value where~$T_n=nT_1$. As is explained in the main text, $\beta^{n;c}$ increases grows monotonically and without bound as we increase~$n$.}
    \label{tab:a2}
\end{table}

\subsection{Stability of Axially Symmetric Strings}\label{sec:stability}

Even the lightest axially symmetric~$n$-strings of flux~$n \geq 2$ (with tension~\eqref{Tndef}) might be unstable and fission into strings of smaller flux. A simple decay channel is~$T_n \to n T_1$, but in principle one needs to check all possible decays allowed by flux conservation. 

In general, stability of~$T_n$ requires the function~$T_n$ to be convex down. This is particularly simple to check in the large-$n$ limit, where we can essentially treat~$n$ as a continuous variable. Then~$T_n$ is convex down if and only if 
\begin{equation}
    \frac{d^2T_n}{dn^2} < 0~.
\end{equation}
If this is true for all~$n$, then all~$T_n$ are stable. Conversely, if~$T_n$ is convex up, 
\begin{equation}
     \frac{d^2T_n}{dn^2} > 0~,
\end{equation}
then the~$T_n$ are unstable. We will now check these conditions in the conventional and degenerate AHMs.

\subsubsection{Conventional Abelian Higgs Model}

Here it is well known that the stability of~$T_n$ is entirely dictated by the parameter~$\beta$, and hence by the bulk spectrum of the model. (See~\cite{Manton:2004tk,Weinberg:2012pjx} for a review.) In the type-I regime, 
\begin{equation}
    T_n < nT_1~, \qquad \beta < 1~, 
\end{equation}
so that~$T_n$ is stable. By contrast, in the type-II regime, it fissions into~$n$ fundamental strings, \
\begin{equation}
    T_n > nT_1~, \qquad \beta > 1~.
\end{equation}
When~$\beta = 1$, the BPS bound~\eqref{BPST} implies that
\begin{equation}\label{bpstnnt1}
    T_n = nT_1~, \qquad \beta=1~,
\end{equation}
so that the strings are exactly marginal at this point.\footnote{~A BPS $n$-vortex has~$2n$ real zero modes, corresponding to the transverse positions of the~$n$ constituent fundamental vortices, which behave like free particles. This was conjectured in~\cite{Weinberg:1979er} and proved in~\cite{Taubes:1979tm}.}

In the large-$n$ limit, these results follow simply from the expression~\eqref{finaltension} for the string tension, leading to 
 \begin{equation}
        \frac{1}{2\pi}\frac{d^2T_n}{dn^2} =  - \frac{\sigma(\beta)}{2\sqrt{2}\beta^{1/4}} \frac{1}{n^{3/2}} \quad 
        \begin{cases}
            < 0 \qquad \beta < 1 \\ 
            =0 \qquad\beta=1 \\
            >0 \qquad \beta > 1 
        \end{cases}~,
\end{equation}
where the sign is determined by the surface tension~$\sigma(\beta)$ (see table~\ref{tab:sigma}) of the planar domain wall separating the Coulomb phase with constant magnetic field in the core and the Higgs vacuum in the exterior. The fact that~$\sigma > 0$ in the type-I regime and~$\sigma < 0$ in the type-II regime is well known in the context of superconductors. 

\subsubsection{Degenerate Model}

At large-$n$, the tension is given by~\eqref{uTdw} and scales as~$T_n~\sim n^{2/3}$, which is a convex function of~$n$,
\begin{equation}\label{stabledeg}
    \frac{d^2T_n}{dn^2} < 0~,
\end{equation}
for all values of~$\beta$. Thus all giant axially symmetric~$n$-strings are stable. 

By contrast, studying the stability of~$T_n$ at finite~$n$ leads to a more complicated picture:\footnote{~We have analyzed this pattern at small values of~$n$, and we conjecture that it persists at higher $n$.} for each~$n$, there is a critical~$\beta = \beta^{n;c}$, at which the~$n$-string is marginally stable to decay into fundamental strings,
\begin{equation}
    T_n = nT_1~, \qquad \beta = \beta^{n;c}~.
\end{equation}
These critical values of~$\beta$ lie deep in the type-II regime and increase with flux~$n$, e.g.
\begin{equation}\label{critical2and3}
    \beta^{2;c} = 28.25 \qquad , \qquad \beta^{3;c} = 49.05~.
\end{equation}
We find that~$T_n$ is stable at sufficiently small~$\beta$, 
\begin{equation}
    T_n < nT_1~, \qquad \beta < \beta^{n;c}~,
\end{equation}
while it is unstable at sufficiently large~$\beta$, 
\begin{equation}
    T_n > nT_1 \qquad , \qquad \beta >  \beta^{n;c}~.
\end{equation}
This is consistent with the stability of giant vortices discussed around~\eqref{stabledeg} above, because the critical value~$\beta^{n; c} \to \infty$ as~$n \to \infty$. Thus all strings become stable if we take the large-$n$ limit at fixed~$\beta$ (as we have done throughout). 

\begin{figure}[t!]
    \centering
    \includegraphics[width=0.49\textwidth]{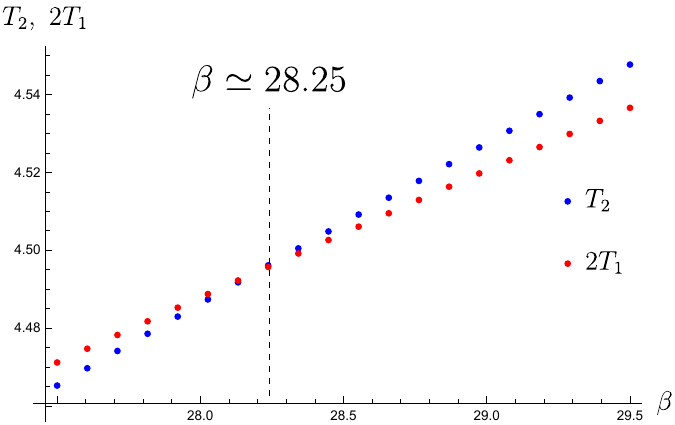}
    \includegraphics[width=0.49\textwidth]{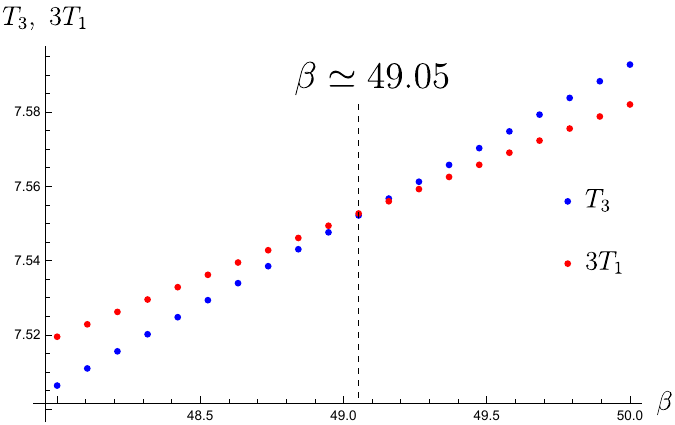}
    \includegraphics[width=0.49\textwidth]{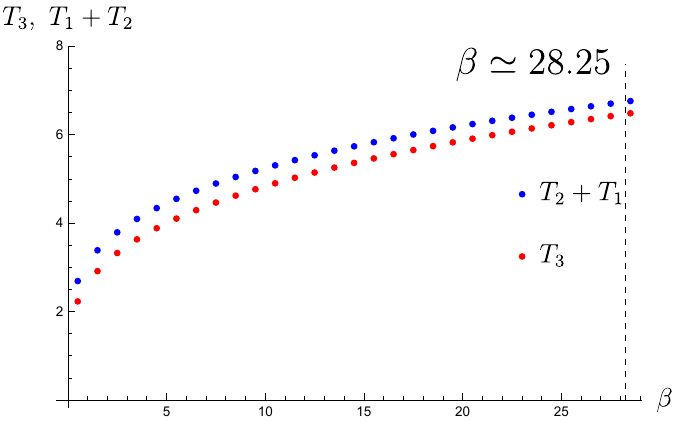 }
    \caption{\label{fig:tnversusnt1}
    In the top panel we compare~$T_n$ and~$n T_1$ for~$n= 2,3$ as a function of~$\beta$, identifying the critical values~$\beta^{n;c}$ in~\eqref{critical2and3} below which~$T_n$ is stable against decay into~$nT_1$. In the bottom panel we show that the candidate decay channel~$T_3 \to T_1 + T_2$ is never realized, because~$T_3 < T_1 + T_2$ for~$\beta < \beta^{2;c} < \beta^{3;c}$.}
\end{figure}

In figure~\ref{fig:tnversusnt1} (top panel) we explicitly compare~$T_n$ and~$n T_1$ for~$n = 2$, $n = 3$ as functions of~$\beta$, which allows us to extract the critical values~\eqref{critical2and3}. Note that in addition to the fission channel~$T_3 \to 3 T_1$, the $n = 3$ string has another possible fission channel~$T_3 \to T_1 + T_2$. This only becomes available for~$\beta < \beta^{2;c}$, where~$T_2$ is stable. We show in the bottom panel of figure~\ref{fig:tnversusnt1} that~$T_3$ is always stable in this range, and hence for all~$\beta<\beta^{3;c}$. 

\subsection{Force Between Parallel Fundamental Strings}\label{sec:interactionenergybetweensepfund}

Here we discuss the force between two parallel~$n = 1$ strings, separated by a transverse distance~$R$. When~$R$ is sufficiently large, the strings are separated by a large region of Higgs vacuum, and the massive Higgs and vector particles in this vacuum mediate attractive and repulsive Yukawa-type interactions between the two strings, respectively. This leads to an interaction potential of the form 
 \begin{equation}\label{eq:int}
    V_{\text{int}}(R) \simeq -A^2 K_0 (m_H R) + B^2 K_0 (m_V R)~, \qquad m_{H, V} R \gg 1~,
\end{equation}
where the constants~$A,~B$ are determined by~$a_\infty,~\varphi_\infty$ in \eqref{eq:ahmasym} via
\begin{equation}\label{AB}
    A^2 = 4 \sqrt{2\beta} \varphi_\infty^2 v^2~, \qquad B^2 = 2\sqrt{2} a_\infty^2 v^2~.
\end{equation}
We only need the modified Bessel function of the second kind~$K_0(x)$ in the asymptotic region,
\begin{equation}\label{besselasym}
    K_0(x) \simeq \sqrt{\frac{\pi}{2}} \frac{e^{-x}}{\sqrt{x}}~, \qquad x\rightarrow \infty~.
\end{equation}
The vortex interaction~\eqref{eq:int} has been derived in~\cite{Speight_1997,Bettencourt_1995} for the conventional AHM. In appendix~\ref{sec:ahminteraction} we present a brief derivation that is valid in both AHMs.

In the type-I regime~$\beta <1$, the first term in~\eqref{eq:int} dominates and the strings attract; in the type-II regime~$\beta>1$, the second term dominates and the strings repel. Thus the question of whether widely separated fundamental strings attract or repel is entirely governed by the mass spectrum in the Higgs vacuum. At~$\beta=1$, the masses of the Higgs boson~$m_H$ and the vector boson~$m_V$ are the same, so that we must compare the coefficients~$A,~B$ in~\eqref{eq:int}. We do this separately for the two different AHMs below. 

In general, knowing the force between two well-separated fundamental~$n = 1$ strings is not sufficient to determine whether they will form a bound state, or whether that bound state will preserve the axial symmetry.\footnote{~See for instance the related quantum mechanical discussion in~\cite{Jackiw:1991je}.} To answer these questions we need to know the full interaction potential~$V_{\text{int}}(R)$. Note that~$R = 0$ corresponds to an axially symmetric $n = 2$ string,\footnote{~We will expand on the definition of~$V_\text{int}(R)$ at short distances below.} whose binding energy is given by 
 \begin{equation}\label{bindingzerosep}
     V_{\text{int}}(R=0) = T_2 - 2 T_1 ~.
 \end{equation}
This binding energy was discussed in section~\ref{sec:stability} (see also table~\ref{tab:a2}). We will make some comments about~$V_\text{int}(R)$ in the different AHMs below.

\subsubsection{Conventional Abelian Higgs Model}\label{convint}

Let us make some comments that are specific to the conventional AHM:
\begin{itemize}
    \item At the BPS point~$\beta = 1$, the constants~$\varphi_\infty,a_\infty$ are related to each other via~\eqref{eqn:bpsasym}, so that  the constants~$A,~B$ in~\eqref{eq:int} satisfy
   \begin{equation}\label{bpsAB}
    A = B~, \qquad \beta = 1~.
\end{equation}
This was also shown analytically in~\cite{Tong:2002rq}, and it is consistent with the fact that BPS strings exert no force on each other. 

\begin{figure}[t!]
    \centering
    \includegraphics[width=0.49\textwidth]{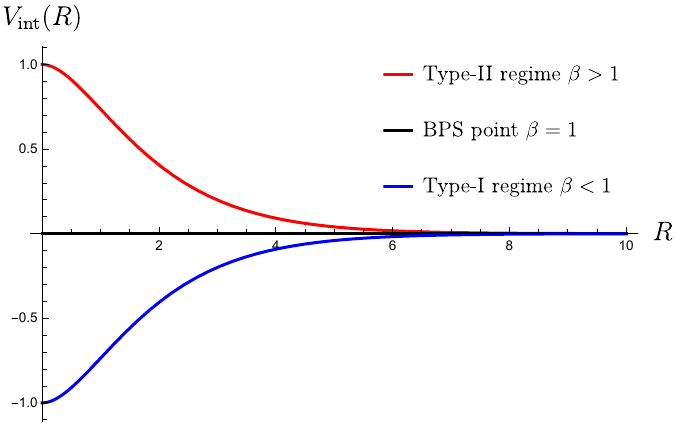}
    \includegraphics[width=0.49\textwidth]{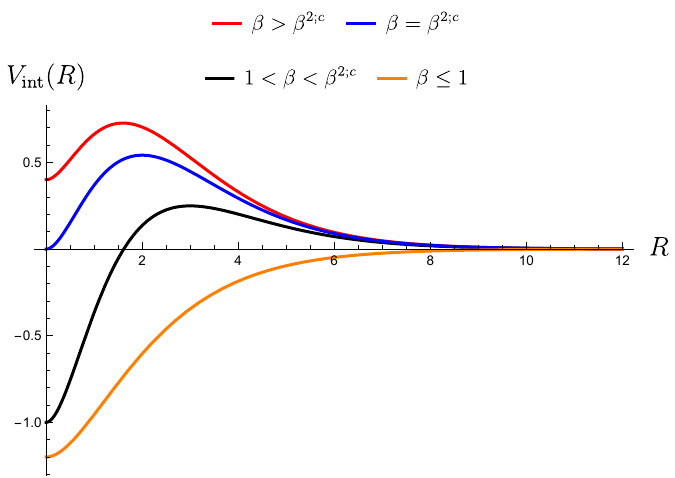}
  \caption{Schematic renditions of the interaction potentials $V_{\text{int}}(R)$ between two parallel~$n = 1$ strings separated by a transverse distance $R$ for the two AHMs under consideration. The left panel schematically depicts the results of~\cite{Jacobs:1978ch} for the conventional AHM; the right panel sketches an analogous conjectured interaction potential for the degenerate AHM.}
  \label{pol}
\end{figure}

\item A variational calculation of~$V_{\text{int}}(R)$ in the conventional AHM was carried out in~\cite{Jacobs:1978ch};\footnote{~A key feature of this analysis was the idea that the location of the vortices is characterized by the vanishing of the Higgs field.} we schematically depict their results in the left panel of figure~\ref{pol}. At large~$R$, the behavior of~$V_{\text{int}}(R)$ is consistent with~\eqref{eq:int}. Remarkably, the large-$R$ behavior persists monotonically all the way to~$R = 0$. As already discussed around~\eqref{bindingzerosep}, $V_{\text{int}}(R=0)$ is the binding energy of a axially-symmetric~$n = 2$ string. It is positive in the type-II regime ($\beta > 1$), where the 2-string is unstable, and negative in the type-I regime ($\beta < 1$), where the 2-string is stable. 

\item The convexity of~$V_{\text{int}}(R)$ near the origin is determined by the perturbative stability of the axially-symmetric~$n = 2$ string. A general conjecture about the stability of axially-symmetric vortices was made in~\cite{Jaffe:1980mj}, and proven in~\cite{gustafson1999stabilitymagneticvortices}:\footnote{~Earlier work~\cite{Bogomolny:1975de} had already shown that~$n\geq 2$ vortices are unstable in the type-II regime.} 
\begin{itemize}
    \item[(i)] Rotationally symmetric~$n=1$ strings are perturbatively stable for all values of~$\beta$.
    \item[(ii)]  For~$n\geq 2$, rotationally invariant strings are stable in the type-I~$\beta <1$ regime.
    \item[(iii)]  For~$n\geq 2$, rotationally invariant strings are unstable for type-II~$\beta >1$ regime. 
\end{itemize}
\noindent This is consistent with the fact that~$V_\text{int}(R \simeq 0)$ in figure~\ref{pol} is convex down for~$\beta > 1$, and convex up for~$\beta < 1$. A direct diagnostic is the fluctuation spectrum around an axially-symmetric~$n = 2$ string; its instability when~$\beta > 1$ portends a tachyon.

\end{itemize}

\subsubsection{Degenerate Abelian Higgs Model}\label{degenint}

Here we will make some observations about the degenerate AHM:
\begin{itemize}
    \item Consider the interaction energy~\eqref{eq:int} when there is a large separation between the two fundamental strings at~$\beta=1$. We can  numerically evaluate~$\varphi_\infty$ and~$a_\infty$, and using~\eqref{AB} also the constants~$A,~B$, at this value of~$\beta$, 
\begin{equation}\label{ABdegen}
    A^2 = 782.5243~v^2~, \qquad  B^2 = 38.4418~v^2~, \qquad \beta=1~.
 \end{equation}
Since~$A>B$, it follows from~\eqref{eq:int} that well-separated~$n = 1$ strings in the degenerate AHM attract at~$\beta = 1$. 

 \item  No calculation of~$V_{\text{int}}(R)$ along the lines of~\cite{Jacobs:1978ch} has been carried out in the degenerate model. Here we briefly comment on some of its features that can be inferred from the information that we have assembled about the strings in this model, which we schematically depict in the right panel of figure~\ref{pol}: 
 \begin{itemize}
     \item[(i)] The large-$R$ behavior is given by~\eqref{eq:int}, and as in the abelian model it is determined by the bulk spectrum through~$\beta$. A point of difference, discussed around~\eqref{ABdegen} above, is that strings at~$\beta = 1$ in the degenerate model attract at long distances. 

     \item[(ii)] According to~\eqref{bindingzerosep},~$V_{\text{int}}(R=0)$ is the binding energy of the axially-symmetric~$n =2$ string. It follows from the discussion in section~\ref{sec:stability} that~$V_{\text{int}}(R=0)$ changes sign at the critical~$\beta = \beta^{2;c}$, being negative for smaller~$\beta$ and positive for larger~$\beta$. 

     \item[(iii)] The numerical study~\cite{DGL} of perturbative fluctuation around an axially-symmetric~$n = 2$ string reveals no tachyons for any value of~$\beta$. This string is therefore perturbatively stable for all values of~$\beta$, so that~$V_{\text{int}}(R=0)$ must be a local minimum. 
 \end{itemize}

\item  The resulting schematic depiction of~$V_\text{int}(R)$ in figure~\ref{pol} is thus not always monotonic. In particular, for~$1<\beta \leq \beta^{n;c}$,  widely separated fundamental strings repel, but there is a stable axially-symmetric $n = 2$ bound state. This is a scenario that does not arise in the conventional AHM. Models with multiple Higgs fields and non-monotonic~$V_{\text{int}}(R)$ (known as type-1.5 superconductors) have been studied in~\cite{PhysRevB.83.020503,Babaev_2017,Speight_2021,Babaev_2012,Carlstr_m_2011,Babaev_2005,Timoshuk_2024}. The degenerate model analyzed here is a particularly simple example with just a single Higgs field. 
\end{itemize}

\bigskip


\section*{Acknowledgments}

\noindent The authors thank Yanyan Li for collaboration on related questions.  TD is grateful to  Philipp Dumitrescu and Zohar Komargodski for discussions. AG acknowledges discussions with Per Kraus and Lukas Lindwasser. The authors acknowledge support from the Mani~L.~Bhaumik Institute for Theoretical Physics, the U.S. Department of Energy through award DE-SC0009937, as well as the Simons Foundation through the Collaboration on Global Categorical Symmetries.


\appendix


\section{Numerics}\label{appnum}
Through examples, we walk through how the numerics are performed in the paper. All the numerics in the paper have been done using packages in {\tt Mathematica}. 

Consider the vortex equations~\eqref{eqn:ahmeom}. We use the finite element method to solve the equations. We solve the equations over a finite domain~$[0,L]$ by approximating~$\infty$ by a finite cutoff~$L$. We divide the finite domain into a discretized set. For instance we can consider~$P$ points, and we solve the equations over a discretized domain given by
\begin{equation}
    S= \Big\{0,\frac{L}{P-1},\frac{2 L}{P-1},\dots , \frac{(P-2) L}{P-1},L\Big\}~.
\end{equation}
The mesh size~$h$ for our domain is given by
\begin{equation}\label{meshsize}
    h = \frac{L}{P-1}~.
\end{equation}
Instead of solving our differential equations over a continuous domain spanning the radial coordinate~$u\in[0,\infty)$, we solve for the functions evaluated at each discretized point in the set~$S$. For the two functions,~$a(u),~\varphi(u)$, we refer to the values the functions take at these discretized points as~$a_i,~\varphi_i$ for~$i\in\{1,\dots,P\}$. Thus,~$a_1 = a(0)$,~$a_2 = a(L/(P-1))$, etc.~up to~$a_P = a(L)$.

To solve for~$a_i,~\varphi_i$, we convert our differential equations into finite difference equations. We have~$2P$ finite difference equations and~$2P$ variables. We organize them into a matrix equation, and solve it to obtain~$a_i,~\varphi_i$. Things are simplified a bit by the fact that~$a_1,~\varphi_1,~a_P,~\varphi_P$ are determined by the boundary conditions  given in~\eqref{eq:ahmbc},
\begin{equation}
    a_1 = 1~, \quad \varphi_1 = 0~, \quad  a_P = 0~, \quad  \varphi_P=1 ~,
\end{equation}
and thus we just have to deal with~$2(P-2)$ difference equations.

We solve this matrix equation using the {\tt FindRoot} function in {\tt Mathematica}. This function requires an initial 'seed' value in the vicinity of which it can search for a solution to the required equation. For the case at hand, we need to specify an initialization value for~$a_i,~\varphi_i$ for~$i\in \{2,\dots,P-1\}$. Once we solve the finite difference matrix equation correctly, we have determined~$a(u),~\varphi(u)$ over the discretized domain. We can then use an interpolating function to construct a smooth solution over the entire domain~$[0,L]$. This completes the process of numerically solving the differential equations.

Let us refer to the solutions obtained via this numerical algorithm as~$a_{(\text{num})},~\varphi_{(\text{num})}$. We can make an error estimate as to how well our numerics have solved the equations. To do this, we can substitute our numerical solutions~$a_{(\text{num})},~\varphi_{(\text{num})}$ back into the differential equations and check how well they solve the respective equations. We define
\begin{equation}\label{E}
    \begin{split}
        E_1(u) =~ & u^2 \varphi''_{(\text{num})}+u \varphi'_{(\text{num})} - n^2a_{(\text{num})}^2\varphi_{(\text{num})} - u^2 ~\beta\varphi_{(\text{num})}(\varphi_{(\text{num})}^2-1) ~,\\ 
        E_2(u) =~ & u a_{(\text{num})}'' - a_{(\text{num})}' -2 u a_{(\text{num})}\varphi_{(\text{num})}^2~, \\ 
    \end{split}
\end{equation}
where,~$E_1(u)=E_2(u)=0 $ over the entire radial domain~$u\in[0,\infty)$ when evaluated on the true solution. By substituting~$a_{(\text{num})}(u),~\varphi_{(\text{num})}(u)$, we get a simple quantitative measure to check how far we deviate from the true solution.

The exact solution to the equations is not known. Say the true solution to the equations is given by~$\varphi^*(u),~a^*(u)$. The numerical solution we obtain depends on the mesh size $h$. We denote this numerical solution as~$a_{(\text{num})}(h),~\varphi_{(\text{num})}(h)$. Let us assume that the error as a function of the mesh size~$h$ takes the following form,
\begin{equation}\label{ABnum}
    \begin{split}
        \varphi^* = &  \varphi_{(\text{num})}(h) + A (u)h^k + \CO(h^{k+1}) ~, \\
        a^* = & a_{\text{num}}(h) + B(u) h^k + \CO(h^{k+1}) ~,
    \end{split}
\end{equation}
where~$k$ is some positive integer and~$\CO(h^k)$ represents the truncation error of the numerical solution~$\varphi_{(\text{num})}(h)$. Here~$A(u),~B(u)$ are the leading-order relative residual functions that quantify how the numerical solutions for the Higgs and gauge fields deviate from the true solution at~$\CO(h^k)$.

To determine the order of the truncation error~$k$, we focus on the Higgs field~$\varphi(u)$; the same analysis will apply for the gauge field~$a(u)$. For mesh sizes~$h$ and~$h/2$, the numerical solution admits an expansion of the forms respectively given by
\begin{equation}
    \begin{split}
        \varphi^* &= \varphi_{(\text{num})}(h) + A(u)\,h^k + \mathcal{O}(h^{k+1}) ~, \\  
        \varphi^* &= \varphi_{(\text{num})}(h/2) + A(u)\,\frac{h^k}{2^k} + \mathcal{O}(h^{k+1}) ~, \\
    \end{split}
\end{equation}
Consequently, by eliminating~$\varphi^*(u)$, we obtain
\begin{equation}\label{ratio}
    \frac{\varphi_{(\text{num})}(h/2)-\varphi_{(\text{num})}(h/4)}
         {\varphi_{(\text{num})}(h)-\varphi_{(\text{num})}(h/2)} = 2^{-k} ~.
\end{equation}
A simple consistency check of the numerics is therefore to perform the computation at three different mesh sizes~$h,~h/2,~h/4$ and verify that the ratio in~\eqref{ratio} approaches~$2^{-k}$. This allows us to read off the order~$k$ of the truncation error. 

 \begin{figure}[t!]
    \centering
    \includegraphics[width=0.49\textwidth]{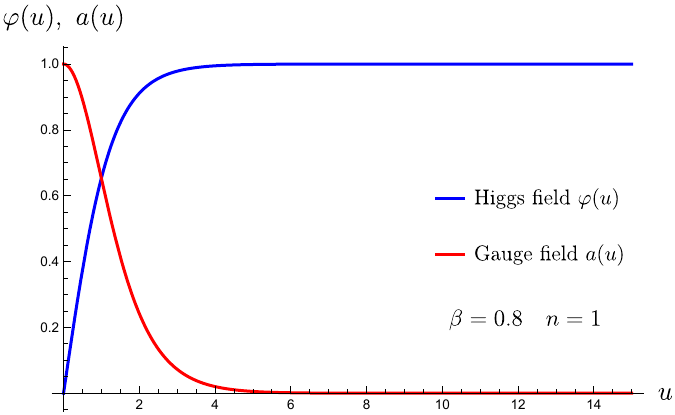}
    \includegraphics[width=0.49\textwidth]{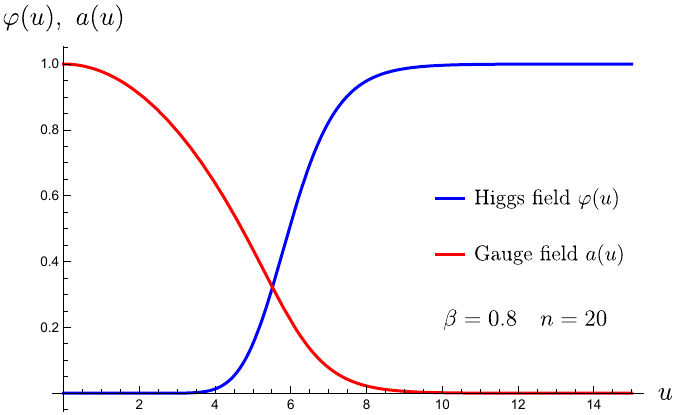}
  \caption{\label{1} Numerical field profiles for the Higgs field~$\varphi(u)$ (shown in blue) and the gauge field~$a(u)$ (shown in red) at~$\beta=0.8$. The left panel shows the fundamental string~$n=1$ and the right panel shows the string solution at~$n=20.$ }
  \label{}
\end{figure}
 
\begin{figure}[t!]
    \centering
    \includegraphics[width=0.49\textwidth]{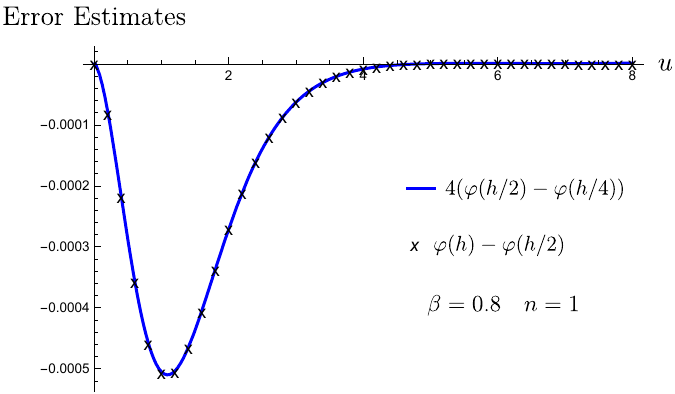}
    \includegraphics[width=0.49\textwidth]{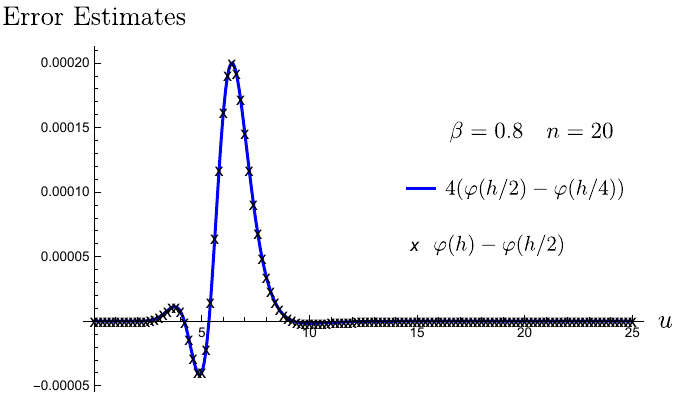}
  \caption{ \label{3} Convergence test for the truncation error using the ratio~\eqref{ratio}. At fixed~$P=500$, the domain size~$L$ (and hence the mesh size~$h$, defined in~\eqref{meshsize}) is varied. The left panel shows the~$n=1$ vortex with~$L=10,~20,~40$ and the right panel the~$n=20$ vortex with~$L=25,~50,~100$. We determine that the truncation order is~$k=2$ and that the numerical solutions deviate from the true solution at~$\CO(h^2)$. }
  \label{}
\end{figure}

\begin{figure}[t!]
        \centering
        \includegraphics[width=0.49\textwidth]{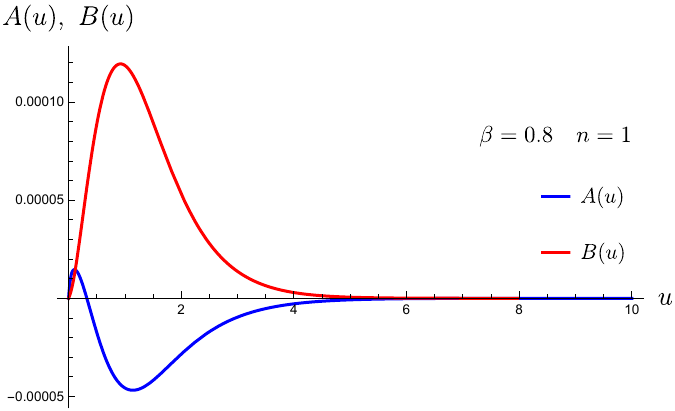}
         \includegraphics[width=0.49\textwidth]{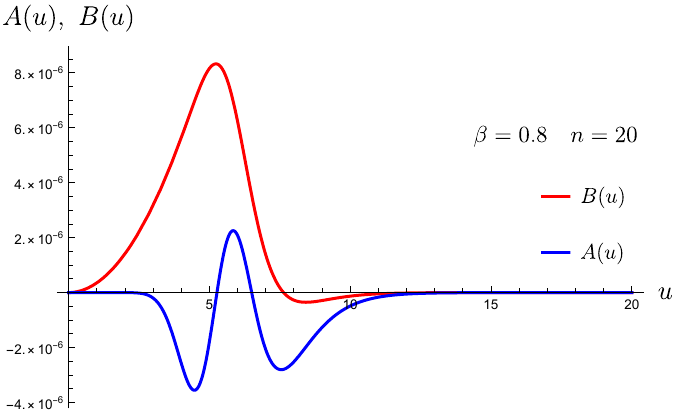}
        \caption{\label{4} The figure illustrates the relative errors,~$A(u),~B(u)$ defined in~\eqref{ABnum} for~$\beta=0.8$. The left panel shows the~$n=1$ string and the right panel, the~$n=20$ string. In both cases, the functions~$A(u),~B(u)$ remain small across the domain, demonstrating the high accuracy of the numerical solutions for both the Higgs~$\varphi(u)$ and the gauge~$a(u)$ fields.}
    \end{figure}

 Below we illustrate the full numerical procedure through two representative examples, first for the fundamental string~$n=1$, and then for a string of higher vorticity~$n=20$.
\begin{itemize}
    \item For the fundamental~$n=1$ string, at~$\beta=0.8$, we work with a discretized domain consisting of~$P=500$ points, and use the simple initialization,
    \begin{equation}
        a_i = \varphi_i = 0.6~, \qquad i = 2, \dots., P-1~.
    \end{equation}
    The resulting field profiles are shown in the left panel of figure~\ref{1}, while the corresponding numerical error functions~$E_1(u),~E_2(u)$, defined in~\eqref{E}, appear in the left panel of figure~\ref{2}.  
    By repeating the computation at several mesh sizes, achieved by varying the domain length~$L$,
    \begin{equation}
        L = 10, ~20,~40~, \qquad P=500~,
    \end{equation}
    and noting that changing~$L$ changes the discretization error~\eqref{meshsize}, we verify that the truncation error behaves as,
    \begin{equation}
        \varphi^* = \varphi_{\text{num}}(h) + \CO(h^2)~, \qquad k=2 ~,
    \end{equation}
    as demonstrated by the ratio test~\eqref{ratio}. This is verified in the left panel of figure~\ref{3}. In figure~\ref{4}, we also plot the leading~$\CO(h^2)$ coefficient functions~$A(u),~B(u)$, defined in~\eqref{ABnum}. We see that both the error functions remain small across the entire radial domain, indicating excellent numerical accuracy. 

     \begin{figure}[t!]
    \centering
    \includegraphics[width=0.49\textwidth]{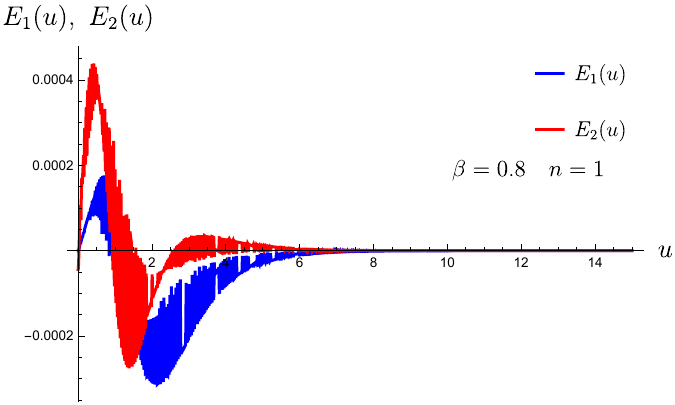}
    \includegraphics[width=0.49\textwidth]{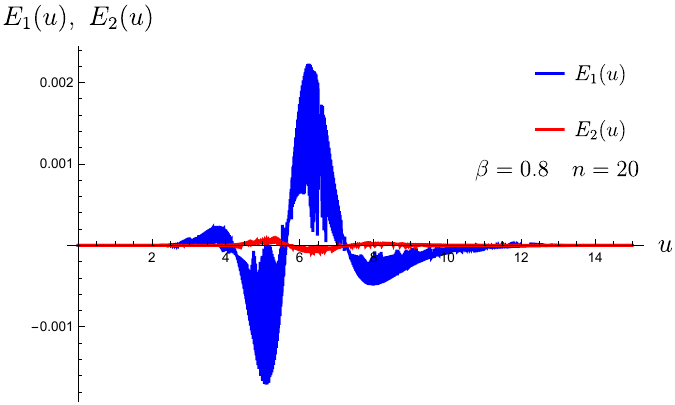}
  \caption{ \label{2} The figure shows the error functions~$E_1(u)$ (shown in blue) and~$E_2(u)$ (shown in red), defined in~\eqref{E}, evaluated on  the numerical solutions at~$\beta=0.8$. The left panel corresponds to the fundamental~$n=1$ string and the right panel corresponds to the~$n=20$ string solution. These curves quantify how well the numerical profiles satisfy the differential equations.}
  \label{}
\end{figure}

    \item For the higher charge case, with~$n>1$, the initialization values~$a_i$ and~$\varphi_i$ (for~$i=2,\dots,P-1$) are generated iteratively. The solution at flux~$n-1$ serves as the initialization to solve for the string solution at flux~$n$. 

    For illustration, we consider the parameter values~$n=20$ at~$\beta=0.8$. The field profiles appear in the right panel of figure~\ref{1} and the corresponding error functions~$E_1(u),~E_2(u)$, defined in~\eqref{E}, appear in the right panel of figure~\ref{2}. The convergence behavior, again demonstrating~$\CO(h^2)$ truncation error, is verified in the right panel of figure~\ref{3}, where the mesh size is varied by choosing,
    \begin{equation}
        L = 25,~50,~100~, \qquad P=500~. 
    \end{equation}
    The associated relative error functions~$A(u),~B(u)$ are plotted in the right panel of figure~\ref{4}, and again we observe that the errors remain small, even for a string solution at higher flux.

\end{itemize}

\newpage


\section{Fundamental Strings at Large Separation}\label{sec:ahminteraction}

For the Abelian Higgs models discussed in this paper, we present a derivation of the interaction energy~\eqref{eq:int} between two parallel fundamental strings separated by a large transverse distance~$R$,
\begin{equation}
    V_{\text{int}}(R) \simeq -A^2 K_0 (m_H R) + B^2 K_0 (m_V R)~, \qquad R\gg {1 \over m_{V,H}}~.
\end{equation}
In particular, we will give explicit formulas for the constants~$A,~B$ in terms of~$\varphi_\infty,~a_\infty$ (see~\eqref{eq:ahmasym}). Here~$K_0(x)$ is a modified Bessel function of the second kind, whose large-$x$ asymptotics are shown in~\eqref{besselasym}.

The string asymptotics in the Higgs vacuum are given by
\begin{equation}\label{app:asym}
   v\varphi(r) \simeq v-\frac{c}{4\pi} K_0(m_Hr)~, \qquad a(r) \simeq \frac{de^2m_V}{2\pi} r K_1 (m_V r)~, \qquad r \rightarrow \infty~,
\end{equation}
where the constants~$c,~d$ are related to~$\varphi_\infty,~a_\infty$, defined in~\eqref{eq:ahmasym}, via
\begin{equation}\label{cd}
    \varphi_\infty = \frac{c}{4\pi v} \sqrt{\frac{\pi}{2}} \frac{1}{(2\beta)^{1/4}} \qquad , \qquad a_\infty = \frac{de^2}{2\pi} \sqrt{\frac{\pi}{\sqrt{2}}} ~.
\end{equation}
At long distances, the theory is Higgsed; the linearized Lagrangian in the Higgs vacuum is 
\begin{equation}\label{lin}
    \SL_{\text{lin}} = -\frac{1}{4e^2} f_{\mu\nu}f^{\mu\nu} - v^2 a_\mu a^\mu - v^2\partial_\mu\varphi_1\partial^\mu\varphi_1 -m_H^2 v^2 \varphi_1^2~,
\end{equation}
where~$\varphi=\varphi_1+i\varphi_2$, and~$\varphi_2$ is the unphysical Goldstone mode that is eaten by the vector boson.  From afar, a static vortex appears as a solution to~\eqref{lin} with a singular point source at the vortex center. If we can identify suitable $\delta$-function sources that replicate the far-field behavior, we can model our vortex as a composite source, valid in the approximation where vortices are sufficiently separated and the fields match their asymptotic profiles. 

The strings are static and translationally invariant along the~$z$‑axis, so the problem reduces to the transverse~$xy$‑plane. Placing one fundamental string at the origin and the other at transverse displacement~$\vec{R}$, the corresponding sources are given by,
\begin{equation}\label{source}
    \begin{split}
         J(\vec{x}) = & c \delta^{(2)} (\vec{x}) + c \delta^{(2)}(\vec{x}-\vec{R}) ~, \\
        j^i(\vec{x}) = & d \ep^{ij} \partial_j \delta^{(2)} (\vec{x}) + d \ep^{ij}\partial_j \delta^{(2)}(\vec{x}-\vec{R}) ~, \\ 
    \end{split}
\end{equation}
where~$i=(x,y)$.  These sources couple to~$\varphi_1$ and~$a_i$ through the following terms in the Lagrangian,
\begin{equation}
    \SL_{\text{source}} = -Jv\varphi_1 - j_i a_i~,
\end{equation}
The strings interact by exchanging the massive Higgs scalar and the massive gauge boson, each exchange contributing separately to the inter‑string potential. The cleanest way to compute these contributions is in Euclidean signature. For time-independent configurations, the partition function~$Z[J] = e^{-TE[J]}$, where~$E$ is the energy of the configuration.  Because the linearized theory is Gaussian, the path integral can be
evaluated exactly,
\begin{equation}\label{pathintegral}
Z[J,j^{\,i}]
   \;=\;
   \int\!\mathcal D\varphi\,\mathcal DA_\mu\,
        \exp\!\Bigl[
           -\!\int d^{4}x\,
                 \bigl(\SL_{\text{lin}}+ \SL_{\text{source}}\bigr)
        \Bigr]~.
\end{equation}
Carrying out the Gaussian integral~$\eqref{pathintegral}$ gives us,
\begin{equation}\label{completepi}
Z[J,j^{\,i}]
   =\exp\!\Biggl\{
       -\!\int d^{4}x\,d^{4}y\;
       \Bigl[
            \tfrac14\, J(\vec{x})\,G(\vec{x}-\vec{y})\,J(\vec{y})
          + \tfrac{e^{2}}2\,j^{\,i}(\vec{x})\,G_{ij}(\vec{x}-\vec{y})\,j^{\,j}(\vec{y})
       \Bigr]
   \Biggr\}~.
\end{equation}
For static strings the time integrals yield a factor~$T$, and the Green’s
functions reduce to their two‑dimensional forms
\begin{equation}
G(\vec{R})=\frac{1}{2\pi}K_{0}(m_{H}|\vec{R}|)~,\qquad
G_{ij}(\vec{R})=\frac{1}{2\pi}\Bigl(\delta_{ij}-\frac{\partial_{i}\partial_{j}}{m_{V}^{2}}\Bigr)
                 K_{0}(m_{V}|\vec{R}|)~.
\end{equation}
Inserting the sources~\eqref{source} of the two strings (one at the origin, one at~$\vec{R}$) then yields,
\begin{equation}
Z[J,j^{\,i}]
   =\exp\Bigl\{
       -T\bigl[E_{\text{self}}
               -\tfrac{c^{2}}{8\pi}K_{0}(m_{H}R)
               +\tfrac{e^{2}d^{2}m_{V}^{2}}{4\pi}K_{0}(m_{V}R)
       \bigr]
   \Bigr\}   ~,
\end{equation}
so that the interaction energy of the two strings is
\begin{equation}
V_{\text{int}}(R)=-\frac{c^{2}}{8\pi}K_{0}(m_{H}R)
     +\frac{e^{2}d^{2}m_{V}^{2}}{4\pi}K_{0}(m_{V}R)~.
\end{equation}
The self-energy, which is independent of $\vec{R}$, is denoted by~$E_{\text{self}}$.  Using~\eqref{cd}, we recover the expression~\eqref{AB} and hence~\eqref{eq:int}. We note that the exchange of the Higgs field leads to an attractive force between the two strings, while the exchange of massive vector bosons leads to a repulsive force. Note that when~$\beta>4$, the asymptotics for the Higgs field are different. (See the discussion around~\eqref{eq:ahmasym}.) In this regime the interaction energy is always dominated by vector-boson exchange (see also~\cite{Manton:2004tk}). 

For completeness, the fields generated by the sources at a point~$\vec{x}$ in the~$xy-$plane are
\begin{equation}
\varphi(\vec{x})=
   -\frac{c}{4\pi v}\Big(K_{0}(m_{H}|\vec{x}|) +K_{0}(m_{H}|\vec{x}-\vec{R}|)\Big)~,
\end{equation}
\begin{equation}
a_{\theta}(\vec{x})=
   \frac{e^{2}d\,m_{V}}{2\pi}
   \Big( |\vec{x}|K_{1}(m_{V}|\vec{x}|)
        +|\vec{x}-\vec{R}|~K_{1}(m_{V}|\vec{x}-\vec{R}|\Big)~.
\end{equation}
When compared to the asymptotic forms in~\eqref{app:asym}, we see that at any point~$\vec{x}$ in the transverse~$xy$-plane, each field is a linear superposition of the contribution from each individual string.

\newpage

\bibliographystyle{utphys.bst}
\bibliography{bibliography.bib}

@article{Linde:1975sw,
    author = "Linde, Andrei D.",
    editor = "Einhorn, M. B.",
    title = "{Dynamical Symmetry Restoration and Constraints on Masses and Coupling Constants in Gauge Theories}",
    reportNumber = "LEBEDEV-75-123",
    journal = "JETP Lett.",
    volume = "23",
    pages = "64--67",
    year = "1976"
}

@article{Weinberg:1976pe,
    author = "Weinberg, Steven",
    title = "{Mass of the Higgs Boson}",
    doi = "10.1103/PhysRevLett.36.294",
    journal = "Phys. Rev. Lett.",
    volume = "36",
    pages = "294--296",
    year = "1976"
}

@article{Dumitrescu:2011iu,
    author = "Dumitrescu, Thomas T. and Seiberg, Nathan",
    title = "{Supercurrents and Brane Currents in Diverse Dimensions}",
    eprint = "1106.0031",
    archivePrefix = "arXiv",
    primaryClass = "hep-th",
    reportNumber = "PUPT-2372",
    doi = "10.1007/JHEP07(2011)095",
    journal = "JHEP",
    volume = "07",
    pages = "095",
    year = "2011"
}

@inproceedings{Cordova:2022ruw,
    author = "Cordova, Clay and Dumitrescu, Thomas T. and Intriligator, Kenneth and Shao, Shu-Heng",
    title = "{Snowmass White Paper: Generalized Symmetries in Quantum Field Theory and Beyond}",
    booktitle = "{Snowmass 2021}",
    eprint = "2205.09545",
    archivePrefix = "arXiv",
    primaryClass = "hep-th",
    month = "5",
    year = "2022"
}

@article{McGreevy:2022oyu,
    author = "McGreevy, John",
    title = "{Generalized Symmetries in Condensed Matter}",
    eprint = "2204.03045",
    archivePrefix = "arXiv",
    primaryClass = "cond-mat.str-el",
    doi = "10.1146/annurev-conmatphys-040721-021029",
    journal = "Ann. Rev. Condensed Matter Phys.",
    volume = "14",
    pages = "57--82",
    year = "2023"
}

@book{Shifman:2012zz,
    author = "Shifman, M.",
    title = "{Advanced topics in quantum field theory.}: {A lecture course}",
    isbn = "978-1-139-21036-2, 978-0-521-19084-8",
    publisher = "Cambridge Univ. Press",
    address = "Cambridge, UK",
    month = "2",
    year = "2012"
}

@article{Lee:1990it,
    author = "Lee, Choon-kyu and Lee, Ki-Myeong and Weinberg, Erick J.",
    title = "{Supersymmetry and Selfdual {Chern-Simons} Systems}",
    reportNumber = "CU-TP-459",
    doi = "10.1016/0370-2693(90)90964-8",
    journal = "Phys. Lett. B",
    volume = "243",
    pages = "105--108",
    year = "1990"
}

@article{Dasgupta:1981zz,
    author = "Dasgupta, C. and Halperin, B. I.",
    title = "{Phase Transition in a Lattice Model of Superconductivity}",
    doi = "10.1103/PhysRevLett.47.1556",
    journal = "Phys. Rev. Lett.",
    volume = "47",
    pages = "1556--1560",
    year = "1981"
}

@article{Cuomo:2022kio,
    author = "Cuomo, Gabriel and Komargodski, Zohar",
    title = "{Giant Vortices and the Regge Limit}",
    eprint = "2210.15694",
    archivePrefix = "arXiv",
    primaryClass = "hep-th",
    doi = "10.1007/JHEP01(2023)006",
    journal = "JHEP",
    volume = "01",
    pages = "006",
    year = "2023"
}

@article{Cuomo:2023vvd,
    author = "Cuomo, Gabriel and Komargodski, Zohar and Zhong, Siwei",
    title = "{Chiral modes of giant superfluid vortices}",
    eprint = "2312.06095",
    archivePrefix = "arXiv",
    primaryClass = "cond-mat.quant-gas",
    doi = "10.1103/PhysRevB.110.144514",
    journal = "Phys. Rev. B",
    volume = "110",
    number = "14",
    pages = "144514",
    year = "2024"
}

@article{Peskin:1977kp,
    author = "Peskin, Michael E.",
    title = "{Mandelstam 't Hooft Duality in Abelian Lattice Models}",
    reportNumber = "HUTP-77/A083",
    doi = "10.1016/0003-4916(78)90252-X",
    journal = "Annals Phys.",
    volume = "113",
    pages = "122",
    year = "1978"
}

@article{Coleman:1985ki,
    author = "Coleman, Sidney R.",
    title = "{Q-balls}",
    reportNumber = "HUTP-85/A050",
    doi = "10.1016/0550-3213(86)90520-1",
    journal = "Nucl. Phys. B",
    volume = "262",
    number = "2",
    pages = "263",
    year = "1985",
    note = "[Addendum: Nucl.Phys.B 269, 744 (1986)]"
}

@article{Berry:1972na,
    author = "Berry, Michael V. and Mount, K. E.",
    title = "{Semiclassical approximations in wave mechanics}",
    doi = "10.1088/0034-4885/35/1/306",
    journal = "Rept. Prog. Phys.",
    volume = "35",
    pages = "315",
    year = "1972"
}

@article{Jackiw:1991je,
    author = "Jackiw, R.",
    title = "{Delta function potentials in two-dimensional and three-dimensional quantum mechanics}",
    reportNumber = "MIT-CTP-1937",
    pages = "35--53",
    month = "1",
    year = "1991"
}

@article{Jackiw:1990pr,
    author = "Jackiw, R. and Lee, Ki-Myeong and Weinberg, Erick J.",
    title = "{Selfdual Chern-Simons solitons}",
    reportNumber = "CU-TP-473, MIT-CTP-1876, BUHEP-90-20",
    doi = "10.1103/PhysRevD.42.3488",
    journal = "Phys. Rev. D",
    volume = "42",
    pages = "3488--3499",
    year = "1990"
}

@article{Jacobs:1978ch,
    author = "Jacobs, Laurence and Rebbi, Claudio",
    title = "{Interaction Energy of Superconducting Vortices}",
    reportNumber = "Print-78-1156 (BNL), BNL-25246",
    doi = "10.1103/PhysRevB.19.4486",
    journal = "Phys. Rev. B",
    volume = "19",
    pages = "4486--4494",
    year = "1979"
}

@article{Abrikosov:1956sx,
    author = "Abrikosov, A. A.",
    title = "{On the Magnetic properties of superconductors of the second group}",
    journal = "Sov. Phys. JETP",
    volume = "5",
    pages = "1174--1182",
    year = "1957"
}

@article{Nielsen:1973cs,
    author = "Nielsen, Holger Bech and Olesen, P.",
    editor = "Taylor, J. C.",
    title = "{Vortex Line Models for Dual Strings}",
    doi = "10.1016/0550-3213(73)90350-7",
    journal = "Nucl. Phys. B",
    volume = "61",
    pages = "45--61",
    year = "1973"
}

@article{Bettencourt_1995,
	doi = {10.1103/physrevd.51.1842},
  
	url = {https://doi.org/10.1103%2Fphysrevd.51.1842},
  
	year = 1995,
	month = {feb},
  
	publisher = {American Physical Society ({APS})},
  
	volume = {51},
  
	number = {4},
  
	pages = {1842--1853},
  
	author = {L. M. A. Bettencourt and R. J. Rivers},
  
	title = {Interactions between U(1) cosmic strings: An analytical study},
  
	journal = {Physical Review D}
}

@article{Speight_1997,
	doi = {10.1103/physrevd.55.3830},
  
	url = {https://doi.org/10.1103%2Fphysrevd.55.3830},
  
	year = 1997,
	month = {mar},
  
	publisher = {American Physical Society ({APS})},
  
	volume = {55},
  
	number = {6},
  
	pages = {3830--3835},
  
	author = {J. M. Speight},
  
	title = {Static intervortex forces},
  
	journal = {Physical Review D}
}

@article{PhysRevLett.125.251601,
  title = {What Becomes of Giant Vortices in the Abelian Higgs Model},
  author = {Penin, Alexander A. and Weller, Quinten},
  journal = {Phys. Rev. Lett.},
  volume = {125},
  issue = {25},
  pages = {251601},
  numpages = {5},
  year = {2020},
  month = {Dec},
  publisher = {American Physical Society},
  doi = {10.1103/PhysRevLett.125.251601},
  url = {https://link.aps.org/doi/10.1103/PhysRevLett.125.251601}
}

@article{Penin_2021,
	doi = {10.1007/jhep08(2021)056},
  
	url = {https://arxiv.org/pdf/2105.12137.pdf},
  
	year = 2021,
	month = {aug},
  
	publisher = {Springer Science and Business Media {LLC}
},
  
	volume = {2021},
  
	number = {8},
  
	author = {Alexander A. Penin and Quinten Weller},
  
	title = {A theory of giant vortices},
  
	journal = {Journal of High Energy Physics}
}

@article{Bolognesi_2006,
   title={Multi-vortices are wall vortices: A numerical proof},
   volume={741},
   ISSN={0550-3213},
   url={http://dx.doi.org/10.1016/j.nuclphysb.2006.01.038},
   DOI={10.1016/j.nuclphysb.2006.01.038},
   number={1–2},
   journal={Nuclear Physics B},
   publisher={Elsevier BV},
   author={Bolognesi, Stefano and Gudnason, Sven Bjarke},
   year={2006},
   month=may, pages={1–16} }

@article{Bolognesi_2005,
   title={Domain walls and flux tubes},
   volume={730},
   ISSN={0550-3213},
   url={http://dx.doi.org/10.1016/j.nuclphysb.2005.09.032},
   DOI={10.1016/j.nuclphysb.2005.09.032},
   number={1–2},
   journal={Nuclear Physics B},
   publisher={Elsevier BV},
   author={Bolognesi, Stefano},
   year={2005},
   month=dec, pages={127–149} }

@book{Jaffe:1980mj,
    author = "Jaffe, Arthur M. and Taubes, Clifford Henry",
    title = "{VORTICES AND MONOPOLES. STRUCTURE OF STATIC GAUGE THEORIES}",
    year = "1980"
}

@article{Bogomolny:1975de,
    author = "Bogomolny, E. B.",
    title = "{Stability of Classical Solutions}",
    reportNumber = "PRINT-76-0543 (LANDAU-INST.)",
    journal = "Sov. J. Nucl. Phys.",
    volume = "24",
    pages = "449",
    year = "1976"
}

@misc{gustafson1999stabilitymagneticvortices,
      title={The Stability of Magnetic Vortices}, 
      author={S. Gustafson and I. M. Sigal},
      year={1999},
      eprint={math/9904158},
      archivePrefix={arXiv},
      primaryClass={math.AP},
      url={https://arxiv.org/abs/math/9904158}, 
}

@book{Manton:2004tk,
    author = "Manton, N. S. and Sutcliffe, P.",
    title = "{Topological solitons}",
    doi = "10.1017/CBO9780511617034",
    isbn = "978-0-521-04096-9, 978-0-521-83836-8, 978-0-511-20783-9",
    publisher = "Cambridge University Press",
    series = "Cambridge Monographs on Mathematical Physics",
    year = "2004"
}

@article{Weinberg:1979er,
    author = "Weinberg, Erick J.",
    title = "{Multivortex Solutions of the Ginzburg-landau Equations}",
    reportNumber = "CU-TP-144",
    doi = "10.1103/PhysRevD.19.3008",
    journal = "Phys. Rev. D",
    volume = "19",
    pages = "3008",
    year = "1979"
}

@article{Taubes:1979tm,
    author = "Taubes, Clifford Henry",
    title = "{Arbitrary N: Vortex Solutions to the First Order Landau-Ginzburg Equations}",
    reportNumber = "HUTMP 79/B73",
    doi = "10.1007/BF01197552",
    journal = "Commun. Math. Phys.",
    volume = "72",
    pages = "277--292",
    year = "1980"
}

@article{Tong:2002rq,
    author = "Tong, David",
    title = "{NS5-branes, T duality and world sheet instantons}",
    eprint = "hep-th/0204186",
    archivePrefix = "arXiv",
    reportNumber = "MIT-CTP-3266",
    doi = "10.1088/1126-6708/2002/07/013",
    journal = "JHEP",
    volume = "07",
    pages = "013",
    year = "2002"
}

@article{Taubes:1979ps,
    author = "Taubes, Clifford Henry",
    title = "{On the Equivalence of the First and Second Order Equations for Gauge Theories}",
    reportNumber = "HUTMP 79/B78",
    doi = "10.1007/BF01212709",
    journal = "Commun. Math. Phys.",
    volume = "75",
    pages = "207",
    year = "1980"
}

@article{Perivolaropoulos_1993,
   title={Asymptotics of Nielsen-Olesen vortices},
   volume={48},
   ISSN={0556-2821},
   url={http://dx.doi.org/10.1103/PhysRevD.48.5961},
   DOI={10.1103/physrevd.48.5961},
   number={12},
   journal={Physical Review D},
   publisher={American Physical Society (APS)},
   author={Perivolaropoulos, Leandros},
   year={1993},
   month=dec, pages={5961–5962} }

@article{DGL,
 author={Dumitrescu, T. and Gaikwad, A. and Li, Y},
  title   = {To appear},
year={2025}
}

@article{DG,
 author={Dumitrescu, T. and Gaikwad, A. },
  title   = {To appear},
year={2025}
}

@book{Weinberg:2012pjx,
    author = "Weinberg, Erick J.",
    title = "{Classical solutions in quantum field theory}: {Solitons and Instantons in High Energy Physics}",
    doi = "10.1017/CBO9781139017787",
    isbn = "978-0-521-11463-9, 978-1-139-57461-7, 978-0-521-11463-9, 978-1-107-43805-7",
    publisher = "Cambridge University Press",
    series = "Cambridge Monographs on Mathematical Physics",
    month = "9",
    year = "2012"
}

@article{PhysRevB.83.020503,
  title = {Giant vortices, rings of vortices, and reentrant behavior in type-1.5 superconductors},
  author = {Dao, V. H. and Chibotaru, L. F. and Nishio, T. and Moshchalkov, V. V.},
  journal = {Phys. Rev. B},
  volume = {83},
  issue = {2},
  pages = {020503},
  numpages = {4},
  year = {2011},
  month = {Jan},
  publisher = {American Physical Society},
  doi = {10.1103/PhysRevB.83.020503},
  url = {https://link.aps.org/doi/10.1103/PhysRevB.83.020503}
}

@article{Gaiotto:2014kfa,
    author = "Gaiotto, Davide and Kapustin, Anton and Seiberg, Nathan and Willett, Brian",
    title = "{Generalized Global Symmetries}",
    eprint = "1412.5148",
    archivePrefix = "arXiv",
    primaryClass = "hep-th",
    doi = "10.1007/JHEP02(2015)172",
    journal = "JHEP",
    volume = "02",
    pages = "172",
    year = "2015"
}

@article{Babaev_2017,
   title={Type-1.5 superconductivity in multicomponent systems},
   volume={533},
   ISSN={0921-4534},
   url={http://dx.doi.org/10.1016/j.physc.2016.08.003},
   DOI={10.1016/j.physc.2016.08.003},
   journal={Physica C: Superconductivity and its Applications},
   publisher={Elsevier BV},
   author={Babaev, E. and Carlström, J. and Silaev, M. and Speight, J.M.},
   year={2017},
   month=feb, pages={20–35} }

@article{Speight_2021,
   title={Intervortex forces in competing-order superconductors},
   volume={103},
   ISSN={2469-9969},
   url={http://dx.doi.org/10.1103/PhysRevB.103.014514},
   DOI={10.1103/physrevb.103.014514},
   number={1},
   journal={Physical Review B},
   publisher={American Physical Society (APS)},
   author={Speight, Martin and Winyard, Thomas},
   year={2021},
   month=jan }

@article{Babaev_2012,
   title={Type-1.5 superconductivity in multiband systems: Magnetic response, broken symmetries and microscopic theory – A brief overview},
   volume={479},
   ISSN={0921-4534},
   url={http://dx.doi.org/10.1016/j.physc.2012.01.002},
   DOI={10.1016/j.physc.2012.01.002},
   journal={Physica C: Superconductivity},
   publisher={Elsevier BV},
   author={Babaev, E. and Carlström, J. and Garaud, J. and Silaev, M. and Speight, J.M.},
   year={2012},
   month=sep, pages={2–14} }

@article{Carlstr_m_2011,
   title={Type-1.5 superconductivity in multiband systems: Effects of interband couplings},
   volume={83},
   ISSN={1550-235X},
   url={http://dx.doi.org/10.1103/PhysRevB.83.174509},
   DOI={10.1103/physrevb.83.174509},
   number={17},
   journal={Physical Review B},
   publisher={American Physical Society (APS)},
   author={Carlström, Johan and Babaev, Egor and Speight, Martin},
   year={2011},
   month=may }

@article{Babaev_2005,
   title={Semi-Meissner state and neither type-I nor type-II superconductivity in multicomponent superconductors},
   volume={72},
   ISSN={1550-235X},
   url={http://dx.doi.org/10.1103/PhysRevB.72.180502},
   DOI={10.1103/physrevb.72.180502},
   number={18},
   journal={Physical Review B},
   publisher={American Physical Society (APS)},
   author={Babaev, Egor and Speight, Martin},
   year={2005},
   month=nov }

@article{Timoshuk_2024,
   title={Microscopic solutions for vortex clustering in two-band type-1.5 superconductors},
   volume={110},
   ISSN={2469-9969},
   url={http://dx.doi.org/10.1103/PhysRevB.110.064509},
   DOI={10.1103/physrevb.110.064509},
   number={6},
   journal={Physical Review B},
   publisher={American Physical Society (APS)},
   author={Timoshuk, Igor and Babaev, Egor},
   year={2024},
   month=aug }

@article{Plohr:1981cy,
    author = "Plohr, B.",
    title = "{The Behavior at Infinity of Isotropic Vortices and Monopoles}",
    doi = "10.1063/1.524774",
    journal = "J. Math. Phys.",
    volume = "22",
    pages = "2184--2190",
    year = "1981"
}

\end{document}